\documentclass{amsart}
\usepackage{amssymb}

\swapnumbers
\theoremstyle{plain}
\newtheorem{thm}{Theorem}[section]
\newtheorem{lem}[thm]{Lemma}
\newtheorem{prop}[thm]{Proposition}
\newtheorem{cor}[thm]{Corollary}

\newtheorem*{OpQu*}{Open question}
\theoremstyle{definition}
\newtheorem{defn}[thm]{Definition}
\theoremstyle{remark}
\newtheorem{note}[thm]{Note}
\newtheorem*{note*}{Note}
\newtheorem{exmp}[thm]{Example}
\newtheorem*{exmp*}{Example}

\numberwithin{equation}{thm}

\DeclareMathOperator{\ord}{ord}
\DeclareMathOperator{\wt}{wt}

\DeclareMathOperator{\Coef}{Coef}

\DeclareMathOperator*{\Wr}{\wr}

\DeclareMathOperator{\XOR}{\scriptstyle{\mathsf{XOR}}}
\DeclareMathOperator{\OR}{\scriptstyle{\mathsf {OR}}}
\DeclareMathOperator{\AND}{\scriptstyle{\mathsf {AND}}}
\DeclareMathOperator{\NEG}{\scriptstyle{\mathsf {NEG}}}

\newcommand{\Z}{\mathbb Z}
\newcommand{\Q}{\mathbb Q}
\newcommand{\N}{\mathbb N}

\renewcommand{\:}{\colon}
\renewcommand{\>}{\rightarrow}

\usepackage[backref,
pagebackref,%
bookmarks=true,%
colorlinks=true]%
{hyperref}

\hypersetup{%
pdfauthor={Vladimir Anashin},
pdftitle={Pseudorandom generation with p-adic ergodic
transformations}
pdfsubject={Pseudorandom generators}
pdfkeywords={pseudorandom, p-adic, ergodic}}
\begin{document}

\hyphenation{appli-cat-ions cryp-to-gra-phy com-ple-xi-ty com-po-si-ti-ons 
dis-tan-ce ad-di-ti-on in-effect-ive multi-pli-cat-ion con-junct-ion 
com-pos-it-ion funct-ions Mau-rer ge-ne-ra-li-z-ed equi-pro-b-ab-le}
%
%
\def\huh{\hbox{\vrule width 2pt height 8pt depth 2pt}}
\def\eqnum#1{\eqno (#1)}
\def\cwdash{\relbar\joinrel}
\def\fnote#1{\footnote}

\title[Pseudorandom generators] {Pseudorandom number generation by $p$-adic ergodic
transformations}
\author{Vladimir Anashin}

\address{Faculty of Information Security, 
Russian State University for the Humanities,\\
Kirovogradskaya Str., 25/2, Moscow 113534, Russia}

\email{anashin@rsuh.ru, vladimir@anashin.msk.su}

\begin{abstract}
The paper study counter-dependent pseudorandom generators;
the latter are generators such that their state
transition function (and output function) is being modified 
dynamically while working:
For such a generator
the recurrence sequence
of states satisfies a congruence
$x_{i+1}\equiv f_i(x_i)\pmod{2^n}$, while its output sequence is of the
form $z_{i}=F_i(u_i)$. The paper introduces techniques and constructions
that enable one to compose
generators that output uniformly distributed sequences
of a maximum period length and with high linear and $2$-adic spans. The corresponding
stream chipher is provably strong against a known plaintext attack (up to
a
plausible conjecture).
Both state
transition function and output function could be key-dependent, so the only
information available to a cryptanalyst is
that these functions belong to some (exponentially large) class.
These functions are compositions of standard machine instructions (such
as addition, multiplication, bitwise logical operations, etc.) The compositions
should satisfy rather loose conditions; so the corresponding generators are 
flexible enough
and could be easily implemented as computer programs.
\end{abstract}

\maketitle

\section {Introduction}
\label{Intro}

The study of ergodic, measure-preserving and equiprobable functions on
the space $\mathbb{Z}_p$ of $p$-adic integers in ~\cite{me-1, me-2, me-conf,
me-ex} was mainly motivated
by possible applications to pseudorandom
number generation for cryptography and simulation. In the present paper
we
consider generators based on these functions, {\slshape prove} that the produced
sequences have
some  (properly defined below) `features of randomness', and calculate {\slshape exact}
values
of certain (crucial for cryptographic security) parameters of these generators. 
Namely, we characterize
all possible output sequences in the class of all sequences, calculate
exact lengths of their periods, distribution of overlapping and non-overlapping
$k$-tuples, linear complexity, and $p$-adic span. Also, we demonstrate
that with the use of these functions it is possible to construct
a stream cipher such that to recover its key is an infeasible problem
(up to some plausible conjectures). 


In fact, the paper introduces certain techniques and constructions that
enable one to design
stream ciphers with both state transition and output functions depending on
key; yet {\slshape independently of key choice} the corresponding 
generator always provides predefined 
values of output sequence parameters, which are
mentioned above. These functions are (key-dependent) compositions of (standard) 
machine instructions:
arithmetic ones, such as addition and
multiplication (exponentiation and raising to negative powers as well), 
logical ones, such as $\XOR$, $\OR$, $\AND$, $\NEG$, etc., and others
(e.g., shifts, masking). Thus, generators of this kind admit quite natural implementation
as a computer program. Such generators are rather flexible: To obtain due performance
a programmer could vary length of the composition and choice of
machine instructions {\slshape without} affecting the above mentioned probabilistic
and cryptographic characteristics.   

Further, focusing on these ideas we introduce counter-dependent generators;
the latter are generators such that their state
transition function (and output function) is being modified dynamically while working.
To be more exact, for these generators
the recurrence sequence
of states satisfies a congruence
$x_{i+1}\equiv f_i(x_i)\pmod{2^n}$, while their output sequence is of the
form $z_{i}=F_i(u_i)$. Note that both state transition function $f_i$ and output
function $F_i$ depend on the number $i$ of a step; yet newertheless the output
sequence is purely periodic, its period length is a multiple of $2^n$,
distribution of  $k$-tuples, $k\le n$ is uniform, its linear complexity is
high, etc.
Moreover, not only $f_i$ and $F_i$ themselves
could be keyed, but also the order they are used during encryption.
\footnote{The notion of a counter-dependent generator was 
originally introduced in \cite{ShTs}. However, in
our paper we consider this notion in a broader  sense: In our counter-dependent
generators
not only the state
transition function, but also the output function depends on $i$. Moreover,
in \cite{ShTs} only a particular case of counter-dependent generators
is studied; namely, counter-assisted generators and their cascaded and two-step
modifications. A state transition function of a counter-assisted generator  is
of the form $f_i(x)=i\star
h(x)$, where $\star$ is a binary quasigroup operation (in particular, group
operation, e.g., $+$ or
$\XOR$), and $h(x)$ does not depend on $i$. 
An output function of a counter-assisted generator does not depend
on $i$ either. The main security notion studied in \cite{ShTs} is diversity, which
generalizes a concept of long cycles. Note that all our generators achieve
maximum possible total diversity, which is equal to the order of the output set.} 

To give an idea of how these schemes look like, consider the following
example of a counter-dependent generator modulo $2^n$. Take arbitrary $m\equiv3\pmod 4$,
then take $m$ {\slshape arbitrary} compositions $v_0(x),\ldots,v_{m-1}(x)$ of the above
mentioned machine instructions (addition, multiplication, $\XOR$, $\AND$,
etc.) and constants, then take another $m$ {\slshape arbitrary} compositions 
$w_0(x),\ldots,w_{m-1}(x)$ of this kind. Arrange two arrays $V$ and $W$
writing these $v_j(x)$ and $w_j(x)$ to memory in {\slshape arbitrary} order. 
Now choose arbitrary $x_0\in\{0,1,\ldots 2^n-1\}$ as a seed. The generator
calculates the recurrence
sequence of states $x_{i+1}=(i+x_i+2\cdot(v_i(x_i+1)-v_i(x_i)))\bmod 2^n$
and outputs the sequence 
$z_{i}=(1+\pi(x_i)+2\cdot (w_i(\pi(x_i+1))-w_i(\pi(x_i))))\bmod 2^n$,
where $\pi$ is a bit order reversing permutation, which reads an $n$-bit
number
$z\in\{0,1,\ldots, 2^{n}-1\}$ in a reverse bit order; e.g., $\pi(0)=0, \pi(1)=2^{n-1},
\pi(2)=2^{n-2},\pi(3)=2^{n-2}+2^{n-1}$, etc. Then the sequence $\{x_i\}$
is a purely periodic sequence of period length $2^nm$ of $n$-bit numbers, and each
number of $\{0,1,\ldots, 2^{n}-1\}$ occurs at the period exactly $m$ times.
Moreover, if we consider $\{x_i\}$ as a binary sequence of period length
$2^nmn$, then the frequency each $k$-tuple ($0< k\le n$) occurs in the
sequence is exactly $\frac{1}{2^k}$. The output sequence $\{z_i\}$ is also
purely
periodic of period length $2^nm$, and each number of $\{0,1,\ldots, 2^{n}-1\}$
occurs at the period exactly $m$ times either. Moreover, 
every binary sequence obtained by reading each $s$\textsuperscript{th}
bit $\delta_s(z_i)$  ($0\le s\le n-1$) of the output sequence is purely
periodic; its  period length is a multiple of $2^n$, hence its linear complexity
(as well as the one of the whole sequence $\{z_i\}$)
exceeds $2^{n-1}$.

In fact, for such stream encryption schemes the only information available
to a cryptanalist is
that both the output and the state transition functions
belong to a certain (exponentially large) class of functions, and practically
nothing
more. 
Thus, practical attacks to such stream encryption scheme seem to be ineffective.

We must immediately note here that, strictly speaking, all these results
give some evidence, yet {\slshape not the proof} of cryptographic security of
these ciphers. We recall, however, that today for no stream cipher based on
deterministic algorithm there exists an unconditional mathematical proof of security.
We ought to emphasize also that  the study of stream encryption schemes 
below should not be considered as an exaustive cryptographic analysis.
The latter one implies a study of attacks against a particular scheme, which numerical
parameters have exact predefined values. Loosely speaking, further
results could be considered as a `toolkit' for a stream cipher designer,
but {\slshape not} as 'make-it-yourself kit': The latter implies detailed
`assemble instructions'; following them guarantees an adequate quality of the whole
thing. No such instructions are given in the present paper, only some ideas
and hints.

The paper is organized as follows:
\begin{itemize}
\item In Section \ref{Prelm} we introduce some basic notions, consider
standard machine instructions as continous $2$-adic mappings, describe their
properties and prove that under certain very loose conditions the output
sequence will be uniformly distributed.
\item In Section \ref{sec:Tool} we state a number of results that enable
one to construct permutations with a single cycle and equiprobable functions
out of standard machine instructions. Moreover, as examples of how these techniques
work we reprove some of 
known results in this area, as well as establish new ones.
\item In Section \ref{sec:Constr} we outline several ways of combining
functions described in Section \ref{sec:Tool} 
in automaton that generates uniformly distributed sequence.
There we introduce a new construction (called wreath product of automata,
by analogy with a corresponding group theory construction)
that enables one to build counter-dependent generators with uniformly
distributed output sequences of a maximum period length.
\item  In Section \ref{sec:Prop} we study complexity and distribution
of output sequences
of automata introduced in Section \ref{sec:Constr}: Linear and $2$-adic spans
of these sequences, 
their structure, distribution of $k$-tuples in them, etc. In particular, we prove that distribution
of (overlapping) $k$-tuples is strictly uniform; namely, that these
output sequences have a property that could be called a generalized
De Bruijn: Being considered as binary sequences, they
are purely periodic, their period lengths are multiples of $2^n$, and each
$k$-tuple ($k\le n$) occurs at the period the same number of times. From
here we deduce
that a large class of
these sequences satisfy Knuth's criterion Q1
\footnote{See \cite[Section 3.5, Definition Q1]{Knuth} } of randomness.
\item In Section \ref{sec:Predict} we demonstrate how to construct a stream
cipher with intractable key recovery problem conjecturing that a set
of $k$
multivariate Boolean polynomials define a one-way function (it is known that to determine
whether a system of $k$ Boolean polynomials in $n$ variables has a common
zero
is an NP-complete
problem \footnote{See e.g. \cite[Appendix
A, Section A7.2, Problem ANT-9]{GJ}}). 

\end{itemize}


\section{Preliminaries}
\label{Prelm}    

Basically, a generator we consider in the paper is a finite automaton
${\mathfrak A}=\langle N,M,f,F,u_0\rangle $ with a finite state set $N$, state
transition function
$f:N\rightarrow N$, finite output alphabet $M$, output function  $F:N\rightarrow M$
and an initial state (seed) $u_0\in N$. Thus, this generator produces a sequence
$$\mathcal S=\{F(u_0), F(f(u_0)), F(f^{(2)}(u_0)),\ldots, F(f^{(j)}(u_0)),\ldots\}$$ 
over
the set $M$, where 
$$f^{(j)}(u_0)=\underbrace{f(\ldots f(}_{j
\;\text{times}}u_0)\ldots)\ \ (j=1,2,\ldots);\quad f^{(0)}(u_0)=u_0.$$ 
Automata of the form $\mathfrak
A$ will be considered either as pseudorandom generators per se, or as components
of more complicated pseudorandom generators, which are introduced in Section
\ref{sec:Constr}; the latter produce pseudorandom
sequences 
$\{z_0,z_1,z_2,\ldots\}$ over $M$ according to the rule
$$z_0=F_0(u_0),u_1=f_0(u_0);\ldots
z_{i}=F_i(u_i), u_{i+1}=f_i(u_i);\ldots$$ 
That is, at the $(i+1)$\textsuperscript{th} step the automaton 
$\mathfrak A_i=\langle N,M,f_i,F_i,u_i\rangle $
is applied to the state $u_i\in N$, producing a new state $u_{i+1}=f_i(u_i)\in
N$, and
outputting a symbol $z_{i}=F_i(u_i)\in M$. 

Quite often in the paper we assume 
that  $N=\mathbb I_n(p)=
\{0,1,\ldots,p^n\nobreak-\nobreak
1\}$, $M=\mathbb I_m(p)$,  $m\le n$,  where $p$ is (usually a 
prime) positive rational integer greater than 1. 
Moreover, mainly we are focused on the case $p=2$ as the
most convenient for computer implementations, and use a shorter notation
$\mathbb I_n$ instead of $\mathbb I_n(2)$. As a rule, further we formulate results 
mainly for this case, making brief remarks for those of them that remain true
for arbitrary $p$.    

Now let $n=km>1$ (may be, $k=1$) be a positive rational integer. Let the
state set $N$ of the above mentioned automaton $\mathfrak A$ be 
$\mathbb I_n=\{0,1,\ldots,2^n\nobreak-\nobreak
1\}$. Further we will identify the set ${\mathbb I}_n$ either with the set of
all elements of the residue class ring $\mathbb Z/2^n$ of integers modulo $2^n$, 
or with a set $\mathbb
W_n(2)$ of all $n$-bit words in the alphabet ${\mathbb I}=\mathbb I_1=\{0,1\}$, 
or with
a set of all elements of a direct product
$$(\mathbb Z/2^m)^{(k)}=\underbrace{\mathbb Z/2^m\times\cdots\times  \mathbb Z/2^m}_{k
\;\text{times}}$$ 
of $k$ copies of the residue class ring $\mathbb Z/2^m$, or with a set $\mathbb W_k(2^m)$ of all words of length
$k$ in the alphabet $\mathbb I_m$. In other words, if necessary, we may treat
a number $i\in\{0,1,\ldots, 2^n-1\}$ either as an $n$-bit word, or as a $k$-tuple
of numbers of $\{0,1,\ldots,2^m-1\}$, or as a $k$-tuple of $m$-bit blocks. 

To be more exact, let $\delta_j^{m}(i)\in\mathbb I_m$ be the $j$\textsuperscript{th} digit of
a number $i$ in its base-$2^m$ expansion: that is, if
$i=i_0+i_1\cdot 2^m +i_2\cdot (2^m)^2+\ldots$, where $i_j\in\mathbb I_m$,
$j=0,1,2,\ldots$, then, by definition, $\delta_j^{m}(i)=i_j$. (For $m=1$
we usually 
omit the superscript, when this does not lead to misunderstanding).
With these notations, if $i\in\mathbb I_n$, then the word
$w_k(i)\in\mathbb W_k(2^m)$ is a concatent
$\delta_0^{m}(i)\ldots\delta_{k-1}^{m}(i)$, and a corresponding element
$r_k(i)\in(\mathbb Z/2^m)^{(k)}$ is 
$r_k(i)=(\delta_0^{m}(i),\ldots,\delta_{k-1}^{m}(i))$.
Thus, for each
$i\in\mathbb I_n$ and for arbitrary mappings
$F:(\mathbb Z/2^m)^{(k)}\rightarrow \mathbb Z/2^m$ and $G:\mathbb W_n(2)\rightarrow \mathbb
W_k(2^m)$ the expressions $F(i)$ and $G(i)$ are correctly defined: namely, 
$F(i)$ stands for $F(r_k(i))$, $G(i)$ stands for $G(w_k(i))$. In view of the above mentioned bijections
between $\mathbb I_m$ and $\mathbb Z/2^m$, both $F(i)$ and $G(i)$ may be considered
as elements of $\mathbb I_m$ and $\mathbb I_n$, respectively.

We will need a particular mapping
$\pi_s^t :\mathbb W_s(2^t)\rightarrow \mathbb W_s(2^t)$, an order reversing
permutation:
$\pi_s^t(u_0u_1\ldots
u_{s-1})=u_{s-1}u_{s-2}\ldots u_0$, where $u_0,\ldots,u_{s-1}\in\mathbb I_t$.
In view of the above conventions, for each $i\in\mathbb I(2^n)$ 
the following expressions are well
defined: 
$\pi_k^m(i),\pi_n^1(i)\in\mathbb I_n$ and $\pi_m^1(\delta_j^{m}(i))\in\mathbb I_m$.
In other words, $\pi_n^1(i)$ reads base-2 expansion of $i$  in reverse order,
while $\pi_k^m(i)$ reads base-$2^m$ expansion of $i$ in reverse order; e.g. 
$\pi_4^1(7)=14$, $\pi_2^2(7)=13$. Often, when it is clear within
a context, we omit a superscript (sometimes together with a subscript) in $\pi_k^m$.  

Note that functions $\pi_k^m,\pi_n^1, \delta_j^{m}$, being compositions
of arithmetic and logical operators, are easily programmable: so
$\delta_j^{m}(i)=\frac{i\AND(2^{mj}(2^m-1))}{2^{mj}}$ (in particular
$\delta_j^{1}(i)=\frac{i\AND(2^{j})}{2^{j}}$) 
 is a composition of
$\AND$ (bitwise logical multiplication, bitwise conjunction) and left and right shifts, $\pi_n^1(i)=\delta_{n-1}^1(i)+\delta_{n-2}^1(i)\cdot 2+\cdots+\delta_0^1(i)\cdot
2^{n-1}$. Note that for certain $m,n$ both $\delta_j^{m}(i)$ and $\pi_n^1(i)$
are just a machine instruction (e.g., `read $j$\textsuperscript{th} memory cell', the latter
assumed to be $m$-bit) or with
use of writing to and reading from memory. For instance, byte order reversing
permutation
$\pi_k^8$ could be implemented with the use of stack writing-reading, whereas
$\pi_8^1$ could be stored in memory as one-dimensional byte array (the $i$\textsuperscript{th} byte
is $\pi_8^1(i)$); then $\pi_k^8$ and $\pi_8^1$ could be combined in an
easy program to obtain $\pi_n^1$. Also we notice that in fact
one uses
the mapping $\pi_n^1$ in simulation tasks when he converts integer output $s_0,s_1,\ldots$
$(s_i\in\{0,1,\ldots, 2^n-1\})$ of a pseudorandom number generator into real
numbers $\{\frac{s_0}{2^n},\frac{s_1}{2^n},\ldots\}$ of unit interval.

It worth mentioning here that, according to the above settled conventions,
we can consider bitwise logical operators (such as $\XOR$, $\AND$, etc.)
as functions defined on the set $\mathbb N_0=\{0,1,2,\ldots\}$: We merely represent
variables in their base-2 expansions (e.g., $1\XOR 3=2$, $1\AND 3=1$).
An $m$-bit right shift is just a multiplication by $2^m$, whereas an $m$-bit
left shift is integer division by $2^m$, i.e., $\lfloor\frac{\cdot}{2^m}\rfloor$,
with $\lfloor\alpha\rfloor$ being the greatest rational integer that does
not exceed $\alpha$. Note that throughout the paper we represent integers $i$
in reverse bit order ---
less significant bits left, according to their occurrences in $2$-adic
canonical representation
of $i=\delta_0(i)+\delta_1(i)\cdot 2+\delta_2(i)\cdot 4+\ldots$; so
$0011$ is $12$, and not $3$. 

Functions $\pi_s^t$ together with arithmetic operations (addition and multiplication)
as well as bitwise logical operations (such as $\XOR$, $\AND$) and other
``machine"
ones (such as left and right shifts) are ``building blocks" of pseudorandom
generators studied below, so for reader's convenience we list the corresponding 
operators
here,
supplying them by definitions and comments, if necessary. 

Bitwise logical operators are defined by the following congruences, which
must hold for all $u,v\in\mathbb
N_0$ (or, equivalently, for all $u,v\in\mathbb Z_2$) and for all $j=0,1,2,\ldots$.
\begin{equation}
\label{eq:opBinLog}
\begin{split}
&\XOR,\ {\text {\rm or}}\ \oplus\, {\text{\rm , a bitwise `exclusive or' operator:}}\
\delta_j(u \XOR v)\equiv\\ 
&\delta_j(u)+\delta_j(v)\pmod 2;\\
&\AND,\ {\text {\rm or}}\ \wedge {\text 
{\rm , a bitwise `and' operator, bitwise conjunction:}}\ \delta_j(u \AND v)\equiv\\
&\delta_j(u)\cdot\delta_j(v)\pmod 2;\\
&\OR,\ {\text {\rm or}}\ \vee {\text 
{\rm , a bitwise `or' operator, bitwise disjunction:}}\ \delta_j(u \OR v)\equiv\\
&\delta_j(u)+\delta_j(v)+\delta_j(u)\cdot\delta_j(v) \pmod 2;\\
&\NEG,\ {\text {\rm or}}\ \neg\,  
{\text{\rm , a bitwise negation:}}\ \delta_j(\NEG(u))\equiv\\
&\delta_j(u)+1\pmod 2.
\end{split}
\end{equation}
%
The other bitwise logical operators (originating from e.g. implication,
etc.) could be defined by the analogy. 

Note that all these operators are
defined on the set $\mathbb N_0$ of non-negative rational integers. Moreover,
they are defined on the set $\mathbb Z_2$ of all $2$-adic integers (see \cite{me-1,
me-2}). The latter ones within the context of this paper could be thought
of as countable infinite binary sequences with members indexed by $0,1,2,\ldots$
. Sequences with only finite number of $1$'s correspond to non-negative
rational integers in their base-2 expansions, sequences with only finite
number of $0$'s correspond to negative rational integers, while eventually periodic 
sequences correspond to rational numbers represented by irreducible fractions
with odd denominators: 
for instance, $3=11000\ldots$, $-3=10111\ldots$, $\frac{1}{3}=11010101\ldots$, 
$-\frac{1}{3}=101010\ldots$. So $\delta_j(u)$ for $u\in\mathbb Z_2$ is merely
the $j$\textsuperscript{th} member of the corresponding sequence. 

Arithmetic operations
(addition and multiplication) with these sequences could be defined via standard
algorithms of addition and multiplication of natural numbers represented in base-2
expansions: Each member of a sequence, which corresponds to a sum (respectively,
to product) of two given sequences, will be calculated 
by these algorithms within a finite number of steps. 

Thus, $\mathbb Z_2$ is
a commutative ring with respect to the so defined addition and multiplication.
It is a metric space with respect to the distance $d_2(u,v)$ defined by the following
rule:  $d_2(u,v)=\|u-v\|_2=\frac{1}{2^n}$, where $n$ is the smallest non-negative
rational integer such that
$\delta_n(u)\ne\delta_n(v)$, and $d_2(u,v)=0$ if no such $n$ exists (i.e.,
if $u=v$). For instance $d_2(3,\frac{1}{3})=\frac{1}{8}$. 
With the use
of this distance it is possible to define convergent sequences, limits,
continuous functions and derivatives in $\mathbb
Z_2$. 

For instance, with respect to the so defined distance, the folowing sequence 
tends
to $-1$,
$$1,3,7,15,31,\ldots,2^n-1,\ldots\xrightarrow[d_2]{}-1,$$
bitwise logical operators (such as $\XOR, \AND$) define continuous
functions in two variables, the function $f(x)=x \XOR a$ is differentiable
everywhere on $\mathbb Z_2$ for every rational integer $a$: Its derivative
is $-1$ for negative $a$, and $1$ in the opposite case (see \ref{DerLog}
for other examples of this
kind and more detailed calculations). 

Reduction
modulo $2^n$ of a $2$-adic integer $v$, i.e., setting all members of the
corresponding sequence with indexes greater than $n-1$ to zero (that is,
taking the first $n$ digits in the representation of $v$)  is just an approximation 
of a $2$-adic integer
$v$ by a rational integer with accuracy $\frac{1}{2^n}$: This approximation
is an $n$-digit positive rational integer $v \AND (2^n-1)$; the latter will
be denoted also as $v\bmod{2^n}$. For formal introduction to $p$-adic
analysis, precise notions and results see e.g. \cite{Mah} or \cite{Kobl}. 
%

Arithmetic and bitwise logical operations are not independent:
Some of them could be expressed via the others. For instance, for all
$u,v\in\mathbb Z_2$
\begin{equation}
\label{eq:id}
\begin{split}
&\NEG (u)=u \XOR (-1);\\
&\NEG (u)+u=-1;\\
&u \XOR v = u+v-2(u \AND v);\\ 
&u \OR v = u+v-(u \AND v);\\
&u \OR v=(u \XOR v)+(u \AND v).
\end{split}
\end{equation}
Proofs of these  
identities \eqref{eq:id} are just an exercise: For example, if 
$\alpha ,\beta\in\{0,1\}$ then $\alpha\oplus \beta=\alpha+\beta -2\alpha\beta$
and $\alpha\vee  \beta=\alpha+\beta -\alpha\beta$. Hence:
\begin{multline*}
u \XOR v =\sum_{i=0}^\infty 2^i(\delta_i (u)\oplus \delta_i (v))= \sum_{i=0}^\infty
\sum_{i=0}^\infty 2^i(\delta_i (u)+\delta_i (v)-2\delta_i (u)\delta_i (v))=\\
\sum_{i=0}^\infty 2^i(\delta_i (u))+\sum_{i=0}^\infty 2^i(\delta_i (v))-
2\cdot\sum_{i=0}^\infty 2^i(\delta_i (u)\delta_i (v))=u+v-2(u \AND v).
\end{multline*}
Proofs of the rest identities could be made by analogy
and thus are omitted. Right shift (towards more
significant digits), as well as masking and reduction modulo $2^m$ could
be derived from the above operations: An $m$-step shift of $u$ is $2^m u$;
masking of $u$ is $u \AND M$, where $M$ is an integer, which base-2 expansion
is a mask (i.e., a string of $0$'s and $1$'s); reduction modulo $2^m$, i.e.,
taking the least non-negative residue of $u$ modulo $2^m$ is $u \bmod 2^m=u
\AND (2^m-1)$.

A common feature  the above mentioned arithmetic, bitwise logical and mashine
operations share is that they all, with the only exception of shifts towards less significant
bits, are {\it compatible}, i.e. $\omega(u,v)\equiv\omega(u_1,v_1)\pmod{2^r}$
whenever both congruences $u\equiv u_1\pmod{2^r}$  and $v\equiv v_1\pmod{2^r}$ hold
simultaneously. The notion of a compatible mapping could be naturally generalized
to mappings $(\mathbb Z/p^l)^{(t)}\rightarrow(\mathbb Z/p^l)^{(s)}$ and 
$(\mathbb Z_p)^{(t)}\rightarrow(\mathbb Z_p)^{(s)}$; compatible
mappings of the latter kind could be also considered as those satisfying
Lipschitz condition with coefficient 1 (with respect to $p$-adic distance),
see \cite{me-2}.
Obviously, a composition of compatible mappings
is a compatible mapping. We list now some important examples of compatible
operators $(\mathbb Z_p)^{(t)}\rightarrow(\mathbb Z_p)^{(s)}$, $p$ prime (see
\cite{me-2}). Part of them originates from arithmetic operations:

\begin{equation}
\begin{split}
& {\text {\rm multiplication,}}\ \cdot:\ (u,v)\mapsto uv;\\ 
& {\text {\rm addition,}}\ +:\ (u,v)\mapsto u+v; 
\\
& {\text {\rm subtraction,}}\ -:\ (u,v)\mapsto u-v;
\\
& {\text {\rm exponentiation,}}\ \uparrow_p:\ (u,v)\mapsto u\uparrow_p v=(1+pu)^v;
\ {\text{\rm in particular,}}
\\ 
& {\text {\rm raising to negative powers}},\ u\uparrow_p(-r)=(1+pu)^{-r}, r\in\mathbb N;
\ {\text{\rm and}}
\\ 
& {\text {\rm division,}}\ /_p: u/_pv=u\cdot (v\uparrow_p(-1))=\frac{u}{1+pv}. 
\label{eq:opAr}
\end{split}
\end{equation}

The other part originates from digitwise logical operations of $p$-valued logic:
\begin{equation}
\label{eq:opLog} 
\begin{split}
& {\text {\rm digitwise multiplication}}\ u\odot_p v: \delta_j(u\odot_p
v)\equiv \delta_j (u)\delta_j (v)\pmod p;\\ 
& {\text {\rm digitwise addition}}\ 
u\oplus_p v: \delta_j(u\oplus_p
v)\equiv \delta_j (u)+\delta_j (v)\pmod p;\\ 
& {\text {\rm digitwise subtraction}}\
u\ominus_p v: \delta_j(u\ominus_p
v)\equiv \delta_j (u)-\delta_j (v)\pmod p.
\end{split}
\end{equation}
Here 
$\delta_j(z)$ $( j=0,1,2,\ldots)$
stands for the $j$\textsuperscript{th} digit of $z$ in its base-$p$ expansion.

More compatible mappings could be derived from the above mentioned
ones. For instance, a reduction modulo $p^n$, $n\in\mathbb N$, is $u\bmod p^n= u\odot_p
\frac{p^n-1}{p-1}$, an $l$-step shift towards more significant digits is just
a multiplication by $p^l$, etc. Obviously, $u\odot_2 v=u\AND v$, $u\oplus_2
v=u\XOR v$. 

In case $p=2$ compatible mappings could be characterized in terms of Boolean
functions. Namely, each mapping $T\colon\mathbb Z/2^n\rightarrow\mathbb Z/2^n$ 
could be
considered as an ensemble of $n$ Boolean functions $\tau_i^T(\chi_0,\ldots,\chi_{n-1})$,
$i=0,1,2,\ldots,n-1$,
in $n$ Boolean variables $\chi_0,\ldots,\chi_{n-1}$ by assuming $\chi_i=\delta_i(u)$,
$\tau_i^T(\chi_0,\ldots,\chi_{n-1})=\delta_i(T(u))$
for $u$ running from $0$ to $2^n-1$. The following proposition
holds.
\begin{prop}
\label{Bool} 
{\rm (\cite[Proposition 3.9]{me-1})}
A mapping $T\colon\mathbb Z/2^n\rightarrow\mathbb Z/2^n$ 
{\rm (}accordingly, a mapping $T\colon\mathbb Z_2\rightarrow\mathbb Z_2${\rm
)}
is compatible
iff each Boolean function $\tau_i^T(\chi_0,\chi_{1},\ldots)=\delta_i(T(u))$,
$i=0,1,2,\ldots$,
does not depend on variables $\chi_{j}=\delta_j(u)$ for $j>i$.
\end{prop}
\begin{note*}
Mappings satisfying conditions of the proposition are also known
as {\it triangle} mappings. The proposition after proper restatement (in
terms of functions of $p$-valued logic) also holds for 
odd prime $p$. For multivariate mappings the theorem \ref{Bool} holds either:
a mapping $T=(t_1,\ldots,t_s)\colon\mathbb (Z_2)^{(r)}\rightarrow\mathbb 
(Z_2)^{(s)}$ is compatible
iff each Boolean function $\tau_i^{t_j}(\chi_{1,0},\chi_{1,1},\ldots,
\chi_{r,0},\chi_{r,1},\ldots)=\delta_i(t_k(u,\ldots,u_r))$ ($i=0,1,2,\ldots$,
$k=0,1,\ldots,s$) does not depend on variables $\chi_{\ell,j}=\delta_j(u_{\ell})$ 
for $j>i$ ($\ell=1,2,\ldots,r$).
\end{note*}

Now, given a compatible mapping $T\colon\mathbb Z_2\rightarrow\mathbb Z_2$, one
can define an
induced mapping 
$T\bmod2^n\colon\mathbb Z/2^n\rightarrow\mathbb Z/2^n$
by assuming $(T\bmod 2^n)(z) =T(z)\bmod 2^n=(T(z))\AND(2^n-1)$ 
for $z=0,1,2,\ldots,2^n-1$. The induced mapping is obviuosly a
compatible mapping of the ring $\mathbb Z/2^n$ into itself. For odd prime
$p$, as well as for multivariate case 
$T\colon(\mathbb Z_p)^{(s)}\rightarrow(\mathbb Z_p)^{(t)}$
an induced mapping $T\bmod p^n$ could be defined by the analogy.
\begin{defn}
\label{def:erg}
(See \cite{me-2}). We call a compatible mapping $T\colon\mathbb Z_p\rightarrow\mathbb Z_p$
{\it bijective modulo $p^n$} iff the induced mapping $T\bmod p^n$ is a permutation
on $\mathbb Z/p^n$; we call $T$ {\it transitive modulo $p^n$}, iff $T\bmod p^n$
is a  permutation with a single cycle. We say that $T$ is {\it
measure-preserving}
(respectively, {\it ergodic}),
iff $T$ is bijective (respectively, transitive) modulo $p^n$ for all $n\in\mathbb N$.
We call a compatible mapping
$T\colon(\mathbb Z_p)^{(s)}\rightarrow(\mathbb Z_p)^{(t)}$
{\it equiprobable modulo $p^n$} iff the induced mapping $T\bmod p^n$ maps
$(\mathbb Z/p^n)^{(s)}$ onto $(\mathbb Z/p^n)^{(t)}$, and each element of 
$(\mathbb Z/p^n)^{(t)}$ has the same number of preimages in $(\mathbb Z/p^n)^{(s)}$.
A mapping $T\colon(\mathbb Z_p)^{(s)}\rightarrow(\mathbb Z_p)^{(t)}$ is called
{\it equiprobable} iff it is equiprobable modulo $p^n$ for all $n\in\mathbb
N$. 
\end{defn}
\begin{note*} The terms measure-preserving, ergodic and equiprobable originate
from the theory of dynamical systems. Namely, the compatible mapping 
$T\colon\mathbb Z_p\rightarrow\mathbb Z_p$ defines a dynamics on the measurable
space
$\mathbb Z_p$ with a probabilistic measure that is normalized Haar measure.
The mapping $T$ is, e.g., ergodic with respect to this measure (in the
sence of the theory of dynamical systems) iff it satisfies \ref{def:erg},
see \cite{me-2} for details.
\end{note*}

Both transitive modulo $p^n$ and equiprobable modulo $p^n$ mappings will
be used as building blocks of pseudorandom generators to provide both large
period
length and uniform distribution of output sequences. The following obvious
proposition holds.
\begin{prop}
\label{prop:Auto}
If the state transition function $f$ of the automaton $\mathfrak A$ is
transitive on the state set $N$, i.e., if $f$ is a permutation with a single cycle
of length $|N|$, if, further, $|N|$ is a multiple of $|M|$, and if the output function 
$F:N\rightarrow M$ is equiprobable
{\rm (}i.e.,  $|F^{-1}(s)|=|F^{-1}(t)|$ for all $s,t\in M${\rm )}, then the output sequence
$\mathfrak S$ of the automaton $\mathfrak A$ is purely periodic with period length  $|N|$ 
{\rm (i.e., maximum possible)}, and each element of  
$M$ occurs at the period the same number of times: $\frac{|N|}{|M|}$ exactly. {\rm
That
is, the
output sequence $\mathfrak S$ is uniformly distributed.} 
\end{prop}
\begin{defn}
\label{def:strict}
Further in the paper we call a sequence $\{s_i\in M\}$ over a finite set
$M$ {\it strictly uniformly
distributed} iff it is purely periodic with period length $t$, 
and with every element of $M$ occuring at the period the same number of times,
i.e., exactly $\frac{t}{|M|}$. A sequence $\{s_i\in \mathbb Z_p\}$ of $p$-adic
integers is called {\it strictly uniformly
distributed modulo $p^k$} iff a sequence $\{s_i\bmod p^k\}$ of residues
modulo $p^k$ is strictly uniformly distributed over a residue ring $\mathbb Z/p^k$.
Also, we say that a sequence is purely periodic of period length {\it exactly}
$t$ iff it has no periods of lengths smaller than $t$. In this case $t$
is called the {\it exact period length} of the sequence.\footnote{An exact
period length is also called {\it the smallest period} of a sequence. We
do not use this term to avoid misunderstanding, since we consider a period
as a repeating part of a sequence.} 
\end{defn}
\begin{note*}  A sequence $\{s_i\in \mathbb Z_p\colon i=0,1,2,\ldots\}$ of $p$-adic
integers is uniformly distributed (with respect to a normalized Haar measure
$\mu$   on $\mathbb
Z_p$) \footnote{i.e., $\mu(a+p^k\Z_p)=p^{-k}$ for all $a\in\Z_p$ and all
$k=0,1,2.\ldots$} iff it is uniformly distributed modulo $p^k$ for all $k=1,2,\ldots$;
that is, for every $a\in\mathbb Z/p^k$ relative numbers of occurences 
of $a$ in the initial segment of length $\ell$ in the sequence 
$\{s_i\bmod p^k\}$ of residues
modulo $p^k$ 
are asymptotically equal,
i.e., 
$\lim_{\ell\to\infty}\frac{A(a,\ell)}{\ell}=\frac{1}{p^k}$, where 
$A(a,\ell)=|\{s_i\equiv a\pmod{p^k}\colon i<\ell\}|$(see \cite{KN} for
details). So strictly uniformly distributed sequences are uniformly distributed
in the common sence of theory of distributions of sequences.
\end{note*}

Thus,
putting $N=\mathbb Z/2^n, M=\mathbb Z/2^m, n=km$, and taking as $f$ and $F$
respectively, $f=\overline f=\widetilde f\bmod {2^n}$ and $F=\overline
F=
\widetilde F\bmod{2^m}$,
where the function  $\widetilde f:\mathbb Z_2\rightarrow \mathbb Z_2$ is
compatible and ergodic, and the function 
$\widetilde F:(\mathbb Z_2)^{(k)}\rightarrow \mathbb Z_2$ is compatible
and equiprobable, we obtain an automaton that generates a uniformly
distributed periodic sequence, and the length of a  period of this sequence
is $2^n$. 
That is, each
element of $\mathbb Z/2^m$ occurs at the period the same number of times
(namely,
$2^{n-m}$). Obviously, the conclusion holds if one takes as 
 $F$ an arbitrary composition of the function
$\overline F=\widetilde F\bmod{2^m}$ and an equiprobable function: for
instance, one may put $F(i)=\overline F(\pi_n(i))$ or $F(i)=\delta_j^{m}(i)$,
etc. Also, the assertion is true for odd prime $p$ either. 
Since all the automata considered further in the paper are of this kind,
their output sequences (considered as sequences over $\mathbb Z/p^m$) are
uniformly distributed purely periodic sequences, and the length of their
periods  is $p^n$,
{\slshape independently of choise} both of the function $\widetilde f$ and of
the function $\widetilde F$. So, the proposition \ref{prop:Auto} makes it possible
to vary both the state transition and the output functions (for instance, to
make them key-dependent) {\slshape without} affecting uniform distribution of
the output sequence.

Of course, to make all this practicable, one needs to choose these functions
$f$ and $F$ from suitably large
classes of ergodic and equiprobable functions. In other words, one has
to obtain certain tools to produce a number of various measure preserving,
ergodic, and equprobable mappings out of elementary compatible functions
like \eqref{eq:opBinLog} and \eqref{eq:opAr}. We consider these tools in the
next section, as well as give some estimates of how the produced classes
are big.   

%
%
\section{Tools}
\label{sec:Tool}
In this section we introduce various techniques that enable one to construct
measure preserving and/or ergodic mappings, as well as to verify whether
a given mapping is measure preserving or, respectively, ergodic. We are
mainly focused at the class of compatible mappings.
\subsection*{Using interpolation series and polynomials}

The general characterization of compatible ergodic functions is given by
the following
\begin{thm}
\label{ergBin}
{\rm{(\cite{me-1},\cite{me-conf})}} 
A function $f\colon{\mathbb Z}_{2}\rightarrow {\mathbb Z}_{2}$ 
is compatible iff it could be represented as 
\begin{equation*}
f(x)=c_0+\sum^{\infty }_{i=1}c_{i}\,2^{\lfloor \log_2 i \rfloor}\binom{x}{i}
\qquad (x\in\mathbb Z_2);
\end{equation*}
The function $f$ 
is compatible and measure-preserving iff it could be represented as 
\begin{equation*}
f(x)=c_0+x+\sum^{\infty }_{i=1}c_{i}\,2^{\lfloor \log_2 i \rfloor +1}\binom{x}{i}
\qquad (x\in\mathbb Z_2);
\end{equation*}
The function 
$f$
is compatible and ergodic iff it could be represented as 
\begin{equation*}
f(x)=1+x+\sum^{\infty}_{i=1}c_{i}2^{\lfloor \log_{2}(i+1)\rfloor+1}\binom{x}{i}
\qquad (x\in\mathbb Z_2),
\end{equation*}
where $c_0,c_1, c_2 \ldots \in {\mathbb Z}_2$.
\end{thm}
Here, as usual, 
\begin{equation*}
\binom{x}{i}=
\begin{cases}\dfrac{x(x-1)\cdots  (x-i+1)}{i!}, 
& \text {for $i=1,2,\ldots$};\cr 1, & \text{for $i=0$},
\end{cases}
\end{equation*}
and $\lfloor\alpha\rfloor$ is the integral part of $\alpha$, i.e.,
the largest rational integer not exceeding $\alpha$.
%
\begin{note*} 
For odd prime $p$ an analogon of the statement of theorem \ref{ergBin}
provides
only sufficient conditions for ergodicity (resp., measure preservation)
of $f$: namely, if $(c_0,p)=1$,
i.e., if $c$ is a unit (=invertible element) of $\mathbb Z_p$, then the
function
$f(x)=c+x+\sum^{\infty}_{i=1}c_{i}p^{\lfloor \log_{p}(i+1)\rfloor+1}\binom{x}{i}$ 
defines a compatible and ergodic mapping of $\mathbb Z_p$ onto itself,
and 
the
function
$f(x)= c_0+c\cdot x+\sum^{\infty}_{i=1}c_{i}p^{\lfloor \log_{p}i\rfloor+1}\binom{x}{i}$ 
defines a compatible and measure preserving mapping of $\mathbb Z_p$ onto itself
see \cite[Theorem 2.4]{me-2}.
\end{note*}
Thus, in view of theorem \ref{ergBin} one can choose a state transition
function to be a polynomial with rational (not necessarily integer)
key-dependent coefficients setting $c_i=0$ for all but finite number of $i$.
Note that to determine whether a given polynomial $f$ with rational (and not
necessarily integer) coefficients is integer valued (that is, maps $\mathbb
Z_p$ into itself), compatible and ergodic, it is sufficient to determine
whether it
induces a cycle on $O(\deg f)$ integral points. To be more exact, the following
proposition holds.
\begin{prop}
\label{prop:Qpol} 
{\rm(see \cite[Proposition 4.2 (4.7 in preprint)]{me-2})}
A polynomial $f(x)\in {\Q}_{p}[x]$ is integer valued, compatible, and ergodic
{\rm (}resp., measure preserving{\rm)} iff 
$$z\mapsto f(z)\bmod p^{\lfloor
\log_p (\deg f)\rfloor +3},$$ 
where $z$ 
runs through $0,1,\ldots,p^{\lfloor
\log_p (\deg f)\rfloor +3}-1$,  is compatible and transitive 
{\rm (}resp., bijective{\rm)} mapping
of the residue ring $\Z/p^{\lfloor
\log_p (\deg f)\rfloor +3}$ onto itself. 
\end{prop}

Despite
it is not very essential for further considerations, we note, however, that
the series in the statement of \ref{ergBin} and of the note thereafter are
uniformly convergent with respect to $p$-adic distance. Thus
the mapping $f\colon\mathbb Z_p\rightarrow\mathbb Z_p$ is well-defined
and continuous with respect to $p$-adic distance, 
see 
\cite[Chapter 9]{Mah}.

Theorem \ref{ergBin} enables one to use exponentiation in design of  
generators that are transitive modulo $2^n$ for all $n=1,2,3,\ldots$ 
(on exponential generators see e.g. \cite{LinRec}). 

\begin{exmp} 
\label{expGen}
For any odd $a=1+2m$ a function $f(x)=ax+a^x$ defines 
a transitive modulo $2^n$ 
generator $x_{i+1}=f(x_i)\bmod 2^n$.

Indeed, in view of \ref{ergBin} the function $f$ defines a compatible and ergodic
mapping of $\mathbb Z_2$ onto $\mathbb Z_2$
since $f(x)=(1+2m)x+(1+2m)^x=x+2mx+\sum_{i=0}^\infty m^i 2^i\binom{x}{i}=
1+x+4m\binom{x}{1}+
\sum_{i=2}^\infty m^i 2^i\binom{x}{i}$ and $i\ge\lfloor\log_2(i+1)\rfloor+1$
for all $i=2,3,4,\ldots$. 

Such a generator could be of practical value since it uses not more than
$n+1$ multiplications modulo $2^n$ of $n$-bit numbers; of course, one should
use calls to the table 
$a^{2^j}\bmod{2^n}$, $j=1,2,3,\ldots,n-1$. The latter table must be precomputed,
 corresponding calculations involve $n-1$ multiplications modulo $2^n$. Obviously,
one can use $m$ as a long-term key, with the initial state $x_0$ being
a short-term
key, i.e., one changes $m$ from time to time, but uses  new $x_0$ for each
new message. Obviously, without a properly
choosen output function such a generator is not secure. The choice of output
function in more details is discussed further in the paper. 

\end{exmp}
\begin{note*}
A similar argument shows that for every prime
$p$ and  every $a\equiv 1\pmod
p$ the function $f(x)=ax+a^x$ defines a compatible and ergodic mapping
of $\mathbb Z_p$ onto itself.
\end{note*}

For polynomials with  (rational or $p$-adic) integer coefficients 
theorem \ref{ergBin} may be restated in the following form.
\begin{prop}
\label{ergPol}
{\rm (See  \cite[Corollary 4.11]{me-1}, \cite[Corollary 4.7]{me-conf})}
Represent a polynomial $f(x)\in\mathbb Z_2[x]$  in a basis of descending 
factorial powers
$$
x^{\underline 0}=1,\ x^{\underline 1}=x,\ x^{\underline 2}=x(x-1),\ldots,\
x^{\underline i}=x(x-1)\cdots(x-i+1),\ldots,$$
i.e., let
$$f(x)=\sum^{d}_{i=0}c_i\cdot x^{\underline i}$$
for $c_0,c_1,\dots,c_d\in\mathbb Z_2$. Then the polynomial $f$ induces
an ergodic {\rm (and, obviously, a compatible)} mapping of $\mathbb Z_2$ onto
itself iff its coefficients $c_0,c_1,c_2, c_3$ satisfy the following congruences:   
$$c_0\equiv 1\ (\bmod\, 2),\quad c_1\equiv 1\ (\bmod\, 4),\quad c_2\equiv 0\
(\bmod\, 2),\quad c_3\equiv 0\ (\bmod\, 4).$$
The polynomial $f$ induces a measure preserving mapping iff
$$c_1\equiv 1\ (\bmod\, 2),\quad c_2\equiv 0\ (\bmod\, 2),\quad c_3\equiv 0\ (\bmod\, 2).$$
\end{prop}
Thus, to provide ergodicity of the polynomial mapping $f$  
it is necessary and sufficient to hold fixed $6$ bits only, while the other bits of 
coefficients of $f$ may vary (e.g., may be key-dependent). This guarantees transitivity
of the state transition function $z\mapsto f(z)\bmod 2^n$ for each $n$, and hence,
uniform distribution of the output sequence.

Proposition \ref{ergPol} implies that the polynomial $f(x)\in\mathbb
Z[x]$ is ergodic (resp., measure preserving) iff it is transitive modulo 8
(resp., iff it is bijective modulo 4). A corresponding assertion
holds in general case, for arbitrary prime $p$.
\begin{thm}
\label{ergPolGen}
{\rm (See \cite{Lar}, \cite{me-2})} A polynomial $f(x)\in\mathbb Z_p[x]$
induces an ergodic mapping of $\mathbb Z_p$ onto itself iff it is transitive
modulo $p^2$ for $p\ne 2,3$, or modulo $p^3$, for $p=2,3$. The polynomial
$f(x)\in\mathbb Z_p[x]$ induces a measure preserving mapping of 
$\mathbb Z_p$ onto itself iff it is bijective
modulo $p^2$.
\end{thm}

\begin{exmp} 
The mapping $x\mapsto f(x)\equiv x+2x^2\pmod{2^{32}}$ (which is used in
RC6, see \cite{RC6}) is bijective, since it is bijective modulo 4: 
$f(0)\equiv 0\pmod4$, $f(1)\equiv 3\pmod4$, $f(2)\equiv 2\pmod4$, 
$f(3)\equiv 1\pmod4$. Thus, the mapping 
$x\mapsto f(x)\equiv x+2x^2\pmod{2^{n}}$ is bijective for all $n=1,2,\ldots$. 
\end{exmp}
Hence, with the use of the theorem \ref{ergPolGen} it is possible to
obtain transitive modulo $q>1$ mappings for arbitrary natural $q$: one can
just take $f(z)=(1+z+\hat qg(z))\bmod q$, where $g(x)\in\mathbb Z[x]$ is
an arbitrary polynomial, and $\hat q$ is a product of $p^{s_p}$ for all
prime factors $p$ of $q$, where $s_2=s_3=3$, and $s_p=2$ for $p\ne 2,3$. Again,
the polynomial $g(x)$ may be choosen, roughly speaking, `more or less at random',
i.e., it may be key-dependent, but the output sequence will be uniformly
distributed for any choice of $g(x)$. This assertion may be generalized
either.
\begin{prop}
\label{ergAn} {\rm (\cite[Lemma 4.4 and Proposition 4.5; resp., Lemma
4.11 and Proposition 4.12 in the preprint]{me-2})} Let
$p$ be a prime, and let
$g(x)$ be an arbitrary composition of mappings listed in \eqref{eq:opAr}.
Then the mapping $z\mapsto 1+z+p^2g(z)$\ $(z\in\mathbb Z_p)$ is ergodic. 
\end{prop}

In fact, both propositions \ref{ergPol}, \ref{ergAn} and theorem \ref{ergPolGen} 
are 
particular cases of the following general
\begin{thm}
\label{ergAnGen}
{\rm (\cite[Theorem 4.2, or 4.9 in the preprint]{me-2})}
Let $\mathcal B_p$ be a class of all functions defined by series of
a form $f(x)=\sum^{\infty}_{i=0}c_i\cdot x^{\underline i}$, where 
$c_0,c_1,\dots$ are $p$-adic integers, and
$x^{\underline i}$ $(i=0,1,2,\ldots)$  
are descending factorial powers {\rm(see \ref{ergPol})}.
Then
the function $f\in \mathcal B_p$ preserves measure iff it is bijective
modulo $p^2$; $f$ is ergodic iff it is transitive modulo $p^2$
{\rm(}for $p\ne 2,3${\rm)}, or modulo $p^3$ {\rm(}for $p\in\{2,3\}${\rm)}.

\end{thm}
\begin{note*} As it was shown in \cite{me-2}, the class $\mathcal B_p$
contains all polynomial functions over $\mathbb Z_p$,
as well as analytic (e.g., rational, entire) functions that are convergent everywhere
on $\mathbb Z_p$. In fact, every mapping that is  a composition 
of arithmetic operators \eqref{eq:opAr} only belong to $\mathcal B_p$; thus, every
such mapping modulo $p^n$ could be induced by a polynomial with rational
integer coefficients (see the end of Section 4 in \cite{me-2}). For instance,
the mapping $x\mapsto (3x+3^x) \bmod 2^n$ (which is transitive modulo $2^n$,
see \ref{expGen}) could be induced by a polynomial $1+x+4\binom{x}{1}+
\sum_{i=2}^{n-1} 2^i\binom{x}{i}=1+5x+\sum_{i=2}^{n-1} \frac{2^i}{i!} \cdot
x^{\underline i}$ --- just note that $c_i=\frac{2^i}{i!}$ are $2$-adic integers
since the exponent of maximal power of $2$ that is a factor of $i!$ 
is exactly $i-\wt_2i$,
where $\wt_2 i$ is a number of $1$'s in the base-2 expansion of $i$
(see e.g. \cite[Chapter 1, Section 2, Exercise 12]{Kobl}); thus 
$\|c_i\|_2=2^{-\wt_2 i}\le 1$, i.e. $c_i\in \mathbb Z_2$ and so $c_i\bmod
{2^n}\in \mathbb Z$.  
\end{note*}

Theorem \ref{ergAnGen} implies that, for instance, the state transition
function $f(z)=(1+z+\zeta(q)^2(1+\zeta(q)u(z))^{v(z)})\bmod q$ is transitive
modulo $q$ for each natural $q>1$ and arbitrary polynomials $u(x),v(x)\in\mathbb
Z[x]$,  where $\zeta(q)$ is a product of all prime factors of $q$. So the
one can choose as a state transition function not only polynomial functions, but
also rational functions, as well as analytic ones. It should be mentioned,
however, that this is merely a form the function is represented (which
could be suitable for some cases and unsuitable for the others), yet, for a
given $q$, all the
functions of this type may also be represented as polynomials over $\mathbb
Z$ (see \cite[Proposition 4.4; resp., Proposition 4.10 in the preprint]{me-2}). 
For instance, certain generators
of inversive kind (i.e., those using taking the inverse modulo $2^n$) could
be considered in such manner.
\begin{exmp}
\label{Invers}
For $f(x)=\frac{1}{2x-1}-x$ a generator $x_{i+1}=f(x_i)\bmod{2^n}$ is transitve.
Indeed, the function $f(x)=(-1+2x-4x^2+8x^3-\cdots)-x=-1+x-4x^2+8(\cdots)$ 
is analytic
and defined everywhere on $\mathbb Z_2$; thus $f\in\mathcal B_p$. Now the
conclusion follows in view of \ref{ergAnGen} since by direct calculations
it coud be easily verified that the function $f(x)\equiv -1+x-4x^2\pmod
8$ is transitive modulo 8. Note that modulo $2^n$ the mapping $x\mapsto
f(x)\bmod 2^n$ could be induced by a polynomial $-1+x-4x^2+8x^3+\cdots+(-1)^n
x^{n-1}$.
\end{exmp}
\subsection*{Combining operators}
The class of all transitive modulo $q$ mappings, induced by polynomials with
rational integer coefficients, is rather wide: For instance, for $q=2^n$
it contains $2^{O(n^2)}$ mappings (for exact value 
see  \cite[Proposition 15]{Lar}, or \ref{Num} below). However, it could
be widened significantly (up to the class of order $2^{2^n-n-1}$ in case
$q=2^n$), by admitting also operators \eqref{eq:opLog} in the composition.  
It turnes out that there is an easy way to construct a measure preserving or ergodic
mapping out of an arbitrary compatible mapping, i.e., out of an arbitrary
composition of both arithmetic \eqref{eq:opAr} and logical \eqref{eq:opLog}
operators. 
\begin{prop}
\label{Delta} \cite[Lemma 2.1 and Theorem 2.5]{me-2}. Let $\Delta$ be a difference operator, i.e., $\Delta g(x)=g(x+1)-g(x)$
by the definition. Let, further, $p$ be a prime, let $c$ be a coprime with
$p$, $\gcd(c,p)=1$, and let $g\colon\mathbb Z_p\rightarrow
\mathbb Z_p$ be a compatible mapping. Then the mapping $z\mapsto c+z+p\Delta
g(z)\ (z\in\mathbb Z_p)$ is ergodic, and the mapping $z\mapsto d+cx+pg(x)$, 
preserves measure for arbitrary $d$.

Moreover, if $p=2$, then the converse also holds: Each compatible and ergodic
\textup {(}respectively each compatible
and measure preserving \textup {)}
mapping $z\mapsto f(z)\ (z\in\mathbb Z_2)$ could be represented as
$f(x)=1+x+2\Delta g(x)$  \textup {(}respectively as
$f(x)=d+x+2g(x)$\textup {)} for suitable $d\in\mathbb Z_2$ and compatible 
$g\colon\mathbb
Z_2\rightarrow \mathbb Z_2$.
\end{prop}
\begin{note*} The case $p=2$ is the only case the converse of the first
assertion of the proposition \ref{Delta} holds. 
\end{note*}
\begin{exmp}
\label{KlSh-2} 
Proposition \ref{Delta} immediately implies
Theorem 2 of \cite{KlSh}: For any composition $f$ of primitive functions, 
the mapping $x\mapsto x +2f(x )\pmod {2^n}$ is invertible --- just note
that
a composition of primitive
functions is compatible (see \cite{KlSh} for the definition of primitive
functions).\qed
\end{exmp}
Proposition \ref{Delta} is maybe the most important tool in design of pseudorandom
generators such that  both their state transition functions and output functions are
key-dependent.
The corresponding schemes are rather flexible: In fact, one may use nearly
arbitrary composition of arithmetic and logical operators to  produce a
strictly uniformly distributed sequence:
Both for 
$g(x)=x\XOR(2x+1)$ and for 
$$g(x)=\Biggl(1+2\frac{x\AND x^{2}+x^3\OR x^4}{3 + 4(5+6x^5)^{x^6\XOR x^7}}\Biggr)^{7+\frac{8x^8}{9+10x^9}}$$   
a sequence $\{x_i\}$ defined by recurrence relation
$x_{i+1}=(1+x_i+2(g(x_i+1)-g(x_i)))\bmod {2^n}$ is strictly uniformly distributed
in $\mathbb Z/2^n$  for each $n=1,2,3\ldots$, i.e., the sequence $\{x_i\}$ 
is 
purely periodic with {\slshape period length exactly} $2^n$, and  {\slshape each} element
of 
$\{0,1,\ldots,2^n-1\}$ occurs at the period {\slshape exactly once}. We will
demonstrate further that a designer could vary the function $g$ in a very
wide scope without worsening prescribed values of 
some important indicators of security. In fact,  choosing the proper operators 
\eqref{eq:opBinLog} and \eqref{eq:opAr} the designer is restricted
only by desirable performance, since any compatible ergodic mapping could
be produced in this way:
\begin{cor}
\label{erg-comp} 
Let $p=2$,
and let $f$
be a compatible
and ergodic mapping of $\mathbb Z_2$ onto itself. Then for each $n=1,2,\ldots$
the state transition function $f\bmod 2^n$ could be represented as a finite
composition of operators \eqref{eq:opBinLog} and \eqref{eq:opAr}.
\end{cor}
\begin{proof}
In view of proposition \ref{Delta} it is sufficient to prove that for
arbitrary compatible $g$ the function $\bar g=g\bmod 2^n$ could be represented
as a finite composition of operators \eqref{eq:opBinLog} and \eqref{eq:opAr}.
In view of \ref{Bool}, one could
represent $\bar g$ as 
$$\bar g(x)=\gamma_0(\chi_0)+2\gamma_1(\chi_0,\chi_1)+\cdots
+2^{n-1}\gamma_{n-1}(\chi_0,\ldots,\chi_{n-1}),$$
where $\gamma_i=\delta_i(\bar g)$, $\chi_i=\delta_i(x)$, $i=0,1,\ldots,n-1$.
Since each $\gamma_i(\chi_0,\ldots,\chi_i)$ is a Boolean function in Boolean
variables $\chi_0,\ldots,\chi_i$, it
could be expressed via finite number of $\XOR$s and $\AND$s of these variables
$\chi_0,\ldots,\chi_i$. Yet each variable $\chi_j$ could be expressed as
$\chi_j=\delta_j(x)=x\AND(2^j)$, and the conclusion follows. 
\end{proof}
\subsection*{Using Boolean representation}
So, in case $p=2$ we have two equivalent descriptions of the class of all
compatible ergodic mappings, namely, theorem \ref{ergBin} and proposition
\ref{Delta}. They enable one to express {\slshape any} compatible and transitive
modulo $2^n$ state transition function either as a polynomial of special kind
over a field
$\mathbb Q$ of rational numbers, or as a special composition of arithmetic
and bitwise logical operations, \eqref{eq:opAr} and \eqref{eq:opBinLog}.
Both these representations are suitable for programming, since they involve
only standard machine instructions. However, we
need one more representation, in a Boolean form (see \ref{Bool}). Despite this
representation is not very convenient for programming,
it will be used further for better understanding of certain important properties
of the considered generators, as well for proving the ergodicity of some
particular
mappings, see e.g. \ref{KlSh-3} below.
The following theorem is just a restatement of a known result from the theory 
of Boolean functions, the so-called bijectivity/transitivity criterion  for triangle
Boolean mappings.
However, the latter belongs to mathematical folklore, and thus it is somewhat
difficult to
attribute it, yet a reader could find a proof in, e.g.,
\cite[Lemma 4.8]{me-1}.
\begin{thm}
\label{ergBool} 
A mapping $T\colon\mathbb Z_2\rightarrow\mathbb Z_2$ is
compatible and measure preserving iff for each $i=0,1,\ldots$ the Boolean function 
$\tau^T_i=\delta_i(T)$
in Boolean variables $\chi_0,\ldots,\chi_{i}$ could be represented as Boolean
polynomial of the form
$$\tau^T_i(\chi_0,\ldots,\chi_i)=\chi_i+\varphi^T_i(\chi_0,\ldots,\chi_{i-1}),$$ 
where $\varphi^T_i$
is a Boolean polynomial. The mapping $T$ is compatible  and ergodic iff,
additionaly, the Boolean function
$\varphi^T_i$ is of odd weight, that is,
takes value $1$ exactly at the odd number of points 
$(\varepsilon_0,\dots,\varepsilon_{i-1})$, where
$\varepsilon_j\in\{0,1\}$ for $j=0,1,\ldots,i-1$. The latter takes place if and only
if $\varphi^T_0=1$, and the degree of the Boolean polynomial $\varphi^T_i$ for
$i\ge 1$ is exactly
$i$, that is, $\varphi^T_i$ contains a monomial
$\chi_0\cdots\chi_{i-1}$.
\end{thm}


\begin{exmp}
\label{KlSh-3} 
With the use of \ref{ergBool} it is possible to give another proof of the
main result of \cite{KlSh}, namely, of Theorem 3:
{\it The mapping $f (x)=x +(x^2\vee C )$ over $n$-bit words is invertible
if and only if the least significant bit of $C$ is 1. For $n\ge 3$ it is a permutation
with a single cycle if and only if both the least significant bit and the third least
significant bit of $C$ are $1$.}

{\it Proof of theorem 3 of \cite{KlSh}.} 
Recall that for $x\in\mathbb Z_2$ and $i=0,1,2,\ldots$ 
we denote $\chi_i=\delta_i(x)\in\{0,1\}$; also we denote $c_i=\delta_i(C)$. 
We will calculate $\delta_i(x+(x^2\vee C))$ as a Boolean
polynomial in $\chi_0,\chi_1,\ldots$ and start with the following easy claims:
\begin{itemize}
\item $\delta_0(x^2)=\chi_0$,\ $\delta_1(x^2)=0$,\ $\delta_2(x^2)=\chi_0\chi_1+\chi_1$,
\item $\delta_n(x^2)=\chi_{n-1}\chi_0+\psi_{n}(\chi_0,\ldots,\chi_{n-2})$ for all
$n\ge 3$, where $\psi_{n}$ is a Boolean function in $n-1$
Boolean
variables $\chi_0,\ldots,\chi_{n-2}$.
\end{itemize}

The first of these claims could be easily verified by direct calculations. To prove
the second one represent $x=\bar x_{n-1}+2^{n-1}s_{n-1}$ for $\bar x_{n-1}=x\bmod
2^{n-1}$ and
calculate $x^2=(\bar x_{n-1}+2^{n-1}s_{n-1})^2=
\bar x_{n-1}^2+2^{n}s_{n-1}\bar x_{n-1}+2^{2n-2}s_{n-1}^2=\bar x_{n-1}^2+2^n\chi_{n-1}\chi_0
\pmod{2^{n+1}}$ for $n\ge 3$ and note that $\bar x_{n-1}^2$ depends only on 
$\chi_0,\ldots,\chi_{n-2}$.

This gives
\begin{enumerate}
\item $\delta_0(x^2\vee C)=\chi_0+c_0+\chi_0c_0$
\item $\delta_1(x^2\vee C)=c_1$
\item $\delta_2(x^2\vee C)=\chi_0\chi_1+\chi_1+c_2+c_2\chi_1+c_2\chi_0\chi_1$
\item $\delta_n(x^2\vee C)=\chi_{n-1}\chi_0+\psi_{n}+c_n+c_n\chi_{n-1}\chi_0+c_n\psi_{n}$
for $n\ge 3$
\end{enumerate} 
From here it follows  
that if $n\ge 3$, then $\delta_n(x^2\vee C)=
\lambda_n(\chi_0,\ldots,\chi_{n-1})$, 
and $\deg \lambda_n\le
n-1$, since $\psi_{n}$ depends only on, may be, $\chi_0,\ldots,\chi_{n-2}$.

Now successively calculate $\gamma_n=\delta_n(x+(x^2\vee C))$ for $n=0,1,2,\ldots$.
We have $\delta_0(x+(x^2\vee C))=c_0+\chi_0c_0$ so necessarily $c_0=1$
since otherwise $f$ is not bijective modulo 2. Proceeding further with
$c_0=1$ we obtain $\delta_1(x+(x^2\vee C))=c_1+\chi_0+\chi_1$, since
$\chi_1$ is a carry. Then $\delta_2(x+(x^2\vee C))=(c_1\chi_0+c_1\chi_1+\chi_0\chi_1)+
(\chi_0\chi_1+\chi_1+c_2+c_2\chi_1+c_2\chi_0\chi_1)+\chi_2=
c_1\chi_0+c_1\chi_1+\chi_1+c_2+c_2\chi_1+c_2\chi_0\chi_1+\chi_2$,
here $c_1\chi_0+c_1\chi_1+\chi_0\chi_1$ is a carry. From here in view of \ref{ergBool}
we immediately have $c_2=1$ since otherwise $f$ is not transitive
modulo 8. 
Now for $n\ge 3$ 
one has $\gamma_n=\alpha_{n}+\lambda_n
+\chi_n$, where $\alpha_n$ is a carry, and $\alpha_{n+1}=\alpha_n\lambda_n
+\alpha_n\chi_n+\lambda_n\chi_n$. But if $c_2=1$ then $\deg\alpha_3=\deg
(\mu\nu+\chi_2\mu+\chi_2\nu)=3$, where $\mu=c_1\chi_0+c_1\chi_1+\chi_0\chi_1$, 
$\nu=(\chi_0\chi_1+\chi_1+c_2+c_2\chi_1+c_2\chi_0\chi_1)=
0$. This implies inductively in view of (4) above that 
$\deg\alpha_{n+1}=n+1$ and that $\gamma_{n+1}=\chi_{n+1}+\xi_{n+1}(\chi_0,\ldots,\chi_{n})$,
$\deg\xi_{n+1}=n+1$. So the conditions of \ref{ergBool} are satisfied, thus
finishing the proof of theorem 3 of \cite{KlSh}.\qed
\end{exmp}

There are some more appications of Theorem \ref{ergBool}.
\begin{prop}
\label{compBool}
Let 
$F\colon\mathbb Z_2^{n+1}\rightarrow\mathbb Z_2$ be a compatible mapping
such that for all $z_1,\ldots,z_n\in\mathbb
Z_2$ the mapping $F(x,z_1,\ldots,z_n)\colon\mathbb Z_2\rightarrow \mathbb
Z_2$ is  measure preserving. Then $F(f(x),2g_1(x),\ldots,2g_n(x))$ preserves
measure for all compatible $g_1,\ldots,g_n\colon\mathbb Z_2\rightarrow \mathbb
Z_2$ and all compatible and measure
preserving $f\colon\mathbb Z_2\rightarrow \mathbb
Z_2$. Moreover, if 
$f$ is ergodic then $f(x+4g(x))$, $f(x\oplus (4g(x)))$, $f(x)+4g(x)$, and
$f(x)\oplus  (4g(x))$ are ergodic for any compatible $g\colon\mathbb Z_2\rightarrow \mathbb
Z_2$
\end{prop}
\begin{proof} Since the function $F$ is compatible, $\delta_i(F(u_0,u_1,\ldots,u_n)$
does not depend on $\delta_j(u_k)=\chi_{j,k}$ for $j>i$ (see \ref{Bool}
and note thereafter). Represent 
$$\delta_i(F(u_0,u_1,\ldots,u_n))=\chi_{0,i}\Psi_i(u_0,u_1,\ldots,u_n)+
\Phi_i(u_0,u_1,\ldots,u_n),$$
where Boolean polynomials 
$\Psi_i(u_0,u_1,\ldots,u_n)$, $\Phi_i(u_0,u_1,\ldots,u_n)$ do not depend on 
$\chi_{0,i}$; 
that is, they depend only on, may be, 
$$\chi_{0,0},\ldots,\chi_{0,i-1},
\chi_{1,0},\ldots,\chi_{1,i},\ldots,\chi_{n,0},\ldots,\chi_{n,i}.$$
In view of \ref{ergBool} it follows that $\Psi_i=1$ since $F(x,z_1,\ldots,z_n)$ 
preserves measure for all $z_1,\ldots,z_n\in\mathbb Z_2$. Moreover, then
$\Phi_i(f(x),2g_1(x),\ldots,2g_n(x))$ does not depend on $\chi_i=\delta_i(x)$
since $\delta_j(2g(x))$ does not depend on $\chi_i$ for all $j=1,2,\ldots,n$.
Now, in view of \ref{ergBool} one has 
$\delta_i(f(x))=\chi_i+\xi_i(f(x))$, where $\xi_i(f(x))$ does not depend
on $\chi_i$ since $f$ preserves measure. 
Finally, 
$$\displaylines{\delta_i(F(f(x),2g_1(x),\ldots,2g_n(x)))=\delta_i(f(x))+
\Phi_i(f(x),2g_1(x),\ldots,2g_n(x))=\hfill\cr
\hfill\chi_i + \xi_i(f(x))+
\Phi_i(f(x),2g_1(x),\ldots,2g_n(x))=\chi_i+\Xi_i,\cr}$$ 
where 
the Boolean polynomial
$\Xi_i$ depends only on, may be, $\chi_0,\ldots,\chi_{i-1}$. 
This proves the 
first assertion of \ref{compBool} in view of \ref{ergBool}.

We prove the second assertion along the similar lines.  For $z\in\mathbb Z_2$
and $i=0,1,2,\ldots$
let $\zeta_i=\delta_i(z)$. Thus one can consider $\delta_i(z\oplus 4g(z))$ and
$\delta_i(z+ 4g(z))$ as 
Boolean
polynomials in Boolean variables $\zeta_0,\zeta_1,\ldots,\zeta_i$. Note that
$\delta_i(z\oplus 4g(z))=\zeta_i+\lambda_i(z)$, where $\lambda_i(z)=0$
for $i=0,1$ and $\deg\lambda_i(z)\le i-1$ for $i>1$, since for $i>1$ the Boolean
polynomial $\lambda_i(z)$ depends, may be, only on $\zeta_0,\ldots,
\zeta_{i-2}$. 

Next, we claim that 
$\delta_i(z+ 4g(z))=\delta_i(z)+\mu_i(z)$, where 
$\mu_i(z)=\mu_i^g(z)$ is 0 for $i=0,1$ and $\deg\mu_i(z)\le i-1$ for $i>1$. 
Indeed, $\mu_i(z)=\lambda_i(z)+\alpha_i(z)$, where the Boolean polynomial
$\alpha_i(z)$ is a carry. Yet $\alpha_i(z)=0$ for $i=0,1,2$, 
and 
$\alpha_i(z)=\zeta_{i-1}\lambda_{i-1}(z)+\zeta_{i-1}\alpha_{i-1}(z)+
\lambda_{i-1}(z)\alpha_{i-1}(z)$ for $i\ge 3$, and $\alpha_i(z)$ depends
only on, may be, $\zeta_0,\ldots,\zeta_{i-1}$ since $\alpha_i(z)$
is a carry. However, $\deg\alpha_3(z)=2$ and if $\deg\alpha_{i-1}(z)\le
i-2$ then $\deg\delta_{i-1}(z)\alpha_{i-1}(z)\le i-1$, 
$\deg\lambda_{i-1}(z)\alpha_{i-1}(z)\le i-1$, and 
$\deg\zeta_{i-1}\lambda_{i-1}(z)\le i-1$ since $\alpha_{i-1}(z)$ depends
only on, may be, $\zeta_0,\ldots,\zeta_{i-2}$ and
$\lambda_{i-1}(z)$ depends, may be, only on $\zeta_0,\ldots,
\zeta_{i-3}$. Thus $\deg\alpha_i(z)\le i-1$ and hence $\deg\mu_i(z)\le
i-1$.

Now, since $f(x)$ is
egodic, $\delta_{i}(f(x))=\chi_i+\xi_i(x)$, where the Boolean polynomial
$\xi_i$ depends only on, may be, $\chi_0,\ldots,\chi_{i-1}$ and,
additionally, $\xi_0=1$, and $\deg\xi_i=i$ for $i>0$ (see \ref{ergBool});
i.e. $\xi_i(x)=\chi_0\chi_1\cdots\chi_{i-1}+\vartheta_i(x)$,
where $\deg\vartheta_i(x)\le i-1$ for $i>0$.
Hence, for $\ast\in\{+,\oplus\}$ one has
$\delta_{i}(f(x\ast 4g(x)))=\delta_i(x\ast 4g(x))+
\delta_0(x\ast 4g(x))\delta_1(x\ast 4g(x))\cdots\delta_{i-1}(x\ast 4g(x))+
\vartheta_i(x\ast 4g(x))$; thus $\delta_{i}(f(x\ast 4g(x)))=\chi_i+
\chi_0\cdots\chi_{i-1}+ \beta_i^\ast(x)$, where $\deg\beta_i^\ast(x)\le
i-1$ for $i>0$, and $\delta_0(f(x\ast 4g(x))=\delta_0(x\ast 4g(x))+1=\chi_0+1$.
Finally, $f(x\ast 4g(x))$ for $\ast\in\{+,\oplus\}$ is ergodic in view of
\ref{ergBool}.

In a similar manner it could be demonstrated that $f(x)\ast 4g(x)$ is ergodic
for $\ast\in\{+,\oplus\}$: $\delta_i(f(x)\ast 4g(x))=\delta_i(f(x))$ 
for $i=0,1$ and thus satisfy the conditions of \ref{ergBool}. For $i>1$
on has $\delta_i(f(x)\oplus 4g(x))=\chi_i+\xi_i(x)+\delta_{i-2}(g(x))$;
but $\delta_{i-2}(g(x))$ does not depend on $\chi_{i-1},\chi_{i}$.
Thus the Boolean polynomial $\xi_i(x)+\delta_{i-2}(g(x))$ in variables
$\chi_0,\ldots,\chi_{i-1}$ is of odd weight, since $\xi_i(x)$
is of odd weight, thus proving that $f(x)\oplus 4g(x)$ is ergodic.

Now  represent $g(x)=g(f^{-1}(f(x)))=h(f(x))$, where $f^{-1}(x)$ is the
inverse mapping for $f$. Clearly, $f^{-1}(x)$ is well defined since 
the mapping $f\colon\mathbb Z_2\rightarrow\mathbb Z_2$ is bijective; 
moreover $f^{-1}(x)$
is compatible and ergodic. Finally  
$\delta_i(f(x)+ 4g(x))=\delta_i(f(x))+\mu_i^\prime(f(x))$, 
where the Boolean polynomial
$\mu_i^\prime(x)=\mu_i^{h}(x)$ in Boolean variables 
$\chi_0,\ldots,\chi_{i-1}$ does not contain a monomial 
$\chi_0\cdots\chi_{i-1}$ (see the claim above). This implies that the Boolean polynomial
$\mu_i^\prime(f(x))$ in Boolean variables $\chi_0,\ldots,\chi_{i-1}$
does not contain a monomial $\chi_0\cdots\chi_{i-1}$ either,
since $\delta_j(f(x))=\chi_j+\xi_j(x)$ and $\xi_j(x)$ depend only,
may be, on $\chi_0,\ldots,\chi_{j-1}$ for $j=2,3,\ldots$. Hence,
$\delta_i(f(x)+ 4g(x))=\chi_i+\xi_i(x)+\mu_i^\prime(f(x))$ and the
Boolean polynomial $\xi_i(x)+\mu_i^\prime(f(x))$ in Boolean variables
$\chi_0,\ldots,\chi_{i-1}$ is of odd weight. This finishes the
proof in view of \ref{ergBool}. 
\end{proof}
\begin{exmp}
\label{XOR}
With the use of \ref{compBool} it is possible to construct very fast generators
$x_{i+1}=f(x_i)\bmod 2^n$ that are transitive modulo $2^n$. 
For instance, take
$$f(x)=(\ldots((((x+c_0)\oplus d_0)+c_1)\oplus d_1)+\cdots +c_m)\oplus
d_m,$$
where $c_0\equiv 1\pmod 2$, and the rest of $c_i,d_i$ are 0 modulo 4.
By the way, this generator, looking somewhat `linear', is as a rule rather
`nonlinear': the corresponding polynomial over $\mathbb Q$ is of high degree.
The general case of these functions $f$ (for arbitrary $c_i, d_i$) 
was studied by the author's student Ludmila Kotomina: She proved that such
a function
is ergodic iff it is transitive modulo 4.
\end{exmp} 
\subsection*{Counting the number of transitive mappings} 
The preceeding results enable us to calculate the number of 
all compatible transitive modulo $2^n$ mappings
of $\mathbb Z/2^n$ onto itself and
the number of them 
that are induced by {\it polynomial mappings over} $\mathbb Z$, i.e., 
that could be expressed as polynomials
with rational integer coefficients.
\begin{prop}
\label{Num}
There are exactly $2^{2^n-n-1}$ compatible and transitive modulo
$2^n$ mappings $T\colon\mathbb Z/2^n\rightarrow\mathbb Z/2^n$. For $n\le
3$ all of them could be represented as polynomials over $\mathbb Z$; if
$n>3$, then exactly $2^{\sum_{i=0}^{\rho(n)}(n-i+\wt_2i)-6}$ 
of them could be represented
as polynomials over $\mathbb Z$ {\rm (see \ref{ergPol}).} Moreover,
$\sum_{i=0}^{\rho(n)}(n-i+\wt_2i)-6\sim \frac{1}{2}n^2$ as $n\to\infty$.
{\rm Here $\wt_2i$ is the binary weight of non-negative rational integer $i$
{\rm (}i.e., the number of $1$'s in base-$2$ expansion of $i${\rm )}, 
and $\rho(n)$ is the biggest natural number $k$ such that $k-\wt_2k<n$.} 
\end{prop}
\begin{proof}
The first assertion is an easy consequence of \ref{ergBool}: obviously,
the number
of Boolean functions of odd weight in $i$ variables is exactly $2^{2^i-1}$,
and the result follows.

To prove the second assertion we first note that each integer-valued polynomial
$f(x)\in\mathbb Q_p[x]$ 
over a field $\mathbb Q_p$ of $p$-adic numbers (that is, a polynomial, which takes
values in $\mathbb Z_p$ at each point of $\mathbb Z_p$) admits a unique
representation
\begin{equation}
\label{eq:Bin}
f(x)=\sum_{i=0}^{\infty}a_i\binom{x}{i}
\end{equation}
for suitable $a_0,a_1,a_2,\dots\in\mathbb Z_p$, with only finite number
of non-zero $a_0,a_1,a_2,\dots$ (see e.g. \cite{Mah}). Further, the polynomial
\eqref{eq:Bin} is identically zero modulo $2^n$ iff $a_i\equiv 0\pmod{2^n}$
for all $i=0,1,2,\dots$ (see proposition 4.2 of \cite{me-1}). Lastly, the
polynomial \eqref{eq:Bin} is a polynomial over $\mathbb Z_2$ iff it could
be represented in the form of \ref{ergPol}, i.e., iff 
$a_i\equiv 0\pmod{2^{\ord_2i!}}$ for all $i=0,1,2,\dots$. Here and after
$\ord_p
q$ stands for the greatest power of a prime $p$, which is a factor of $q\in\mathbb
N$: $p^{\ord_p q}\mid q$, but $p^{1+\ord_p }\nmid q$; it is well known
that $\ord_p i!=\frac{1}{p-1}(i-\wt_pi)$, see e.g. \cite{Kobl}, Chapter
1, Section 2, Exercise 13. 

Thus, each mapping of $\mathbb Z/2^n$ onto $\mathbb Z/2^n$ that is
induced by polynomial over $\mathbb Z$
admits a unique representation by polynomial \eqref{eq:Bin} of degree not greater
than $\rho(n)$, and with $a_0,a_1,a_2,\dots\in\mathbb Z/2^n$ such that $a_i\equiv
0\pmod{2^{i-\wt_2i}}$ for $i=2,3,\dots$. In view of \ref{ergBin}, the latter polynomial
is transitive modulo $2^n$ iff $a_0\equiv 1\pmod{2}$, $a_1\equiv 1\pmod
4$, and $a_i\equiv 0\pmod{2^{\lfloor\log_2(i+1)\rfloor+1}}$ for $i=2,3,\dots$.
Since $i-\wt_2i<\lfloor\log_2(i+1)\rfloor+1$ iff $i=0,1,2,3$, the
number of all transitive modulo $2^n$ mappings of $\mathbb Z/2^n$
into $\mathbb Z/2^n$ that are induced by polynomials over $\mathbb Z$ is exactly
$2^{\eta(n)}$, where $\eta(n)=4n-8+\sum_{i=4}^{\rho(n)}(n-i+\wt_2i)=
-6+\sum_{i=0}^{\rho(n)}(n-i+\wt_2i)$ for $n>3$, and $\eta(1)=1$, $\eta(2)=2$,
$\eta(3)=16$.

Now, to finish the proof of proposition \ref{Num} we only have to
demonstrate that $\lim_{n\to\infty}\frac{2\eta(n)}{n^2}=1$. We start with
estimating $\rho(n)$. 

Represent $n$ as $n=2^k+t$ where $0\le t<2^k$. Verify that $\rho(2^{k+1}-1)=2^{k+1}-1$
by direct calculations. So, $\rho(n)=n$, if $n=2^{k+1}-1$ (i.e., if $t=2^k-1$), 
and $\rho(n)=2^k+s$ for certain $s\ge 0$, in the opposite case (i.e., if $t<2^k-1$). 
We claim that $s<2^k$. Indeed, the function $k-\wt_2k$, and hence, the function
$\rho(n)$ are nondecreasing; thus, $s\le2^k$. However, assuming $s=2^k$
we get a contradiction: On the one hand, 
$2^k+t=n>\rho(n)-\wt_2\rho(n)=2^k+2^k-\wt_2(2^k+2^k)=2^{k+1}-1$, 
but $t<2^k-1$ on the other. Thus for $t<2^k-1$, i.e., for $n\ne
2^{k+1}-1$,  we have that $\rho(n)=2^k+s$ for some $t\le s\le 2^k-1$ since
obviously $\rho(n)\ge n$. Hence 
$n=2^k+t>\rho(n)-\wt_2(\rho(n))=2^k+s-1-\wt_2s$; consequently
$s=\max\{r\in\mathbb N : s-\wt_2s<t+1\}=\rho(t+1)$
by definition of the function $\rho$. Thus we proved the formula
\begin{equation*}
\rho(n)=\rho(2^k+t)=
\begin{cases}
2^k+t, &\text{if $t=2^k-1$, i.e., if $n=2^{k+1}-1$};\cr
2^k+\rho(t+1), &\text{if $t<2^k-1$, i.e., if $n\ne 2^{k+1}-1$}.
\end{cases}
\end{equation*} 
This implies an obvious recursive procedure for calculating $\rho(n)$, which
halts not later than in $k$ steps; mind that $k+1$ is the number of digits
in base-$2$ expansion of $n$.  We conclude finally 
that $n\le\rho(n)\le n+\lfloor\log_2n\rfloor$ since
the number of digits in base-$2$ expansion of $n$ is exactly 
$\lfloor\log_2n\rfloor+1$ and $2^{r}-1=\underbrace{11\ldots1}_{r}$.

Now we succesively calculate 
$\eta(n)=\sum_{i=0}^n(i+\wt_2i)+\sum_{j=n+1}^{\rho(n)}(n-j+\wt_2j)-6=
\frac{n(n+1)}{2}+\sum_{i=1}^n\wt_2i-\frac{(\rho(n)-n)(\rho(n)-n+1)}{2}+
\sum_{j=1}^{\rho(n)-n}\wt_2(n+j)-6$. Finally, taking into the account that 
\begin{multline*}
\sum_{i=1}^n\wt_2i\le
\sum_{i=1}^{2^{\lfloor\log_2n\rfloor+1}-1}\wt_2i=\sum_{i=1}^{\lfloor\log_2n\rfloor+1}
i\binom{\lfloor\log_2n\rfloor+1}{i}\\
=(\lfloor\log_2n\rfloor+1)2^{\lfloor\log_2n\rfloor}
\le(1+\log_2n)n
\end{multline*}
and also that $\rho(n)-n\le\log_2n$, $\wt_2(a+b)\le \wt_2a+\wt_2b$,
$\wt_2a\le 1+\log_2a$, 
we conclude that 
$\lim_{n\to\infty}\frac{2\eta(n)}{n^2}=1$.
\end{proof}
\begin{note}
\label{Num-1} 
During the proof of proposition \ref{Num} we have demonstrated
that each mapping of $\mathbb Z/2^n$ onto $\mathbb Z/2^n$
induced by a polynomial over $\mathbb Z$
could be represented by a polynomial of degree not greater
than $\rho(n)\le n+\log_2n$, and this estimate is sharp. Moreover, 
from the final part of the proof it could be deduced that the number of 
transitive
mappings of $\mathbb Z/2^n$ onto itself that are induced by polynomials
over $\mathbb Z$ is $O(2^{\frac{1}{2}n(n+1)+n(1+\log_2n)+
\frac{1}{2}(1+\log_2n)\log_2n+(1+\log_2\log_2n)\log_2n})$. The case 
$n=2^k$ is of special interest since usually the word length of contemporary
processors is a power of $2$. In this case $\rho(n)=n+1$, and for $k\ge 2$ direct
calculations of $\eta(n)$ (see 
the proof of \ref{Num}) imply that the number
of transitive modulo $2^n$ mappings of $\mathbb Z/2^n$ onto itself that
are induced by polynomials over $\mathbb Z$ is exactly 
$2^{2^{2k-1}+(k+1)2^{k-1}-4}$. For instance, in the case $n=32$ this makes $2^{604}$
transitive mappings; all of them are induced by polynomials over $\mathbb
Z$ of degree $\le 33$, i.e, could be expressed via arithmetic operations
\eqref{eq:opAr}. Yet for $n=8$ this makes only $2^{44}$ polynomials of
degree not exceeding $9$. By the use of bitwise logical operations 
\eqref{eq:opBinLog} along with arithmetic operations  
one could significantly increase the number of transitive mappings,
up to $2^{2^{n}-n-1}$. Each of these mappings could be expessed
as a polynomial over $\mathbb Q$ (see \ref{ergBin}), yet the bound for
its degree $d$ raises significantly either. 
Namely, from the
proof of \ref{Num} it follows that $\lfloor\log_2(d+1)\rfloor+1<n$ for $n>2$,
i.e., $d\le 2^{n-1}-2$, and this bound is sharp. For $n=8$, e.g.,
this makes $2^{247}$ transitive polynomials over $\mathbb Q$ of degree
$\le 126$. Note that for each $1\le d\le \rho(n)$ (resp., for each 
$1\le d\le 2^{n-1}-2$) there exist an ergodic polynomial over $\mathbb Z$
(resp., a compatible and ergodic polynomial over $\mathbb Q$) of degree
exactly $d$. The number of pairwise distinct modulo $2^n$ mappings induced by these
polynomials may also be calculated using the ideas of the proof of \ref{Num}.
We omit details. 
\end{note}
\subsection*{Using uniform differentiability}
Now we are going to give general descriptions of equiprobable
(in particular, multivariate measure-preserving) mappings following
\cite[section 3]{me-2}, \cite[Section 5]{me-conf}, \cite[Section 5]{me-1}. 
These mapping could be used as output functions
of the generators assuring uniform distribution of the produced sequence,
see \ref{prop:Auto}.

To describe equiprobable (and, in particular, measure preserving) mappings
we need $p$-adic differential calculus techniques as well as certain notions introduced 
in \cite{me-1, me-2, me-conf}.
\begin{defn}
\label{def:Der}
A function $F=(f_{1},\ldots  ,f_{m})\colon{\mathbb Z}^{(n)}_{p}\rightarrow {\mathbb Z}^{(m)}_{p}$
is said to be {\it differentiable modulo $p^k$} at the point 
$ \mathbf u=(u_{1},\ldots  ,u_{n})\in {\mathbb Z}^{(n)}_{p}$
if there exists a positive
integer rational
$N$ and $n\times m$ matrix $F^{\prime}_{k}(\mathbf u)$ over ${\mathbb Q}_{p}$
(called {\it the Jacobi matrix modulo} $p^{k}$ of the function $F$ at the
point
$\mathbf u$) such that for every positive rational integer 
$K\ge N$ and every $ \mathbf h=(h_{1},\ldots  ,h_{n})\in {\mathbb Z}^{(n)}_{p}$  
the inequality 
$\|\mathbf h\| _{p}\le     p^{-K}$ implies that
\begin{equation}
\label{Der} 
F( \mathbf u+\mathbf h)\equiv F(\mathbf u)+ 
\mathbf hF^{\prime}_{k}(\mathbf u)\pmod{p^{k+K}}.
\end{equation}
 In case $m=1$ the
Jacobi matrix modulo $p^k$ is called a {\it differential modulo $p^k$}. In
case $m=n$ a determinant of Jacobi matrix modulo $p^k$ is called a {\it Jacobian
modulo $p^k$}. The elements of Jacobi matrix modulo $p^k$
are called {\it partial derivatives modulo} $p^k$ of the function $F$ at
the point $\mathbf u$.
\end{defn} 
A partial derivative (respectively, a differential) modulo $p^k$ are
sometimes  denoted as 
$\frac{\partial_k f_i (\mathbf u)}{\partial_k x_j}$ (respectively, as
$d_{k}F(\mathbf u)=\sum^n_{i=1} \frac {\partial_k F(\mathbf u)}{\partial_k x_i}d_{k}x_{i}$).
\par
The definition immediately implies that partial derivatives 
modulo $p^k$ of the function $F$ are defined up to the $p$-adic integer
summand whith $p$-adic norm does not exceeding $p^{-k}$. In cases when all partial derivatives
modulo $p^k$ at all points of  
$\mathbb Z_p^{(n)}$ are
$p$-adic integers, we say that the function 
$F$ has {\it integer-valued derivative modulo} $p^k$; 
in these cases we can associate to each partial derivative modulo $p^k$
a unique element of the ring $\mathbb Z/p^k$, 
and a Jacobi matrix modulo $p^k$ 
at each point $\mathbf u\in \mathbb Z_p^{(n)}$ 
thus can be considered as a matrix over a ring $\mathbb Z/p^k$. It turnes
out that this is exactly the case for compatible $F$. Namely, the following
proposition holds.
\begin{prop}
\label{intDer}
{\rm(\cite[Corollary 3.8]{me-1}, \cite[Corollary 3.3]{me-conf})}
Let a compatible function 
$F=(f_{1},\ldots  ,f_{m})\colon{\mathbb Z}^{(n)}_{p}\rightarrow {\mathbb Z}^{(m)}_{p}$ be uniformly
differentiable modulo $p^k$ at the point $\mathbf u\in {\mathbb Z}^{(n)}_{p}$.
Then $\big\|\frac{\partial_k f_i (\mathbf u)}{\partial_k x_j}\big\|_p\le 1$, i.e.,
$F$ has integer-valued derivatives modulo $p^k$.
\end{prop}
For the functions with integer-valued derivatives modulo $p^k$  
the `rules of differentiation
modulo $p^k$' have the same (up to congruence modulo $p^k$ instead of equality)
form as for usual differentiation.
For instance, if both functions 
$G\colon{\mathbb Z}^{(s)}_{p}\rightarrow {\mathbb Z}^{(n)}_{p}$ and
$F\colon{\mathbb Z}^{(n)}_{p}\rightarrow {\mathbb Z}^{(m)}_{p}$ 
are differentiable modulo 
$p^{k}$ at the points, respectively, $\mathbf v=(v_{1},\ldots  ,v_{s})$
and $\mathbf u=G(\mathbf v)$, and their partial derivatives modulo $p^{k}$ at
these points are $p$-adic integers, then a composition 
$F\circ G\colon{\mathbb Z}^{(s)}_{p}\rightarrow {\mathbb Z}^{(m)}_{p}$ 
of these functions is uniformly differentiable modulo $p^{k}$ at the point
$\mathbf v$, all its partial derivatives 
modulo $p^{k}$ at this point are $p$-adic integers, and 
$(F\circ G)^\prime_k (\mathbf v)\equiv G^\prime_k (\mathbf v) F^\prime_k (\mathbf u)\pmod
{p^k}$.

By the analogy with classical case we can give the following
\begin{defn}
\label{def:uniDer}
A function $F\colon{\mathbb Z}^{(n)}_{p}\rightarrow {\mathbb Z}^{(m)}_{p}$ is said
to be 
{\it uniformly differintiable modulo $p^k$ on $\mathbb Z_p^{(n)}$} iff there
exists $K\in\mathbb N$ such that \ref{Der} holds simultaneously for all 
$\mathbf u \in \mathbb Z_p^{(n)}$ as soon as
$\| h_{i}\| _{p}\le     p^{-K}$, $(i=1,2,\ldots  ,n)$. The
least such 
$K\in\mathbb N$
is denoted via $N_k(F)$. 
\end{defn}
We recall that   all  partial derivatives
 modulo $p^k$ of a uniformly differentiable modulo $p^k$ function $F$
 are periodic functions with period  
$p^{N_k(F)}$ (see \cite[Proposition  2.12]{me-1}). 
This in particular implies that each partial derivative modulo
$p^k$ could be considered as a function defined on $\mathbb Z/p^{N_k(F)}$. 
Moreover, if a continuation $\tilde F$ of the function
$F=(f_{1},\ldots , f_{m})\colon{\mathbb N}^{(n)}_{0}\rightarrow {\mathbb N}^{(m)}_{0}$  
to the space $\mathbb Z_p^{(n)}$ is uniformly differentiable modulo $p^k$ on the
$\mathbb Z_p^{(n)}$, then one could continue both the function $F$  and all its
(partial) derivatives modulo $p^k$ to the space $\mathbb Z_p^{(n)}$
simultaneously. This imples that we could study if necessary (partial) 
derivatives modulo $p^k$
of the function $\tilde F$ instead of studying those of $F$ and vise versa.
For example, a partial derivative $\frac{\partial_k f_i (\mathbf u)}{\partial_k x_j}$
modulo $p^k$ vanishes modulo $p^k$ at no point of  $\mathbb Z_p^{(n)}$
(that is,
$\frac{\partial_k f_i (\mathbf u)}{\partial_k x_j}\not\equiv 0\pmod{p^k}$
for all $u\in \mathbb Z_p^{(n)}$, or, the same
$\big\|\frac{\partial_k f_i (\mathbf u)}{\partial_k x_j}\big\|_p> p^{-k}$
everywhere on $\mathbb Z_p^{(n)}$) if and only if 
$\frac{\partial_k f_i (\mathbf u)}{\partial_k x_j}\not\equiv 0\pmod{p^k}$
for all $u\in\{0,1,\ldots,p^{N_k(F)}-1\}$.

To calculate a derivative of, for instance, a state transition function, which
is a composition of `elementary' functions, see \ref{erg-comp},
one needs to know derivatives of these  `elementary' functions, 
such as \eqref{eq:opBinLog}
and \eqref{eq:opAr}. Thus, we briefly introduce a $p$-adic analogon of  
`table of derivatives' of classical Calculus.
\begin{exmp} Derivatives of bitwise logical operations.
\label{DerLog}
\begin{enumerate}
\item {\it a function $f(x)=x\AND c$ is uniformly differentiable on $\mathbb
Z_2$ for any $c\in
\mathbb Z$; $f^\prime(x)=0$ for $c\ge 0$, and $f^\prime(x)=1$ for $c<0$,} since
$f(x+2^ns)=f(x)$, and 
$f(x+2^ns)=f(x)+2^ns$ for $n\ge l(|c|)$, where $l(|c|)$ is the bit length
of absolute value of $c$
(mind that for $c\ge 0$ the $2$-adic representation
of $-c$ starts with $2^{l(c)}-c$ in less significant bits followed by $11\ldots$:
$-1=11\ldots$, $-3=10111\ldots$, etc.).
\item {\it a function $f(x)=x\XOR c$ is uniformly differentiable on $\mathbb
Z_2$ for any $c\in
\mathbb Z$; $f^\prime(x)=1$ for $c\ge 0$, and $f^\prime(x)=-1$ for $c<0$.} This
immediately
follows from (1) since $u\XOR v=u+v-2(x\AND v)$ (see \eqref{eq:id}); thus
$(x\XOR c)^\prime=x^\prime+c^\prime-2(x\AND c)^\prime=1+2\cdot(0,\ \text{for}\
c\ge 0;\ \text{or}\
-1,\ \text{for}\ c<0)$.
\item in the same manner it could be shown that {\it functions $(x\bmod
2^n)$, $\NEG(x)$
and $(x\OR c)$ for $c\in \mathbb Z$ are uniformly differentiable on $\mathbb
Z_2$, and $(x\bmod 2^n)^\prime=0$, $(\NEG x)^\prime=-1$, 
$(x\OR c)^\prime=1$ for $c\ge 0$, 
$(x\OR c)^\prime=0$ for $c< 0$.}
\item {\it a function $f(x,y)=x\XOR y$ is not uniformly differentiable on 
$\mathbb Z_2^{(2)}$,
yet it is uniformly differentiable modulo $2$ on $\mathbb Z_2^{(2)}$};
from (2) it follows that its partial derivatives modulo 2 are 1 everywhere
on $\mathbb Z_2^{(2)}$.
\end{enumerate}
\end{exmp} 

Here how it works altogether.
\begin{exmp*}
A function $f(x)=x+(x^2\OR 5)$ is uniformly differentiable
on $\mathbb Z_2$, 
and $f^\prime (x)=1+2x\cdot
(x\OR 5)^\prime=1+2x$.

A function $F(x,y)=(f(x,y),g(x,y))=
(x \oplus 2(x \wedge y ),(y +3 x^3 )\oplus x )$ 
is uniformly differentiable modulo $2$ as bivariate
function, and $N_1(F)=1$; namely
$$F(x+2^nt,y+2^ms)\equiv F(x,y)+(2^nt,2^ms)\cdot
\begin{pmatrix}
1&x+1\\
0&1
\end{pmatrix}
\pmod{2^{k+1}}$$ 
for all $m,n\ge 1$ (here $k=\min\{m,n\}$). The matrix
$\begin{pmatrix}
1&x+1\\
0&1
\end{pmatrix}
=F^\prime_1(x,y)$ is Jacoby matrix modulo 2 of $F$; here how we calculate
partial derivatives modulo $2$: for instance,  
$\frac{\partial_1 g(x,y)}{\partial_1 x}=\frac{\partial_1 (y +3 x^3)}{\partial_1 x}
\cdot \frac{\partial_1 (u\oplus x)}{\partial_1 u}\big|_{u=y +3 x^3}+
\frac{\partial_1 x}{\partial_1 x}\cdot 
\frac{\partial_1 (u\oplus x)}{\partial_1 x}\big|_{u=y +3 x^3}=9x^2\cdot 1+1\cdot
1\equiv x+1\pmod 2$.
Note that a partial derivative modulo 2 of the function 
$2(x \wedge y )$ is always $0$ modulo 2 because of the multiplier 2:
the function $x \wedge y$ is not differentiable modulo 2 as bivariate function,
yet $2(x \wedge y )$ is. So the Jacobian of the function $F$ is 
$\det F^\prime_1=1\pmod 2$.
\end{exmp*}
%
Now let  $F=(f_{1},\ldots , f_{m})\colon{\mathbb Z}^{(n)}_{p}\rightarrow {\mathbb Z}^{(m)}_{p}$  
and $f\colon{\mathbb Z}^{(n)}_{p}\rightarrow {\mathbb Z}_{p}$ be compatible
functions, 
which are uniformly differentiable on $\mathbb Z_p^{(n)}$  modulo $p$. This is a
relatively
weak restriction since all uniformly differentiable on $\mathbb Z_p^{(n)}$ functions,
as well as functions, which are uniformly differentiable on $\mathbb Z_p^{(n)}$
modulo $p^k$ for some $k\ge
1$, are uniformly differentiable on $\mathbb Z_p^{(n)}$ modulo $p$;
note that 
$\frac{\partial F}{\partial x_i}\equiv \frac{\partial_k F}{\partial_k x_i}\equiv
\frac{\partial_{k-1} F}{\partial_{k-1} x_i}\pmod{p^{k-1}}$.  
Moreover,
all values of all partial derivatives modulo $p^k$ (and thus, modulo $p$)
of $F$ and $f$ are $p$-adic integers everywhere on, 
respectively, $\mathbb Z_p^{(n)}$ and $\mathbb Z_p$ (see \ref{intDer}),
so to calculate these values one can use the techniques considered above.

\begin{thm}
\label{equi}
{\rm(}\cite[Theorems 3.1 and 3.2; resp., 3.7 and 3.9 in the preprint]{me-2},
\cite[5.2 -- 5.5]{me-conf}, \cite[5.2 -- 5.5]{me-1}{\rm)}
A function  $F\colon{\mathbb Z}^{(n)}_{p}\rightarrow {\mathbb Z}^{(m)}_{p}$ is 
equiprobable whenever it is equiprobable modulo $p^{k}$ for some
$k\ge N_{1}(F)$ and the rank of its Jacobi matrix $F_1^\prime (\mathbf
u)$ modulo
$p$ is exactly $m$ at all points  
$\mathbf u=(u_{1},\ldots  ,u_{n})\in (\mathbb Z/p^{k})^{(n)}$. In case
$m=n$ these conditions are also necessary, i.e., the function $F$ preserves
measure iff it is bijective modulo $p^{k}$ for some
$k\ge N_{1}(F)$ and $\det(F_1^\prime (\mathbf u))\not\equiv 0\pmod{p}$ for all
$\mathbf u=(u_{1},\ldots  ,u_{n})\in (\mathbb Z/p^{k})^{(n)}$. Moreover,
in the considered case these conditions  imply  that $F$ preserves measure
iff it is bijective modulo $p^{N_1(F)+1}$. 
\end{thm}
That is, if the mapping
$\mathbf u\mapsto F(\mathbf u)\bmod p^{N_1(F)}$ is equiprobable, and if
the rank of Jacobi matrix $F_1^\prime (u)$ modulo
$p$ is exactly $m$ at all points  $
\mathbf u\in (\mathbb Z/p^{N_1(F)})^{(n)}$
then 
{\it each} mapping $\mathbf u\mapsto F(\mathbf u)\bmod p^r$ of
$(\mathbb Z/p^r)^{(n)}$ onto $(\mathbb Z/p^r)^{(m)}$
$(r=1,2,3,\ldots)$ is equiprobable (i.e., each point 
$\mathbf u\in (\mathbb Z/p^{r})^{(m)}$ has the same number of preimages
in $(\mathbb Z/p^{r})^{(m)}$, see \ref{def:erg}).
\begin{exmp}
\label{KlSh-ex}
(see \cite{KlSh})
\begin{enumerate}
\item {\it A mapping 
$$(x,y ) \mapsto F(x,y)=(x \oplus 2(x \wedge y ),(y +3 x^3 )\oplus x )\bmod{2^r}$$
of $\mathbb (Z/2^r)^{(2)}$ onto $\mathbb (Z/2^r)^{(2)}$ 
is bijective for all $r=1,2,\ldots$}

Indeed, the function $F$ is bijective modulo $2^{N_1(F)}=2$ (direct verification)
and  $\det(F_1^\prime (\mathbf u))\equiv 1\pmod 2$ for all $\mathbf u\in(\mathbb
Z/2)^{(2)}$ (see \ref{DerLog} and example thereafter).
\item {\it The following mappings of $\mathbb Z/2^r$ onto $\mathbb Z/2^r$ 
are bijective for all $r=1,2,\ldots$}: 
\begin{equation*} 
\qquad \quad x\mapsto (x +2x^2)\bmod{2^r},\ x\mapsto (x +(x^2\vee 1))\bmod{2^r},\ 
x\mapsto (x \oplus (x^2\vee 1))\bmod{2^r}
\end{equation*}

Indeed, all three mappings are uniformly differentiable
modulo 2, and $N_1=1$ for all of them. So it sufficies to prove that
all three mappings are bijective modulo 2, i.e. as mappings of the residue
ring $\mathbb Z/2$ modulo 2 onto itself (this could be checked by direct calculations), 
and that
their derivatives modulo 2 vanish at no point of $\mathbb Z/2$. The latter
also holds, since  the derivatives are, respectively,
$$\qquad\ 1+4x\equiv 1\pmod 2,\ 1+2x\cdot 1\equiv 1\pmod 2,\ 1+2x\cdot 1\equiv
1\pmod 2$$
since $(x^2\vee 1)^\prime=2x\cdot 1\equiv 1\pmod 2$, and $(x\oplus C)^\prime_1\equiv
1\pmod 2$,
(see \ref{DerLog}).
\item {\it The following closely related variants of the previous mappings
of 
$\mathbb Z/2^r$ onto $\mathbb Z/2^r$ 
are NOT bijective for all $r=1,2,\ldots$}:
$$\qquad \quad x\mapsto (x +x^2)\bmod{2^r},\  x \mapsto (x +(x^2\wedge 1))\bmod{2^r},\  
x\mapsto (x +(x^3\vee 1))\bmod{2^r},$$ since they are compatible but
not bijectve modulo 2.
\item (see \cite{Riv}, also \cite[Theorem 1]{KlSh}) {\it Let $P (x )=a_0 +a_1 x + \cdots+a_d x^d$ be a polynomial with integral
coefficients. Then $P (x )$ is a permutation polynomial } (i.e., is bijective)
{\it modulo $2^ n$,
$n>1$ if and
only if $a_1$ is odd, $(a_2 +a_4 + \cdots)$ is even, and $(a_3 +a_5 +\cdots)$
is even.}

In view of \ref{equi} we have to verify whether the two conditions
hold: first, whether $P$ is bijective modulo 2, and second,
whether
$P^\prime(z)\equiv 1\pmod 2$ for $z\in\{0,1\}$.
The first condition gives that $P(0)=a_0$ and $P(1)=a_0+a_1+a_2+\cdots a_d$
must be distinct modulo 2; hence $a_1+a_2+\cdots a_d\equiv 1\pmod 2$. 
The second condition implies that
$P^\prime(0)=a_1\equiv 1\pmod2,\ P^\prime(1)\equiv a_1+a_3+a_5+\cdots\equiv 1\pmod 2$.
Now combining all this together we get $a_2+a_3+\cdots a_d\equiv 0\pmod 2$ and 
$a_3+a_5+\cdots\equiv 0\pmod 2$, hence $a_2 +a_4 + \cdots\equiv 0\pmod 2$.
\item As a bonus, we can use exactly the same proof to
get exactly the same characterization of bijective modulo $2^r$ $(r=1,2,\ldots)$
mappings of the form $x\mapsto P (x )=
a_0\oplus  a_1x\oplus \cdots\oplus  a_dx^d\bmod 2^r$ since $u\oplus v$ is uniformly
differentiable modulo 2 as bivariate function, and its derivative modulo
2 is exactly the same as the derivative of $u+v$, and besides, $u\oplus v\equiv
u+v\pmod 2$. 
\end{enumerate}
\end{exmp}
Note that in general theorem \ref{equi} could be applied to a class of
functions that is narrower than the class of all compatible functions.
However, it turnes out that for $p=2$ this is not the case. Namely, the
following proposition holds, which in fact is just a restatement of a 
corresponding assertion of \ref{ergBool}.
\begin{prop}
\label{mpDer}
{\rm(\cite[Corollary 4.6]{me-1}, \cite[Corollary 4.4]{me-conf})}
If a compatible function $g\colon\mathbb Z_2\rightarrow\mathbb Z_2$ preserves
measure then it is  uniformly differentiable modulo $2$ and has integer derivative
modulo $2$ (which is always $1$ modulo $2$).
\end{prop}

The techniques introduced above could also be applied to characterize ergodic
functions.
\begin{thm}
\label{ergDer}
{\rm (}\cite[Theorem 3.4, resp. 3.14 in the preprint]{me-2}, \cite[Theorem
5.7]{me-conf}, \cite[Theorem 5.7]{me-1}{\rm )} 
Let a compatible function $f\colon{\mathbb Z}_{p}\rightarrow {\mathbb Z}_{p}$ 
be uniformly differentiable modulo $p^{2}$. Then $f$ is
ergodic if and only if it is transitive modulo $p^{N_{2}(f)+1}$ when
$p$ is an odd prime,  or modulo $2^{N_{2}(f)+2}$ when $p=2$. 
\end{thm}
\begin{exmp}
\label{ergKlSh}
In \cite{KlSh} there is stated that ``...neither the invertibility nor 
the cycle structure of
$x +(x^2\vee 5)$ could be determined by his ({\slshape i.e., mine --- V.A.}) techniques.''
See however how it could be immediately done with the use of Theorem
\ref{ergDer}:
The function $f(x)=x+(x^2\vee 5)$ is uniformly differentiable
on $\mathbb Z_2$, thus, it is uniformly differentiable modulo 4 
(see \ref{DerLog} and an example thereafter), and $N_2(f)=3$. Now to
prove that $f$ is ergodic, in view of \ref{ergDer} it sufficies 
to demonstrate that $f$ induces a permutation
with a single cycle on $\mathbb Z/32$. Direct calculations show that a
string
$0,f(0)\bmod 32, f^2(0)\bmod 32=f(f(0))\bmod 32, \ldots, f^{31}(0)\bmod
32$ is a permutation of a string $0,1,2,\ldots,31$, thus ending the proof.
\end{exmp}

Note that both Theorems \ref{equi} and \ref{ergDer} share the same feature:
To prove ergodicity (or measure preservation) of a certain mapping
it sufficies to verify only whether this mapping is transitive (respectively,
bijective) modulo $p^N$ for a certain $N$. The origin of this feature is
a pecularity of the $p$-adic distance; in fact such an effect goes back
to Hensel's lemma. By the way, using this feature, namely, the fact that
a polynomial $f$ with integer coefficients induces an ergodic mapping of $\mathbb
Z_2$ onto itself iff $f$ is transitive
modulo 8 (see \ref{ergPolGen}; note that \ref{ergDer} implies modulo 16), 
M.V.Larin proved the following theorem in a
spirit of one of Rivest's \ref{KlSh-ex}(4).
\begin{thm}
{\rm (\cite[Proposition 21]{Lar})}
Let $P (x )=a_0 +a_1 x + \cdots+a_d x^d$ be a polynomial with integral
coefficients. Then $P (x )$ induces a permutation with a single cycle 
modulo $2^ n$, $n>2$ if and only if the following congruences hold simultaneously:
\begin{gather*}
a_3+a_5+a_7+a_9+\cdots\equiv 2a_2\pmod 4;\\
a_4+a_6+a_8+\cdots\equiv a_1+a_2-1\pmod 4;\\
a_1\equiv 1\pmod 2;\\
a_0\equiv 1\pmod 2.
\end{gather*} 
\end{thm}
It would be of interest to understand whether an
analogon of \ref{KlSh-ex}(5) for ergodic polynomials over $\mathbb Z$ 
could be proved:
A straightforward application of the same ideas does not work since the function
$x\oplus y$ is uniformly differentiable modulo 2, but not modulo 4, cf.
Theorem \ref{ergDer}.  

\section{Constructions}
\label{sec:Constr}

In this section we introduce several constructions that enable one to
built pseudorandom number
generators 
out of `building blocks' based on ergodic and equiprobable mappings.
Output sequences of these generators are always strictly uniformly
distributed.
Other probabilistic and cryptographic properties of these generators are discussed
in further sections. 

Our base construction is 
a finite automaton
${\mathfrak A}=\langle N,M,f,F,u_0\rangle $ such that
\begin{itemize}
\item the state set $N$ is finite;
\item the state transition function
$f:N\rightarrow N$ is transitive (i.e., $f$ is a permutation with a single
cycle); 
\item the output alphabet $M$ is finite, and $|M|$ is a factor
of $|N|$; 
\item  the output function  $F:N\rightarrow M$ is equiprobable, i.e., all
preimages $F^{-1}(z)$, $z\in M$, have the same cardinality $\frac{|N|}{|M|}$;
\item  the initial state (a seed) $u_0$ is an arbitrary element of $N$.
\end{itemize}

Under these conditions the output sequence 
$$\mathcal S(u_0)=
\{F(u_0), F(f(u_0)), F(f^{(2)}(u_0)),\ldots, F(f^{(j)}(u_0)),\ldots\}$$ 
of the automaton $\mathfrak A$ is strictly uniformly distributed
over $M$ i.e., $\mathcal S(u_0)$ is a purely periodic sequence, 
$|N|$ is its period length,
and every element $z\in M$ occurs at the period exactly $\frac{|N|}{|M|}$
times, see \ref{prop:Auto}.
\subsection*{Congruential generator of a maximum period length} This corresponds to a case when $N=M$,
$f$ is compatible and transitive mapping of the residue ring $\mathbb Z/|N|$ onto itself,
and $F$ is an identical transformation (we identify $N$ with $\mathbb Z/|N|$
in an obvious manner).
This generator is said to be
{\it congruential} since the algebraic notion of compatibility just
means that $f$ preserves all congruences of the ring $\mathbb Z/|N|$,
i.e. for all $a,b\in N$, $a\equiv b\pmod d\Rightarrow f(a)\equiv f(b)\pmod
d$ whenever $d\,\big | |N|$.
\begin{note}
\label{note:Congr}
In order to avoid future misunerstanding it is important to emphasize here
that {\slshape our notion of a congruential generator
differs from one of Krawczyk}, \cite{Kr}. According to the latter paper, a (general)
congruential generator is a number generator for which the $i$\textsuperscript{th}
element $s_i$ of the sequence is a $\{0,1,\ldots,m-1\}$-valued number computed
by the congruence
\begin{equation}
\label{eq:Kr}
s_i\equiv\sum_{j=1}^k\alpha_j\Phi_j(s_{-n_0},\ldots,s_{-1},s_0,\ldots,
s_{i-1})\pmod m,
\end{equation}
where $\alpha_j\in\mathbb Z$, $m\in\{2,3,\ldots\}$ and $\Phi_j$, $1\le
j\le k$ is an arbitrary integer-valued function. Note that this definition
could be restated in the equivalent form: a (general) congruential generator
is a number generator for which the $i$\textsuperscript{th}
element $s_i$ of the output sequence is computed by the congruence 
$$s_i\equiv \Phi(s_{-n_0},\ldots,s_{-1},s_0,\ldots,s_{i-1})\pmod m,$$
where, as Krawczyk notes (see \cite[page 531]{Kr}), $\Phi$ is an {\slshape arbitrary} integer-valued function that works on
{\slshape finite sequences} of integers. Thus, {\slshape according to Krawczyk's definition,
an arbitrary infinite sequence over $\{0,1,\ldots,m-1\}$ should be considered
as  a congruential generator}. Such a definition is too general for the purposes
of our paper. Results of \cite{Kr} in connection with a
problem of predictability of the generators considered in this paper will
be discussed later.
\end{note}

So {\it further in the paper a  congruential generator is assumed to be
the automaton
$\mathfrak A$ such that $M=N$, $F:M\rightarrow M$ is a trivial permutation,
and state transition function $f$, being considered
as a mapping of the residue ring $\mathbb Z/|N|$ into itself, preserves all
congruences of this ring}.  

In case the number of states is composite, $|N|=p_1^{n_1}p_2^{n_2}\cdots p_t^{n_t}$,
$p_j$ prime, $j=1,2,\ldots,t$, this generator could 
obviously be represented as a direct product of congruential generators
with prime power state set: 
$\mathbb Z/|N|=\mathbb Z/p_1^{n_1}\times \cdots \times\mathbb Z/p_t^{n_t}$,
and $f=f_1\times\cdots\times f_t$, where $f_j=(\tilde f_j)\bmod p_j^{n_j}$,
$\tilde f_j\colon\mathbb Z_{p_j}\rightarrow\mathbb Z_{p_j}$ is a compatible
and ergodic mapping, $j=1,2,\ldots,t$. 
\begin{exmp*} For $N=10^k=2^k\cdot 5^k$ the mapping
$f(x)=11x+{11^x}$ is transitive modulo
$10^k$ for all  $k=1,2,\ldots$ (see \ref{expGen} and a note thereafter). 
\end{exmp*}
Thus, the case of composite number
of states could be reduced to the case when a number of states is a power
of a prime, i.e., when $|N|=p^n$.
An obvious disadvantage of this congruential generator is that the {\slshape
period length
of the sequence}
$\{\delta_j(f^{(i)}(u_0)): i=0,1,2,\dots\}$  (where  $\delta_j(z)$
stands for the $j$\textsuperscript{th} digit of the base-$p$ expansion
of $z$) {\slshape is exactly $p^{j+1}$, i.e., only the most
significant bit of the output sequence has a maximum period length}, which is
obviously equal to the period of the whole output sequence. 

While being
not very
significant in case the output sequence is applied to simulation 
tasks
(espesially if one uses
the sequence $\Big\{\frac{f^{(i)}(u_0)}{p^n}\Big\}$; 
the latter use is common for numerical experiments), 
this disadvantage in general
leads to a cryptographic insecurity of
the generator  whenever the function $f$ is known
to a cryptoanalyst. 
Indeed, to solve a congruence $z\equiv f(x)\pmod {p^n}$
(and as a result to find a key, which is an initial state $u_0$ in this
case) one might use a version of $p$-adic Newton's method (the latter is  
a base of a canonical
proof of Hensel's lemma). 

Namely, one solves a congruence $z\equiv f(x)\pmod {p}$,
thus finding the least significant digit $\delta_0(x)$ of $x$. Provided 
$\delta_j(x)$ for $j=0,1,\ldots,k-1$ are already found, to find $\delta_k(x)$
one has to find a (unique) solution of a congruense $z\equiv f(\hat x)+
p^k\check f_k(\hat x,\delta_k(x))\pmod
{p^{k+1}}$, where $\hat x=\delta_0(x)+\delta_1(x)\cdot p+\cdots+\delta_{k-1}(x)\cdot
p^{k-1}$ and the mapping $\check f_k(\cdot,\cdot)\colon \mathbb Z/p^k\times
\mathbb Z/p\rightarrow\mathbb Z/p$ is uniquelly determined by $f$. Of course,
to express explicitly $\check f_k(\cdot,\cdot)$ is a separate problem, yet
it is easy in a number of important cases. For instance, 
$\check f_k(\hat x,\delta_k(x))=\delta_k(x)$ in case $p=2$ (see \ref{mpDer}).

We may also consider a case when $f$ is not is known to a cryptoanalyst:
e.g., for $p=2$ one may take 
$f=1+x+4g(x)$, where $g(x)$ is a compatible key-dependent function, which
is not known to a cryptoanalyst. 
Such function $f$ is ergodic, see
\ref{compBool}. 
This situation is a little better in comparison with a known $f$. 
However,
the sequence formed of less significant bits of $f^{(i)}(u_0)$ is predictable
in both directions, i.e. knowing $k$ members of the sequence $\{f^{(i)}(u_0)\}$
a cryptoanalyst finds $\delta_j(f^{(i)}(u_0))$ for all $j<\log_2 k$ and
all $i=0,1,2,\ldots$, stretching
the corresponding periods in both directions. Thus, a good idea is to discard
less significant bits of the output sequence: Note that methods of \cite{Kr},
as it is directly pointed out there, do not apply to generators that output
only parts of the numbers generated. So  we come to the notion of  
\subsection*{Truncated congruential generator of a maximum period length} The
latter is an automaton $\mathfrak A$ such that $|N|=p^n$, $p$ prime, $|M|=p^m$,
$m<n$, $f=(\tilde f)\bmod p^n$, $f$ is a compatible and ergodic mapping
of $\mathbb Z_p$ onto itself, $F(u)=\big\lfloor \frac{u}{p^{n-m}}\big\rfloor$,
$u\in\{0,1,\ldots, p^n-1\}$. 
Note that the function $F$ is not compatible, yet equiprobable, so
the output sequence, considered as a sequence over $\mathbb Z/p^m$, 
is purely periodic with period length exactly $p^n$, and
each element of $\mathbb Z/p^m$ occurs at the period exactly $p^{n-m}$
times.
In this paper we are mainly focused at the case
$p=2$.

%

 
An important example of such an output function $F$ is 
the mapping 
$\delta_j\colon\mathbb Z_2\rightarrow\mathbb Z/2$. It returnes the 
$j$\textsuperscript{th} digit of $z$
and is obviously equiprobable.
We call the corresponding sequence $\{\delta_j(f^{(i)}(z)): i=0,1,2,\dots\}$ 
the {\it
$j$\textsuperscript{th} coordinate sequence}, since the sequence $\{f^{(i)}(z): i=0,1,2,\dots\}$
could be thought of as a sequence of vectors 
$\{(\delta_0(f^{(i)}(z)),\delta_1(f^{(i)}(z)),\dots): i=0,1,2,\dots\}$
over a field $\mathbb Z/2$ of two elements. Of course, the use of $\delta_j$
as an output function of the automaton $\mathfrak A$ significantly reduces
the performance, and the corresponding pseudorandom
generator might be not of much practical value. 
Nonetheless, we have to study coordinate sequences 
to be able to prove certain important properties of output sequences of pseudorandom
generators considered in the paper. In particular, while studying probabilstic 
quality of output sequences of truncated
congruential generators one has to study correlations among coordinate
sequences. We postpone these issues to Section \ref{sec:Prop}.

A truncation usually makes generators slower but more secure: 
general methods that predict truncated congruential generators are not known, see
\cite{Bri-Od},\cite{Menz}. However, such methods exist in some particular
cases, for instance, when $f$ is a polynomial over $\mathbb Z$ of degree
$1$, and/or a relatively small part of less significant bits are discarded,
see \cite{five}. However, in general truncated congruential generators
seem to be rather secure even their state transition function is relatively
simple: For instance, an analysis made in \cite{KlSh-2} shows that for 
$f(x)=(x+(x^2\vee C))\bmod 2^n$ the corresponding stream cipher 
is quite strong against a number of attacks. Note also that in generators
we study here both the state
transition function and output function could be keyed.


\subsection*{Wreath products of congruential generators} This construction enables
one to construct pseudorandom generators such that their state
transition function (and output function) is being modified dynamically while working, 
i.e. generators with recurrence sequence
of states satisfying a congruence
$$x_{i+1}\equiv f_i(x_i)\pmod{2^n}.$$ Such generators  are called 
{\it counter-dependent}, see \cite[Definition 2.4]{ShTs}.
The problem here is how to guarantee period length (and statistical quality)
of this sequence $\{x_i\}$. The construction we introduce below offers a
certain solution to this problem; the idea of the construction goes back
to wreath products of permutation groups. The exact definition (which could
be found in, e.g., \cite{Pas}) is not needed within a context of this paper;
we note, however, that this construction is just a permutation 
that belongs to a wreath product of a Sylow $2$-subgroup of a symmetric
group on $2^n$ elements by a cyclic group.

The idea of the construction is the following:
Consider a (finite or infinite) sequence of automata 
$\mathfrak A_j=\langle N,M,f_j,F_j\rangle$, $j\in J=\{0,1,2,\ldots,\}$
(where $J$ is finite, or $J=\mathbb N_0$). 
Note that all the automata $\mathfrak A_j$ 
have the same state set $N$
and the same output alphabet $M$. Now produce the following sequence 
$\{z_i\colon i=1,2,\ldots\}$:  
Choose an arbitrary $u_0\in N$ and put
$$z_0=F_0(u_0),u_1=f_0(u_0);\ldots
z_{i}=F_i(u_i), u_{i+1}=f_i(u_i);\ldots$$ 
That is, at the $(i+1)$\textsuperscript{th} step the automaton $\mathfrak A_i$
is applied to the state $u_i$ producing a new state $u_{i+1}=f_i(u_i)$ and
outputting a symbol $z_{i}=F_i(u_i)$.


Now we give a more formal
\begin{defn}
\label{def:WP}
Let $\mathfrak A_j=\langle N,M,f_j,F_j\rangle$  be a family of 
automata with the same state set $N$ and the same
output alphabet $M$ indexed by elements of
a non-empty (possibly, countably infinte) set $J$ 
(members of the family are  not necessarily pairwise distinct). 
Let $T\colon J\rightarrow J$ be an arbitrary mapping. A {\it wreath product}
 $\mathfrak A_j\Wr_{j\in J}T$
 of the family $\{\mathfrak A_j\}$ of the automata
 by the mapping $T$ is an automaton with state set $N\times J$, state
transition function $\breve f(j,z)=(f_j(z),T(j))$ and output function 
$\breve F(j,z)=F_j(z)$. The state transition function $\breve f(j,z)=(f_j(z),T(j))$
is called a {\it wreath product of family of mappings $\{f_j\colon j\in
J\}$ by the mapping $T$}; it is denoted as $\breve f=f_j\Wr_{j\in J}T$.
\end{defn}
It worth noticing here  that if $J=\mathbb N_0$ and $F_i$ does not depend on $i$, this construction 
will give us a number of examples of counter-dependent generators 
in a sence of \cite[Definition 2.4]{ShTs}.
Note also that generators we consider in this subsection are counter-dependent 
in a broader sence: Not only 
their state transition functions depend on $i$,
but their output functions as well. 

In fact, we are already familiar with wreath products of mappings: See the
following
\begin{exmp*}
Let $J=\mathbb Z/2^n$, let $T\colon\mathbb Z/2^n\rightarrow\mathbb Z/2^n$
be an arbitrary compatible permutation with a single cycle.
Put $N=\{0,1\}$, $f_z(u)=u\oplus\beta(z)$, where $u\in N$ and 
$\beta(z)=\beta(\delta_0(z),\ldots,\delta_{n-1}(z))$ is a Boolean polynomial
of degree $n$ in $n$ Boolean variables (so $\{f_z\}$ is a family of linear
congruential generators modulo $2$). Then $\breve f=f_z\Wr_{z\in J}T$ could
be considered as a mapping of $\mathbb Z/2^{n+1}$ onto itself (we identify
$(\varepsilon,z)\in N\times J$ with $z+\varepsilon\cdot 2^n\in\mathbb Z/2^{n+1}$);
moreover, $\breve f$ is a compatible permutation on $\mathbb Z/2^{n+1}$
with a single cycle in view of \ref{ergBool}. Thus, every compatible and
ergodic mapping modulo $2^k$ could be obtained by succesive application
of wreath products. In fact, all compatible mappings of $\mathbb Z/2^{n+1}$
onto itself form a group $Syl_2(2^{n+1})$ with respect to a composition. 
This group is a Sylow
$2$-subgroup of a symmetric group $Sym(2^{n+1})$ on $\mathbb Z/2^{n+1}$;
it is known (see e.g. \cite{Pas}) that 
$$Syl_2(2^{n+1})=\underbrace {Sym(2)\wr Sym(2)\wr\cdots\wr Sym(2)}_{\text{$n+1$ factors}}.$$
Here $\wr$ stands for the wreath product of groups.
\end{exmp*} 

A generalization of the above example gives the following
\begin{prop}
\label{WP-even}
Let $T\colon\mathbb Z/2^m\rightarrow\mathbb Z/2^m$, $m\ge 1$, 
be an arbitrary permutation
with a single cycle, 
let $\{c_0,\ldots,c_{2^m-1}\}$ be a finite sequence
of $2$-adic integers, 
and let $\{f_0,\ldots,f_{2^m-1}\}$ be a finite sequence of compatible 
mappings of $\mathbb Z_2$ onto itself. 
Put $H_j(x)=c_j+x+4\cdot f_j(x)$.
Then the wreath product
$H_j\Wr_{j=0}^{2^m-1}T$ defines a bijective mapping 
$W\colon\mathbb Z_2\twoheadrightarrow \mathbb Z_2$ 
$$W(x)=T(x\bmod{2^m})+2^m\cdot H_{x\bmod{2^m}}
\bigg(\Big\lfloor\frac{x}{2^m}\Big\rfloor\bigg);$$
this mapping is asypmtotically compatible and asymptotically ergodic
{\rm (i.e., $a\equiv b\pmod{2^k}\Rightarrow W(a)\equiv W(b)\pmod{2^k}$ and
$W$ is transitive modulo $2^k$ for all sufficiently large $k$; in fact,
for all $k>m$, see \cite{me-conf, me-1, me-2} for definitions)} if and only if
$\sum_{j=0}^{2^m-1}c_j\equiv 1\pmod 2$.

In other words, 
every recurrence sequence $\mathcal U_n=\{x_i\}$ defined by the relation 
$$x_{i+1}=H_{i\bmod{2^m}}(x_i)\bmod{2^n}$$
is  strictly uniformly distributed sequence
over $\mathbb Z/2^n$  of period length exactly $2^{n+m}$
if and only if
$\sum_{j=0}^{2^m-1}c_j\equiv 1\pmod 2$.
\end{prop}
\begin{proof} Since wreath product of permutations on sets $N$ and $M$
is a permutation on the direct product $N\times M$ (see \ref{def:WP}),
the sequence $\mathcal U_n$ is purely periodic. 
Moreover, since the permutations $T$ and $I\colon z\mapsto (z+1)\bmod 2^m$ are
conjugate in $Sym(2^m)$, and thus both wreath products 
$(H_j\bmod 2^n)\Wr_{j=0}^{2^m-1}T$
and $(H_j\bmod 2^n)\Wr_{j=0}^{2^m-1}I$ have the same cycle structure (the same number
of cycles of length $\ell$, for all $\ell=1,2,\ldots$), it is suffisient to study a period of a 
sequence $x_{i+1}=H_{i}(x_i)\bmod{2^n}$, assuming $H_i=H_{i\bmod{2^m}}$
for $i\ge 2^m$. 
Further, since 
$W_n=(H_j\bmod 2^n)\Wr_{j=0}^{2^m-1}I\in Syl_2(2^{n+m})$,
the period length of the sequence $\{x_i\}$ is a power of $2$. Finally,
since the mapping $W_n\colon\mathbb Z/2^{n+m}\rightarrow\mathbb Z/2^{n+m}$
is compatible, it is necessary and sufficient to understand when  $W_n$ is transitive
modulo $2^{n+m}$ for all $k=n+m$. Yet the mapping $W_n$ could be considered
as a function of a variable $z=i+2^m\cdot x\in\mathbb Z/2^{m+n}$, where
$i\in\{0,1,\ldots, 2^m-1\}$ and $x\in\{0,1,\ldots, 2^n-1\}$. 
Thus, we could apply \ref{ergBool} to study transitivity of $W_n$.
Since $W_n(z)\equiv z+1\pmod{2^m}$ by the definition,
we only have to calculate $\delta_j(H_i(x))$.

One has $\delta_0(c_i+x)\equiv \chi_0+\beta(i)\pmod 2$ and
$$\delta_j(c_i+x)\equiv \chi_j+\beta(i)\chi_0\cdots\chi_{j-1}+
\gamma_{ji}(\chi_0,\ldots,\chi_{j-1})\pmod 2 \qquad (j>0),$$
where $\chi_j=\delta_j(x)$, $\beta(i)=\delta_0(c_i)$, 
$\gamma_{ji}(\chi_0,\ldots,\chi_{j-1})$ is a Boolean polynomial of degree
$<j$ in Boolean variables $\chi_0,\ldots,\chi_{j-1}$. Yet
$\delta_i(4\cdot g_j(x))$ is a Boolean polynomial in Boolean variables 
$\chi_0,\ldots,\chi_{j-2}$ for $j\ge 2$, and is $0$ otherwise. Thus,
\begin{equation}
\label{eq:WP-even}
\delta_j(H_i(x))\equiv\chi_j+\beta(i)\chi_0\cdots\chi_{j-1}+
\lambda_{ji}(\chi_0,\ldots,\chi_{j-1})\pmod 2,
\end{equation}
where $\deg\lambda_{ji}<j$, $j=1,2,\ldots$, and 
$\delta_0(H_i(x))\equiv\chi_0+\beta(i)\pmod 2$.

Assuming $\zeta_r=\delta_r(z)$ for $r=0,1,\ldots, m+n-1$ one can consider
$\beta(i)$ for $i\in\{0,1,\ldots, 2^m-1\}$ 
as a Boolean polynomial in Boolean variables
$\zeta_0,\ldots,\zeta_{m-1}$; similarly, $\lambda_{ji}$
could be considered as a Boolean polynomial in Boolean variables
$\zeta_0,\ldots,\zeta_{m+j-1}$. Since the degree of $\lambda_{ji}$ in variables
$\chi_0,\ldots,\chi_{j-1}$ is less than $j$ (see the argument above), the
degree of this polynomial in variables  $\zeta_0,\ldots,\zeta_{m+j-1}$
is less than $m+j$. Thus, in view of \ref{Delta} and \eqref{eq:WP-even},
the mapping $W_n$ is transitive iff $\deg\beta=m$, i.e., iff the Boolean
polynomial $\beta$ is of odd weight. Yet the latter is equivalent to the
condition $\sum_{i=0}^{2^m-1}\beta(i)\equiv 1\pmod 2$. This proves the
proposition since $\sum_{i=0}^{2^m-1}\beta(i)\equiv \sum_{i=0}^{2^m-1}
c_i\pmod 2$.
\end{proof} 

Two important  notes worth being stated here. The first of them concerns further
generalizations
of proposition \ref{WP-even}
\begin{note}
\label{WP-even-more}
The proof of \ref{WP-even}  shows that {\it
the proposition holds if $H_j$ satisfy 
the following conditions: $\sum_{j=0}^{2^m-1}H_j(0)\equiv 1\pmod 2$ and 
$\delta_i(H_j(x))\equiv \delta_i(x)+\rho_i(j;x)\pmod 2$ $(i=0,1,2\ldots)$, where the
Boolean polynomial $\rho_i$ in Boolean variables $\delta_r(j)$, $\delta_s(x)$
$(r\in\{0,1,\ldots,m-1\}$, $s\in\{0,1,\ldots, i-1\})$ is of odd weight
for $i>0$} (see the argument proving \eqref{eq:WP-even} and text thereafter). 
In oder to satisfy the latter condition of these  one can take e.g. $H_j(x)=x+h_j(x)$, 
where every $\delta_i(h_j)$ is a Boolean polynomial of even weight in Boolean
variables $\delta_0(x),\ldots,\delta_{i-1}(x)$ \footnote{Such mappings $h_j$
are called {\it even parameters} in \cite{KlSh-2}}.
Also, one can assume in conditions of \ref{WP-even} that, e.g., 
$H_j=(c_j+x)\oplus(2\cdot g_j(x))$ (or $H_j=c_j+x+2\cdot g_j(x)$) for measure
preserving $g_j$, etc. 
\end{note}
\begin{exmp*}
{\it Let $H_j(x)=c_j+x+(x^2\vee C_j)$, where $\sum_{j=0}^{2^m-1}c_j\equiv 1\pmod 2$
and $C_j\equiv 7\pmod 8$, then the recurrence sequence defined by 
$x_{i+1}= c_{i\bmod 2^m}+x_i+(x_i^2\vee C_{i\bmod 2^m})$ is strictly
uniformly distributed modulo $2^n$}. It is sufficient to note only that $x^2\vee
7$ is an even parameter, see \cite{KlSh-2}. This example is  a variation of
theme of theorem 3 there, which considers similar problem for the sequence
defined by relation $x_{i+1}= (x_i+(x_i^2\vee C_{i\bmod m}))\bmod 2^n$
with odd $m$ (the case when $T$ acts on a set of odd order is discussed below).
\end{exmp*}
The second important note relates wreath products and truncation.
\begin{note}
\label{WP-even-trunc}
From the proof of proposition \ref{WP-even} immediately follows that {\it
each recurrence sequence $\mathcal X_n$ defined by $x_{i+1}=f_{i\bmod 2^m}(x_i)\bmod 2^n$  with
compatible $f_i$ could be
obtained by a truncation of $m$ low order bits of the recurrence sequence 
defined by $z_{i+1}=G(z_i)\bmod 2^{n+m}$ for a suitable compatible mapping
$G\colon\mathbb Z_2\rightarrow\mathbb Z\sb 2$}. However, in practice it
could be more convenient to produce the sequence according to the law 
$x_{i+1}=f_{i\bmod 2^m}(x_i)\bmod 2^n$ than to the law 
$z_{i+1}=G(z_i)\bmod 2^{n+m}$ with further truncation, since the mapping
$G$ could be extremely complicated despite all $f_i$ are relatively simple.
As a bonus we have also that {\it all the results that are established further
in the paper for
truncated congruential generators remain true for generators of form 
$x_{i+1}=f_{i\bmod 2^m}(x_i)\bmod 2^n$}.
\end{note}

Using ideas of proposition \ref{WP-even} it is possible to handle a case
when $T$ acts on a set of odd order.
\begin{prop}
\label{WP-odd}
Let $m>1$ be odd; let, further, $\{f_0,\ldots,f_{m-1}\}$ be a finite sequence of compatible 
and ergodic mappings of $\mathbb Z_2$ onto itself, and let
$\{d_0,\ldots,d_{m-1}\}$  be a finite sequence of $2$-adic 
integers such that 
\begin{itemize}
\item $\sum_{j=0}^{m-1}d_j\equiv
0\pmod 2$, and 
\item the sequence 
$\{d_{i\bmod m}\bmod 2\colon i=0,1,2,\ldots\}$ is purely periodic 
with period length exactly $m$.
\end{itemize}
Put $H_j(x)=d_j\oplus f_j(x)$ {\rm (}respectively, $H_j(x)=d_j+f_j(x)${\rm)}.
Then the wreath product
$(H_j\bmod 2^n)\Wr_{j=0}^{m-1}I$, where $I(j)=(j+1)\bmod m$, 
defines a permutation $W\colon\mathbb Z/2^nm\twoheadrightarrow \mathbb Z/2^nm$
with a single cycle. 

Moreover, a recurrence  sequence $\mathcal W_n=\{x_i\in\mathbb Z/2^n\}$ defined by the relation  
$$x_{i+1}=H_{i\bmod m}(x_i)\bmod 2^n$$ 
is a strictly uniformly distributed 
purely periodic sequence with period length exactly $2^nm$ such that
every element of $\mathbb Z/2^n$ occurs at the period exactly $m$ times.
\end{prop}
Obviously, it is sufficient to prove only the second part of the statement.
We need the following
\begin{lem}
\label{le:WP-odd} 
Let $g_0,\ldots,g_{m-1}$ be a finite sequence of compatible
mappings of $\mathbb Z_2$ onto itself such that 
\begin{itemize}
\item $g_j(x)\equiv x+c_j\pmod 2$ for $j=0,1,\ldots,m-1$, 
\item $\sum_{j=0}^{m-1}c_j\equiv
1\pmod 2$, 
\item the sequence 
$\{c_{i\bmod m}\bmod 2\colon i=0,1,2,\ldots\}$ is purely periodic 
with period length exactly $m$,
\item $\delta_k(g_j(z))\equiv \zeta_k+\varphi_k^j(\zeta_0,\ldots,\zeta_{k-1})\pmod
2$, $k=1,2,\ldots$,
where $\zeta_r=\delta_r(z)$, $r=0,1,2,\ldots$, 
\item for each $k=1,2,\ldots$ an odd number of Boolean polynomials 
$\varphi_k^j(\zeta_0,\ldots,\zeta_{k-1})$ 
in Boolean variables $\zeta_0,\ldots,\zeta_{k-1}$ are of odd weight.
\end{itemize}
Then a recurrence sequence $\mathcal Y=\{x_i\in\mathbb Z_2\}$ defined by a relation 
$x_{i+1}=g_{i\bmod m}(x_i)$ is a strictly uniformly distributed sequence
over $\Z_2$: it is purely periodic modulo $2^k$ for all $k=1,2,\ldots$
with period length exactly $2^km$, and with each element of $\mathbb Z/2^k$ occuring at
the period exactly $m$ times. 
Moreover,
\begin{enumerate}
\item $2^{s+1}m$ is a {\rm (not necessarily exact, see definition \ref{def:strict})} 
period length of the sequence 
$\mathcal D_s=\{\delta_s(x_i)\colon i=0,1,2,\ldots\}$
 $(s=0,1,\ldots, k-1)$, 
\item $\delta_s(x_{i+2^{s}m})\equiv\delta_s(x_{i})+1\pmod
2$ for all $s=0,1,\ldots, k-1$, $i=0,1,2,\ldots$, 
\item for each $t=1,2,\ldots,k$ and each $r=0,1,2,\ldots$ the sequence 
$$x_r\bmod 2^t,x_{r+m}\bmod 2^t,x_{r+2m}\bmod 2^t,\ldots$$
is a purely periodic sequence of period length exactly $2^t$, and each element
of $\mathbb Z/2^t$ occurs at the period exactly once.
\end{enumerate} 
\end{lem}
\begin{note*}
In view of \ref{ergBool} the conditions of the lemma imply that all the
mappings $g_j$ preserve measure.
\end{note*}
\begin{proof}[Proof of lemma \ref{le:WP-odd}]
Since every $g_j$ induces a permutation modulo $2^n$ (see \ref{ergBool}),
the wreath product $(g_j\bmod 2^k)\Wr_{j=0}^{m-1}I$
is a permutation $R_k$ on $\mathbb Z/m\times\mathbb Z/2^k$; hence, the recurrence
sequence $\mathcal
Y_k$ defined by a relation $x_{i+1}=g_{i\bmod m}(x_i)\bmod 2^k$
is purely periodic.

%
%
%
We continue the proof of the lemma with induction on $k$. 
For $k=1$ one has 
$$x_{i+1}=(c_{i\bmod m}+x_i)\bmod 2,$$ 
Thus, $x_{i}\equiv x_0+\sum_{j=0}^{i-1}c_{j\bmod m}\pmod 2$, and we have
to calculate an exact length $P$ of a period  of a sequence 
$b_i=(\sum_{j=0}^{i-1}c_{j\bmod m})\bmod 2$ (see definition \ref{def:strict}). Yet  
$0\equiv\sum_{j=i}^{P+i-1}c_{j\bmod m}\pmod 2$ for all $i$; this means
that the sequence
$\mathcal C=\{c_{j\bmod m}\bmod 2\}$  is a linear recurrence 
sequence over a field $\mathbb
Z/2$
with characteristic polynomial $1+y+\cdots+y^{P-1}\in(\mathbb Z/2)[y]$
(see e.g. \cite{LinRec} for definitions). Since the latter polynomial is a factor
of a polynomial $y^P-1$, $P$ is a period length of the sequence $\mathcal
C$. Yet $m$ is an exact period length of the sequence $\mathcal
C$, so  $m$ must be a factor of $P$. Yet $x_{i+m}\equiv
x_0+\sum_{j=0}^{m-1}c_{j\bmod m}\equiv x_0+1\pmod 2$, and
$x_{i+2m}\equiv
x_0+2\cdot\sum_{j=0}^{m-1}c_{j\bmod m}\equiv x_0\pmod 2$; thus, $P=2m$.
This proves the lemma for $k=1$, since $\mathcal D_0=\mathcal Y_1$ in this
case.

%
Now let the lemma be true for $k=n$; consider $k=n+1$. 
Denote $\delta_n(x_i)=\chi_n^i$, then
\begin{equation}
\label{eq:WP-odd} 
\chi_n^i\equiv\chi_n^0+\sum_{j=0}^{i-1}
\varphi_n^j(\chi_0^j,\ldots,\chi_{n-1}^j)\pmod 2.
\end{equation}
Since by the induction hypothesis the period length of the sequence  $\mathcal
Y_n$ is exactly $2^nm$,
and since all $g_j$ are compatible, the period length of $\mathcal Y_{n+1}$
is a multiple of $2^nm$; thus only two cases are possible: the exact period length of
$\mathcal Y_{n+1}$ is either $2^{n+1}m$, or it is $2^nm$. We shall prove that the
latter case does not take place. To do this we only have to demonstrate
that $\chi_n^{2^mn}\not\equiv \chi_n^0\pmod 2$.
In view of the  induction hypothesis one has
\begin{multline}
\label{eq:WP-odd-1}
\chi_n^{2^nm+r}\equiv\chi_n^r+\sum_{j=r}^{2^nm-1+r}
\varphi_n^j(\chi_0^j,\ldots,\chi_{n-1}^j)\equiv \\
\chi_n^r+\sum_{j=0}^{m-1}\sum_{z\in\mathbb Z/2^n}\varphi_n^j(\zeta_0,\ldots,\zeta_{n-1})\equiv
 \chi_n^r+1 \pmod 2, 
\end{multline}
for all $r=0,1,2,\ldots$,
since an odd  number of Boolean polynomials $\varphi_n^0,\varphi_n^1,\ldots
\varphi_n^{m-1}$ are of odd
weight. This proves (2) of the lemma's statement; also, as  \eqref{eq:WP-odd-1}
implies $\chi_n^{2^mn}\not\equiv \chi_n^0\pmod 2$, the exact period
length of $\mathcal Y_{n+1}$ is $2^{n+1}m$ in view of the above note. Morover,
congruence \eqref{eq:WP-odd-1} 
implies $\chi_n^{2^{n+1}m+r}\equiv \chi_n^{r}\equiv\pmod 2$, thus proving
claim (1) of the lemma. Last, by claim (3) of the induction hypothesis
the following string of $2^nm$ numbers
\begin{equation*}
x_r\bmod 2^n,x_{r+m}\bmod 2^n,x_{r+2m}\bmod 2^n,\ldots,x_{r+(2^n-1)m}\bmod 2^n
\end{equation*}
is a permutation of $0,1,2,\ldots,2^n-1$. Hence, all  the numbers 
$$x_r,x_{r+m},x_{r+2m},\ldots,x_{r+(2^n-1)m}$$
are pairwise distict modulo $2^{n+1}$. Thus, for each $z\in\{0,1,\ldots,2^n-1\}$
among the numbers
\begin{equation}
\label{eq:WP-odd-3}
x_r,x_{r+m},x_{r+2m},\ldots,x_{r+(2^{n+1}-1)m}
\end{equation}
there exist exactly two numbers (say, $x_u$ and $x_v$) such that $u\ne
v$ and  $z\equiv
x_u\equiv x_v\pmod{2^n}$. Thus, $u\equiv v\pmod {2^nm}$ in view of claim
(3) of the induction hypothesis. Hence necessarily $v=u+\cdot 2^nm$. But
then $x_u\not\equiv x_v\pmod{2^{n+1}}$, since $\delta_n(x_v)\equiv\delta_n(x_v)+1\pmod
2$ in view of \eqref{eq:WP-odd-1}. Thus, all $2^{n+1}$ numbers of \eqref{eq:WP-odd-3}
are pairwise distinct modulo $2^{n+1}$. This proves claim (3) of the lemma.

Since, as we have already proved, the sequence $\mathcal Y_{n+1}$ is purely 
periodic with period length exactly $2^{n+1}m$, a finite sequence
$$x_0\bmod 2^{n+1},x_1\bmod 2^{n+1},\ldots,x_{2^{n+1}-1}\bmod 2^{n+1}$$ 
is
a period of $\mathcal Y_{n+1}$. But according to already proven claim 
(3), among these numbers there exist exactly $m$ numbers that are congruent
to $z$ modulo $2^{n+1}$ for each given $z\in\{0,1,\ldots,2^{n+1}-1\}$.
This completes the proof of the lemma.
\end{proof}
\begin{note*} Nowhere in  the proof of lemma \ref{le:WP-odd} we used that
$m$ is odd. Hence, the lemma holds for arbitrary, and not necessarily odd $m>1$.
\end{note*}
\begin{proof}[Proof of proposition \ref{WP-odd}.]
The proof of proposition \ref{WP-odd}  for a case $H_j(x)=d_j\oplus f_j(x)$
is now obvious in view of 
\ref{ergBool} and lemma \ref{le:WP-odd}: Note only that the sequence $\{d_j+1\:
j=0,1,2,\ldots\}$ satisfies conditions of the lemma.
So to finish the proof we only have to consider
a case $H_j=d_j+f_j(x)$.

The proof in the latter case goes along the lines similar to those of lemma \ref{le:WP-odd}.
Namely, for $n=1$ one has $x_{i+1}=(d_{i\bmod m}+x_i+1)\bmod 2$, 
since every ergodic mapping modulo $2$ is equivalent to the mapping $x\mapsto
x+1$, see \ref{Delta};
so putting $c_i=d_i+1$ returns us to the situation of lemma \ref{le:WP-odd}
whenever $n=1$.

Assuming the proposition is true for $n=k$ prove it for $n=k+1$.
In view of \ref{ergBool} we have that for $s>0$ 
$$\delta_s(H_j(x))\equiv \chi_s+(d_j+1)\chi_0\cdots\chi_{s-1}+
\psi_s^j(\chi_0,\ldots,\chi_{s-1})\pmod 2,$$
where $\deg \psi_s^j<s$ (this congruence could be easily proved by induction
on $s$: the coefficient of the monomial $\chi_0\cdots\chi_{s-1}$ in the
Boolean polynomial that represents a carry to $s$\textsuperscript{th} digit
is $\delta_0(d_j)$). Thus, for $k\ge 1$ one obtains
\begin{multline*}
\chi_k^{2^km}\equiv\chi_k^0
+\sum_{j=0}^{2^km-1}(d_{j\bmod m}+1)\chi_0^j\cdots\chi_{k-1}^j
+\sum_{j=0}^{2^km-1}
\psi_k^j(\chi_0^j,\ldots,\chi_{k-1}^j)\equiv \\
\chi_k^0+
\sum_{j=0}^{m-1}(d_j+1)\sum_{z\in\mathbb Z/2^k}\zeta_0\cdots\zeta_{k-1}+
\sum_{j=0}^{m-1}\sum_{z\in\mathbb Z/2^k}\psi_k^j(\zeta_0,\ldots,\zeta_{k-1})\equiv\\
 \chi_k^0+1 \pmod 2, 
\end{multline*}
since all Boolean polynomials $\psi_k^j(\zeta_0,\ldots,\zeta_{k-1})$ are
of even weight. This completes the proof of the proposition.
\end{proof}
\begin{exmp*}
A mapping $g_j(x)=x+(x^2\vee C_j)$ is ergodic
iff $\delta_0(C_j)=1$ and $\delta_2(C_j)=1$ (see  \ref{KlSh-3}). Let
a sequence $\{d_j\colon j=0,1,2,\ldots\}$ satisfy conditions of  proposition
\ref{WP-odd}.
Then the sequence $\{x_{i+1}=x_i+d_i+(x_i^2\vee C_i)\bmod 2^n\colon i=0,1,2,\ldots\}$
is purely periodic modulo $2^k$ for all $k=1,2,\ldots$
with period length $2^km$, and each element of $\mathbb Z/2^k$ occurs at
the period exactly $m$ times.

This is another variation of theme of \cite[Theorem 3]{KlSh-2}. Note that
we prove a somewhat stronger claim: Not only a sequence of pairs $(y_i,
x_i)$ defined by $y_{i+1}=(y_i+1)\bmod m$; $x_{i+1}=(x_i+d_i+(x_i^2\vee C_{y_i}))\bmod 2^n$
is periodic with period length $2^nm$, yet the period length of the sequence
$\{x_i\}$ is $2^nm$. The latter could never be achieved under the conditions
of Theorem 3 of \cite{KlSh-2}: They imply that the period length of the
sequence $\{x_i\pmod 2\}$ is $2$, and not $2m$.
\end{exmp*}
\begin{note*}
Obviously, after corresponding restatement proposition \ref{WP-odd}, as
well as lemma \ref{le:WP-odd}, remain true
for arbitrary permutation $I\colon\mathbb Z/m\twoheadrightarrow\mathbb Z/m$
with a single cycle.  
\end{note*}
In connection with proposition \ref{WP-odd} there arises a natural
question: how to construct a sequence $\{d_j\}$ that satisfies its conditions?
\begin{prop}
\label{prop:WP-odd:constr}
Let $m>1$ be odd, and let $u\:\Z/m\>\Z/m$ be an arbitrary permutation with a
single cycle. Choose
arbitrary $z\in\Z/m$ and
set $d_{i}=u^{(i)}(z)\bmod m$, if $m\equiv 1\pmod 4$, or 
set $d_{i}=(u^{(i)}(z)+1)\bmod m$ otherwise $(i=0,1,2,\ldots)$.  Then 
the  sequence  $\mathcal D=\{d_i\}$ satisfies conditions
of proposition \ref{WP-odd}: that is, $\mathcal D$ is purely periodic with
period length exactly $m$, and $\sum_{j=0}^{m-1}d_j\equiv 0\pmod 2$.
\end{prop}
\begin{proof}
Obviously, the sequence $\mathcal D$ is purely periodic. Let $P$
be the period length  of $\mathcal D$. Thus, $P$ is a factor
of $m$. 
Note that since $m=2s+1$, exactly $s$ numbers of $0,1,\ldots,m-1$ are odd.  
Denote $r_0$ (respectively, $r_1$) the number of even (respectively,
odd) numbers at the period of $\mathcal D$: so $\frac{m}{P}r_1=s$, and
$\frac{m}{P}r_0=s+1$. Thus, $\frac{m}{P}(r_0-r_1)=1$; hence $\frac{m}{P}=1$.
So, the period length of $\mathcal D$ is exactly $m$. The result now follows
since $\sum_{i=0}^{m-1}i\equiv 0\pmod 2$ iff $s\equiv 0\pmod 2$.
\end{proof}
\begin{note}
\label{note:WP-odd:constr}
Thus, to construct a sequence $\{d_j\}$ of proposition \ref{WP-odd} it
is sufficient to construct a permutation with a single cycle modulo $m$.
Of course, this could be done in various ways, depending on extra conditions
the whole generator should satisfy. For instance, if one intends to use maximum
of memory calls instead of computations on the fly,
he can merely take an arbitrary array of $\{0,1,\ldots, m-1\}$ in arbitrary
order. 
On the contrary, if one needs to produce $d_j$ on the fly, he could
construct a corresponding generator modulo $m$ with a compatible state transition
function and a bijective modulo $m$ output function. This could be done e.g. 
with the use of  
\ref{ergPolGen}, \ref{ergAn}, \ref{ergAnGen},  and \ref{Delta}. 
In case
$m=2^k-1$ an alternative 
way is to use linear recurrence sequences of maximum period over $\Z/2$:
note that often  sequences of this kind could be constructed with the use 
of $\XOR$'s and left-right shifts only, see e.g. \cite{Mars}.
\end{note}

The above results of this subsection show how to construct a sequence $x_{i+1}=f_{i\bmod
m}(x_i)\bmod 2^n$ of maximum period length $2^nm$ in two cases: when $m$
is odd, and when $m=2^k$. Now we consider a general case of arbitrary $m>1$.
\begin{thm}
\label{thm:WP}
Let $\mathcal G=\{g_0,\ldots,g_{m-1}\}$ be a finite sequence of 
compatible measure preserving
mappings of $\mathbb Z_2$ onto itself such that
\begin{enumerate}
\item the sequence $\{(g_{i\bmod m}(0))\bmod 2\colon i=0,1,2,\ldots\}$ is a
purely periodic sequence with period length exactly $m$;
\item $\sum_{i=0}^{m-1}g_i(0)\equiv 1\pmod 2$;
\item $\sum_{j=0}^{m-1}\sum_{z=0}^{2^k-1}g_j(z)\equiv 2^{k}\pmod {2^{k+1}}$
for all $k=1,2,\ldots$ .
\end{enumerate}
Then the recurrence sequence $\mathcal Z$ defined by the relation $x_{i+1}=g_{i\bmod
m}(x_i)$ is strictly uniformly distributed modulo $2^n$ for all $n=1,2,\ldots:$
i.e., modulo each $2^n$ it is  a purely periodic sequence with 
period length exactly
$2^nm$ and with each element of $\mathbb Z/2^n$ occuring at the period
exactly $m$ times. 
\end{thm}
\begin{note*}
Since in view of \ref{ergBool}
a compatible mapping $g_i\colon\mathbb Z_2\rightarrow\mathbb Z_2$ preserves
measure iff 
$$\delta_k(g_i(x))\equiv \chi_k+\varphi_k^i(\chi_0,\ldots,\chi_{k-1})\pmod
2,$$
where $\chi_s=\delta_s(x)$ $(s=0,1,2,\ldots)$, the {\it condition } (3)
{\it of theorem \ref{thm:WP} could be replaced by the equivalent condition
$$\sum_{j=0}^{m-1}\wt\varphi_k^j\equiv 1\pmod 2 \qquad (k=1,2,\ldots),$$}
where $\wt\varphi_k^j$ is a weight of the Boolean polynomial $\varphi_k^j$
in variables $\chi_0,\ldots,\chi_{k-1}$. In turn, since for every Boolean
polynomial $\varphi$ in variables $\chi_0,\ldots,\chi_{k-1}$ holds $\wt\varphi\equiv
\Coef_{0,\ldots,k-1}(\varphi)\pmod 2$, where $\Coef_{0,\ldots,k-1}(\varphi)$
stands for a coefficient of the monomial $\chi_0\cdots\chi_{k-1}$ in the
Boolean polynomial $\varphi$, the {\it latter condition could be also replaced
by
$$\sum_{j=0}^{m-1}\Coef_{0,\ldots,k-1}(\varphi_k^j)\equiv 1\pmod 2 \qquad (k=1,2,\ldots),$$
or by
$$\sum_{j=0}^{m-1}\bigg\lfloor\frac{\deg\varphi_k^j}{k}\bigg\rfloor\equiv 1\pmod {2} \qquad (k=1,2,\ldots).$$} 
\end{note*}
\begin{proof}[Proof of theorem \ref{thm:WP}.]
Practically everything is already done during the proof of \ref{le:WP-odd}:
we just note that congruence \eqref{eq:WP-odd-1} now holds in view of condition
(3) of the theorem.
\end{proof}
\begin{note*} For $m=1$ theorem \ref{thm:WP} turns into ergodicity criterion
\ref{ergBool}: so theorem \ref{thm:WP} could be considered as a generalization
of this criterion.
\end{note*}
Theorem \ref{thm:WP} is our main technical tool in constructing automata with 
strictly uniformly
distributed recurrence sequences $x_{i+1}=f_i(x_i)$ of internal
states outputting strictly uniformly distributed sequences
of the form $F_0(x_0), F_1(x_1),\ldots$ .
The above mentioned results (e.g. \ref{WP-even-more}and \ref{WP-odd}) could
be derived from theorem \ref{thm:WP}, as well as new results for even $m$ that is
not power of 2 could also be obtained with the use of it:
\begin{exmp*}
For instance, take odd $s$, $1\le s<m$, and take $s$ arbitrary compatible and ergodic
mappings $g_j\colon\mathbb Z_2\rightarrow\mathbb Z_2$, $(j=0,1,\ldots,s-1)$.
Take $m-s$ arbitrary compatible and measure preserving mappings
$h_k\colon\mathbb Z_2\rightarrow\mathbb Z_2$, and
set $g_k(x)=x\oplus 2h_k(x)$ $(k=s,s+1,\ldots,m-1)$. Then in view of \ref{ergBool}
it is easy to see that a finite sequence $\{g_i\colon i=0,1,\ldots, m-1\}$
satisfies conditions of theorem \ref{thm:WP}, and thus the recurrence sequence
$x_{i+1}=g_{i\bmod m}(x_i)$ is strictly uniformly distributed modulo $2^n$
for all $n=1,2,\ldots$ .
\end{exmp*}
\begin{note}
\label{note:halfper-odd}
During the proof of theorem \ref{thm:WP} and of lemma \ref{le:WP-odd}
we have demonstrated that {\it every 
$j$\textsuperscript{th} coordinate sequence $\mathcal D_j=\{\delta_j(x_i)\colon i=0,1,2,\ldots\}$
$(j=0,1,2,\ldots)$
is a purely periodic binary sequence of period length $2^{j+1}m$, and the
second half of the period is a bitwise negation of the first half}: $\delta_j(x_{i+2^jm})\equiv
\delta_j(x_i)+1\pmod 2$, $i=0,1,2,\ldots$  (see claims (1)--(2) of 
lemma \ref{le:WP-odd}). Note, however, that {\slshape
the exact
period length $P$ of the sequence 
$\{\delta_j(x_i)\colon i=0,1,2,\ldots\}$ could actually be less than $2^{j+1}m$},
i.e., $P\big|2^{j+1}m$, yet not necessarily $P=2^{j+1}m$ (however, $P$
is always a multiple of $2^{j+1}$, see \ref{thm:lincomp:sharp}). Indeed, the sequence
$101010\ldots$ is a purely periodic sequence with period $10$ of length $2$; at
the same time it could be considered as a purely periodic sequence with
period $101010$ of length $6$. Note that in both cases the second half
of the period is a bitwise negation of its first half. Such an effect could
never occur for $j=0$, since $\mathcal D_0=\mathcal Y_1$, and the latter
sequence has period length exactly $2m$ in view of lemma \ref{le:WP-odd}.
However, this effect could occur for senior coordinate sequences. For instanse,
let $\mathcal D_0$ be a purely periodic sequence with period $111000$;
let $\mathcal D_1$ be a purely periodic sequence with period $110011001100$.
The exact period length of $\mathcal D_1$ is $4$; yet it could be considered
as a sequence of period $12$, and the second half of the period is a bitwise
negation of the first half. The sequence $\mathcal Y_2$ in this case is
a purely periodic sequence with period $331022113200$. It is not difficult
to demonstrate that this sequence $\mathcal Y_2$ satisfy lemma \ref{le:WP-odd},
i.e., one could construct mappings $g_0,g_1,g_2$ satisfying the lemma,
such that outputted sequence $\mathcal Y_2$ 
is our sequence with period  $331022113200$. A characterization 
of possible
output sequences is given by theorem \ref{thm:WP:AnyHalfPer} further. 
\end{note}

Finally we consider a case of wreath products of automata with non-identity
output functions.
\begin{cor}
\label{cor:WP}
Let  a finite sequence  of mappings $\{f_0,\ldots,f_{m-1}\}$ of $\mathbb
Z_2$ into itself satisfy conditions
of theorem \ref{thm:WP}, and let $\{F_0,\ldots,F_{m-1}\}$ be an arbitrary
finite sequence of equiprobable {\rm (}and not necessarily compatible{\rm
)} mappings of $\mathbb Z/2^n$ $(n\ge 1)$ onto $\mathbb
Z/2^k$, $1\le k\le n$. Then the sequence
$\mathcal F=\{F_{i\bmod m}(x_i)\colon i=0,1,2\ldots\}$, where $x_{i+1}=f_{i\bmod m}(x_i)\bmod
2^n$, is strictly uniformly distributed over $\mathbb Z/2^k:$ It is purely
periodic with period length $2^nm$, and each element of $\mathbb Z/2^k$
occurs at the period  exactly $2^{n-k}m$ times.
\end{cor}
\begin{proof}
Obvious: combine  claim (3) of lemma
\ref{le:WP-odd} and proposition \ref{prop:Auto}. 
\end{proof}

Note that the results of this subsection  could be extended to cover the
case $p$ odd, that is, to the case of wreath products of the form 
$H_j\Wr_{j=0}^{p^m-1}T$, where $T\:\Z/p^m\>\Z/p^m$ (and even for
$H_j\Wr_{j=0}^{m-1}T$, where $T\:\Z/m\>\Z/m$, $m>1$ arbitrary rational
integer). This case is also of cryptographic importance: the corresponding
techniques could be used e.g. to construct sequences of type $\mathcal D$
of proposition \ref{prop:WP-odd:constr}. However, this is an issue of a
forthcoming paper.
\subsection*{Equalizing period lengths of coordinate sequences.} All the
generators with the identity output function considered above
demonstrate a property, which is already mentioned at the beginning
this section, and 
which in loose terms could
be stated as follows: {\slshape Less significant
bits of output have smaller periods}.  To be more exact, despite
for any of these automata the corresponding 
output sequence $\mathcal S=\{s_0,s_1,\ldots\}$ 
over $\Z/2^n$ 
is always purely periodic 
of period length exactly $2^n\ell$ (where $\ell=2^m$ for sequences outputted
by wreath products of automata described
by \ref{WP-even} or \ref{WP-even-trunc}, $\ell=m$ in case the wreath products
are of \ref{WP-odd}, \ref{le:WP-odd},
or \ref{thm:WP}, and $\ell=1$ for  congruential
generators of a maximum period length), the $j$\textsuperscript{th} coordinate
sequence $\delta_j(\mathcal S)=\{\delta_j(s_0),\delta_j(s_1),\ldots\}$ could
be of smaller period length (see e.g. note \ref{note:halfper-odd} above).
In fact, as it is shown further, the exact period length of the $j$\textsuperscript{th} coordinate
sequence of congruential generator of a maximum period length is $2^{j+1}$
(see \ref{halfper}); it is a factor of $2^{j+1}\ell$ and a multiple of 
(which is possibly equal to) $2^{j+1}$ 
for wreath products of generators (see \ref{thm:lincomp:sharp}). So only senior
coordinate sequence $\delta_{n-1}(\mathcal S)$ may achieve exact period
length $2^n\ell$; at least, the exact period length of it is not less than
$2^n$. 
Nothing more could be said either if we use general non-identity equiprobable
output
functions (see  \ref{prop:Auto} and \ref{cor:WP}). However, such a ``disbalance"
of periods could be cured if we apply non-identity output functions in some
special way.

Namely, let $\pi=\pi_n^1$ be a bit order reversing permutation on $\Z/2^n$,
which was defined in section \ref{Prelm}, and let $h_i$ $(i=0,2,\ldots,m-1)$
be compatible and ergodic mappings of $\Z_2$ onto itself. Then the composition
$F_i(x)\colon x\mapsto (h_i(\pi(x)))\bmod 2^n$ $(x\in\{0,1,\ldots,2^n-1\})$
is a bijective mapping of $\Z/2^n$ onto itself. We argue that if we take
$F_i$ as an output function, then the sequence $\mathcal F$ of \ref{cor:WP}
is free of less significant bit effect mentioned above. To be more exact,
the following proposition holds:
\begin{prop}
\label{prop:reverse}
Let $h_i$, $i=0,1,2,\ldots,m-1$,
be compatible and ergodic mappings of $\Z_2$ onto itself. Define
$F_i\colon\Z/2^n\>\Z/2^n$ by
$F_i(x)=(h_i(\pi(x)))\bmod 2^n$ $(x\in\{0,1,\ldots,2^n-1\})$,
where $\pi=\pi_n^1$ is a bit order reversing permutation on $\Z/2^n$ {\rm(see
Section \ref{Prelm} for the definition of the latter)}. Consider
a sequence $\mathcal F$ over $\Z/2^n$ defined in \ref{cor:WP}.
Then the exact period length of the $j$\textsuperscript{th}
coordinate sequence $\delta_j(\mathcal F)$ $(j=0,1,2,\dots,n-1)$ 
is $2^nk_j$, where $1\le k_j\le\ell$. 

Moreover, the same holds if $m=1$
{\rm (}and whence $\ell=1${\rm )}, i.e., when $\mathcal F$ is an output
sequence of the automaton ${\mathfrak A}=\langle N,M,\bar f,F,u_0\rangle $,
where $N=M=\Z/2^n$, $\bar f=f\bmod 2^n$, $f$ and $h$ are compatible
and ergodic mappings of $\Z_2$ onto itself, 
$F(x)=(h(\pi(x)))\bmod 2^n$, $x\in\{0,1,\ldots,2^n-1\}${\rm :} The exact period length of the $j$\textsuperscript{th}
coordinate sequence $\delta_j(\mathcal F)$ is $2^n$ for all $j=0,1,2,\dots,n-1$.  
\end{prop}
\begin{note*} Hence, $\mathcal F$ is a purely periodic sequence of period
length exactly $2^nm$, and with each element of $\Z/2^n$ occuring at the
period exactly $m$ times (see \ref{cor:WP},\ref{prop:Auto}).
\end{note*}
To prove this proposition we need the following easy
\begin{lem}
\label{le:reverse}
Let $\mathcal X=\{x_i\: i=0,1,2,\dots\}$ and $\mathcal Y=\{y_i\: i=0,1,2,\dots\}$ be purely periodic
sequences
over $\Z/2$ with exact period lengths $2^u$ and $2^v$, respectively, and
let $u>v$. Then the sequence 
$\mathcal X\oplus\mathcal Y=\{x_i\oplus y_i\: i=0,1,2,\dots\}$ is purely
periodic with period length exactly $2^u$.

If, additionally, $x_{i+2^{u-1}}\equiv x_i+1\pmod 2$ for all $i=0,1,2,\ldots$,
and if $\mathcal Y$ is a non-zero sequence, then the sequence
$\mathcal X\odot\mathcal Y=\{x_i\cdot y_i\: i=0,1,2,\dots\}$ is 
purely periodic with period length exactly
$2^u$. 
\end{lem}
\begin{proof}[Proof of lemma \ref{le:reverse}.]
The first assertion of the lemma is obvious. To prove the second one assume
$P$ is the exact period length of the sequence $\{x_i\cdot y_i\: i=0,1,2,\dots\}$.
Then $P=2^s$ for suitable $s\le u$. Yet if $s<u$, then 
$x_{i+2^{u-1}}\cdot y_{i+2^{u-1}}\equiv x_i\cdot y_i\pmod 2$ for all $i=0,1,2,\ldots$;
thus $(x_i+1)\cdot y_i\equiv x_i\cdot y_i\pmod 2$ and hence $y_i\equiv 0\pmod
2$ for all $i=0,1,2,\ldots$. A contradiction.
\end{proof}
\begin{proof}[Proof of proposition \ref{prop:reverse}.] In view of assertions
(2) and (3) of lemma \ref{le:WP-odd}, each subsequence 
$\mathcal F(r)=\{z_{r+tm}\:t=0,1,2,\ldots\}$,  $r=0,1,\ldots,m-1$, of the
sequence $\mathcal F=\{z_i\:i=0,1,2,\ldots\}$ satisfies the following condition:
Each coordinate sequence $\delta_j(\mathcal F(r))$ is a purely periodic
sequence of period length exactly $2^{j+1}$, and the second half of the
period is a bitwise negation of the first half, i.e., $\delta_j(z_{r+(t+2^j)m})\equiv
\delta_j(z_{r+tm})+1\pmod 2$ for all $t=0,1,2,\ldots$. Thus, in view of
theorem \ref{AnyHalfPer}, which is proved further, the sequence $\mathcal F(r)$
is an output sequence of a suitable automaton $\mathfrak B=\langle \Z_2,\Z/2^n,f,\bmod
2^n,z_r\rangle$, where $f$ is a compatible and ergodic mapping of $\Z_2$
onto itself. Thus, the first assertion of the proposition follows from
the second one, i.e., it is sufficient to consider only a case $m=1$.

Now represent $h$ in a Boolean form according to \ref{ergBool}. So,
$$\delta_j(h(x))\equiv \chi_j+\varphi_j(\chi_0,\ldots,\chi_{j-1})\pmod 2,$$
where $\chi_k=\delta_k(x)$, and $\varphi_j$ is a Boolean polynomial of odd
weight in Boolean variables $\chi_0,\ldots,\chi_{j-1}$ for $j>0$, $\varphi_0=1$.
Note that for $j>0$
\begin{multline}
\label{eq:reverse}
\delta_j(h(x))\equiv \chi_j+\chi_0\cdot\chi_1\cdots\chi_{j-1}+
\psi_j(\chi_0,\ldots,\chi_{j-1})\equiv \\
\chi_j+\chi_0\cdot\alpha_j(\chi_1,\ldots,\chi_{j-1})+\beta_j(\chi_1,\ldots,\chi_{j-1})
\pmod 2,
\end{multline}
where $\psi_j,\alpha_j,\beta_j$ are Boolean polynomials of corresponding
Boolean variables, and $\deg\psi_j<j$, so $\alpha_j$ is a non-zero polynomial.  

For binary sequences $\mathcal U, \mathcal V, \mathcal W, \ldots$ (which
could be treated as $2$-adic integers) and a Boolean polynomial 
$\gamma(\upsilon,\nu,\omega,\ldots)$ of Boolean variables $\upsilon,\nu,\omega,\ldots$
denote $\gamma(\mathcal U, \mathcal V, \mathcal W, \ldots)$ a binary sequence
$\mathcal S$ (thus, a $2$-adic integer) such that
$$\delta_j(\mathcal S)\equiv
\gamma(\delta_j(\mathcal U),\delta_j(\mathcal V),\delta_j(\mathcal W),\ldots)\pmod2,$$
for all $j=0,1,2,\ldots$. Loosely speaking, we just substitute, respectively,
$\XOR$ 
and $\AND$ for $+$ and $\cdot$ in the Boolean polynomial $\gamma$ and let
variables $\upsilon,\nu,\omega,\ldots$ run through the space $\Z_2$ of
$2$-adic integers. Thus we obtain a well defined multivariate function $\gamma$
on $\Z_2$ valuated in $\Z_2$. Since there is a natural one-to-one correspondence
between infinite binary sequences and $2$-adic integers, the sequence 
$\gamma(\mathcal U, \mathcal V, \mathcal W, \ldots)$ is well defined. Note
also that treating binary sequences as $2$-adic integers enables one to
produce infinite  sequences of $n$-bit rational integers out of $n$ infinite binary sequences
in an obvious manner: Say, $\mathcal U+2\cdot\mathcal V+4\mathcal W$ is a
sequence $\mathcal N=\{n_0,n_1,\ldots\in\N_0\}$ such that $n_j=\delta_j(\mathcal
U)+2\cdot\delta_j(\mathcal V)+4\cdot\delta_j(\mathcal W)$ for $j=0,1,2\ldots$.
For instance, if $\mathcal U=101\ldots$, $\mathcal V=110\ldots$, and 
$\mathcal W=010\ldots$, then $\mathcal N=361\ldots$ is a sequence over
$\{0,1,\ldots,7\}=\Z/8$.

Proceeding with these conventions, let $\mathcal C_j$ (respectively, $\mathcal D_j$) be the $j$\textsuperscript{th} output sequence of the
automaton $\mathfrak B$ (respectively, $\mathfrak A$). Let $\mathcal E=111\ldots$. Then in view of
\eqref{eq:reverse} one has:
\begin{gather*}
\mathcal D_0=\mathcal C_{n-1}\oplus \mathcal E;\\
\mathcal D_1=\mathcal C_{n-2}\oplus\mathcal C_{n-1}\oplus \mathcal B;\\
\mathcal D_j=\mathcal C_{n-j-1}\oplus\mathcal C_{n-1}\odot
\alpha_j(\mathcal C_{n-2},\ldots,\mathcal C_{n-j})\oplus
\beta_j(\mathcal C_{n-2},\ldots,\mathcal C_{n-j})\qquad(j\le 2),
\end{gather*}
where $\mathcal B=\beta_1\beta_1\beta_1\ldots$ is a constant binary sequence.
Note that  $\mathcal C_i$ is purely periodic
binary sequence of period length exactly $2^{i+1}$, and the second half
of the period is a bitwise negation of the first half, see \ref{halfper}
further. This completes the proof
of proposition \ref{prop:reverse}
in view of lemma \ref{le:reverse} and conventions made above, if we prove
that the sequence
$\alpha_j(\mathcal C_{n-2},\ldots,\mathcal C_{n-j})$, $2\le j\le n-1$,
is a non-zero binary
sequence.

Consider a sequence
$\mathcal G_j=2^{n-2}\cdot\mathcal C_{n-2}+\dots+2^{n-j}\cdot\mathcal C_{n-j}$ 
over $\Z/2^{j-1}$. The latter sequence is just an output sequence of the
automaton $\mathfrak G_j=\langle \Z/2^{n-1},\Z/2^{j-1},f\bmod 2^{n-1},
T_{n-j-1}, u\rangle$, where $T_{n-j-1}$ is a truncation of the first
$n-j$ low order bits: $T_{n-j-1}(z)=\lfloor\frac{z}{2^{n-j}}\rfloor$. 
Thus, $\mathcal G_j$ is a purely periodic sequence
of period length exactly $2^{n-1}$ and with each element of  $\Z/2^{j-1}$
occuring at the period the same number of times. Yet
$\alpha_j$  is a non-zero Boolean polynomial (see above); thus it takes
value $1$ at least at one $(j-1)$-bit word of $\Z/2^{j-1}$. Consequently, at least
one member of the sequence $\alpha_j(\mathcal C_{n-2},\ldots,\mathcal C_{n-j})$
is $1$.
\end{proof}
\begin{note*} There are other methods that improve periods of coordinate
sequences. For insatnce, using the ideas of the proof of \ref{prop:reverse}
it is not difficult to demonstrate that {\it if a recurrence sequence is
defined by a relation $x_{i+1}=f(x_i)$, where $f\:\Z_2\>\Z_2$ is compatible
and ergodic mapping, then a binary sequence 
$\{\delta_k(x_i+2^j\cdot\delta_s(x_i))\:i=0,1,2,\ldots\}$ is purely periodic with
period length exactly $2^s$ whenever $j\le k<s$}. From here it could be
deduced that e.g. the  sequence 
$$\mathcal Z=\Big\{\Big(x_i+\pi_k^1\Big(\Big\lfloor\frac{x_i}{2^k}\Big\rfloor\bmod
2^k\Big)\Big)\bmod 2^k\:i=0,1,2,\ldots\Big\}$$ 
is a purely periodic sequence over $\Z/2^k$ of period length exactly 
$2^{2k}$,
such that each element of $\Z/2^k$ occurs at the period exactly $2^k$ times,
and that each coordinate sequence of $\mathcal Z$ is purely periodic binary sequence of
period length exactly $2^{2k}$. Note that $\mathcal Z$ is obtained according
to a very simple rule: at the $i$\textsuperscript{th} step take $(2k)$-bit output
of congruential generator of a maximum period length with state transition
function $f$, read the second half of this output as a $k$-bit 
number in reverse bit order and
add this number modulo $2^k$ to the $k$-bit number that agrees with the first
half of the output. 
\end{note*}

%
\section{Properties}
\label{sec:Prop}
In this section we study common  probabilistic, cryptographic and other properties
of output sequences of the generators considered in preceeding sections:
Linear and $2$-adic spans
of these sequences, 
their structure, distribution of $k$-tuples in them, etc.
We begin our study with properties of coordinate sequences of the automata
considered above, that is, 
of the sequences $\{\delta_j(s_{i})\colon i=0,1,2,\ldots\}$, where $\{s_i\}$
is the output sequence of the automaton.
\subsection*{Properties of coordinate sequences}
To study coordinate sequences it is convenient to consider an 
automaton $\mathfrak A^\prime$ with a state set $\mathbb Z_2$, compatible
and ergodic state transition function $f\colon\mathbb Z_2\rightarrow\mathbb
Z_2$
and with identity
output function $F(z)=z$. We also consider an automaton $\mathfrak A^\prime_j$
which differs from $\mathfrak A^\prime$ only by the output function, which
is $\delta_j(z)$ in this case. 
Thus the output sequence of $\mathfrak A^\prime_j$ is just 
the $j$\textsuperscript{th} coordinate
sequence $\mathcal S_j=\{s_i=\delta_j(f^{(i)}(z)): i=0,1,2,\dots\}$ of the
automaton $\mathfrak A^\prime$ (here
$z\in\mathbb Z_2$ is the initial state of the automaton $\mathfrak
A^\prime$). Note that since $f$ is compatible, we may assume
if necessary that
$z\in\mathbb Z/2^{j+1}$, i.e., that all but possibly the first $j+1$ junior
bits of $2$-adic representation of $z$ are $0$. That is, the output sequence
of the automaton $\mathfrak A^\prime_j$ is the same as the one of the automaton
$\mathfrak A=\langle \mathbb Z/2^{j+1}, \mathbb Z/2, f\bmod 2^{j+1},\delta_j,z\bmod 2^{j+1}\rangle$,
see Section \ref{Prelm}.

It turnes out that the $j$\textsuperscript{th} coordinate
sequence has rather specific structure. Namely, the following theorem
holds.
\begin{thm}
\label{halfper}
The $j$\textsuperscript{th} coordinate sequence is purely periodic, and
$2^{j+1}$ is the length of its period.
The second half of the period is a bitwise negation of its first half, i.e.,
$s_{i+2^j}\equiv s_{i}+1\pmod 2$
for each $i=0,1,2,\ldots$.  
\end{thm}
\begin{proof} 
Since the mapping $f\colon\mathbb Z_2\rightarrow\mathbb Z_2$ is compatible and ergodic,
the sequence $\{x_{i+1}=f(x_i)\bmod 2^{j+1}:i=0,1,2,\dots\}$ is purely periodic, 
with $2^{j+1}$ being the length of its period, whereas the sequence 
$\{x_{i+1}=f(x_i)\bmod 2^{j}:i=0,1,2,\dots\}$ 
is purely periodic, and the length of its period is exactly
$2^{j}$. Yet $x_{i+1}\bmod 2^{j+1}=x_{i+1}\bmod 2^{j}+2^j\delta_j(x_{i+1})$,
and the first assertion of \ref{halfper} follows.

Supposing $\delta_j(x_{i+1})=\delta_j(x_{i+1+2^j})$ for some $i$, from 
the preceeding equality one obtains $x_{i+1+2^j}\equiv x_{i+1}\pmod {2^{j+1}}$,
and hence $x_{i+t+1+2^j}\equiv f^{(t)}(x_{i+1+2^j})\equiv f^{(t)}(x_{i+1})\equiv
x_{i+t+1}\pmod{2^{j+1}}$ for all $t=0,1,2,\dots$, in view of compatibility of $f$.
So the length of the period of the sequence $\{x_i\bmod 2^{j+1}:i=0,1,2,\dots\}$ 
does not exceed $2^j$, in contradiction with the ergodicity of $f$, see
\ref{def:erg}.
\end{proof}
\begin{note}
\label{note:halfper}
Theorem \ref{halfper} could be generalized in two directions. First, to
output sequences of wreath products of automata (this is already done,
see \ref{note:halfper-odd}),
and second, to the case $p$ odd.

In the latter case provided transformation $f\colon\mathbb Z_p\rightarrow\mathbb Z_p$ is
compatible
and ergodic, the $j$\textsuperscript{th} coordinate sequence $\{\delta_j(f^{(i)}(z)):i=0,1,2,\dots\}$
is purely periodic, with  $p^{j+1}$ being the length of its period (here
and further within this remark
$\delta_j(z)$ stands for the $j$\textsuperscript{th} digit in base-$p$
expansion of $z$). Each subsequence 
$\{\delta_j(f^{(i+p^t)}(z)):t=0,1,2,\dots\}$ is a purely periodic sequence
with  $p$ being the length of period; 
moreover, for $j>0$ it is generated by a linear congruential 
generator modulo
$p$, i.e., by a polynomial $a+x$ for appropriate $a\in\{1,2,\dots,p-1\}$.
So this sequence is
strictly uniformly distributed modulo
$p$: each $u\in\mathbb Z/p$ occurs at the period exactly once. 
%
The generator $\delta_0(f^{(i)}(z))$ is a (generally speaking, nonlinear)
congruential generator of the form $v_{i+1}\equiv g(v_i)\pmod p$ for an
appropriate
transitive modulo $p$
polynomial $g(x)$ over  a field $Z/p$ of residues modulo $p$.  

A proof of this assertion could be deduced from the  proof of theorem
3.4 of \cite{me-2} since in view of the $p$-adic Weierstrass theorem (see
\cite{Mah}) a transformation $z\mapsto f(z)\bmod p^{j+1}$ of the residue
ring $\mathbb Z/p^{j+1}$ may be considered
as a polynomial transformation $z\mapsto w(z)\bmod p^{j+1}$ induced by
an
integer-valued and compatible polynomial $w(x)\in\mathbb Q[x]$, i.e., by
a polynomial of the form mentioned in \ref{ergBin}. Thus the
mapping $z\mapsto f(z)\bmod p^{j+1}$ could be considered as a reduction modulo
$p^{j+1}$ of the compatible and ergodic mapping $w\colon\mathbb Z_p\rightarrow\mathbb
Z_p$; the latter mapping is uniformly differentiable everywhere on $\mathbb Z_p$. Hence
the assumptions of theorem 3.4 of \cite{me-2} are satisfied. We omit
further details.  
\end{note}
We recall that a linear complexity $\Psi_F(\mathcal S)$ 
of the sequence $\mathcal S=\{s_i\:i=0,1,2,\ldots\}$
over a field $F$ is the smallest $n\in\mathbb N$ such that every $n$ succesive
members of the sequence satisfy some non-trivial linear relation of length $n+1$,
i.e., there exist $a_0,a_1,\ldots,a_n$, not all equal to $0$, such that
 $a_0s_i+a_1s_{i+1}+\dots +a_ns_{i+n}=0$ for all $i=0,1,2,\ldots$. In this
 case we also say that
the polynomial $a_0+a_1x+\dots+a_nx^n\in F[x]$ {\it annihilates} $\mathcal S$
\footnote{A polynomial that annihilates $\mathcal S$ is also
called  a {\it characteristic polynomial of the sequence $\mathcal S$}.}.
In other
words, linear complexity is just a degree of the minimal polynomial of $\mathcal
S$ (the minimum degree nonzero polynomial that annihilates $\mathcal
S$; a polynomial $g(x)\in F[x]$ annihilates $\mathcal S$ iff the minimal
polynomial of $\mathcal S$ is a factor of $g(x)$ --- see e.g. \cite{LinRec} or
\cite{LidNied} for 
references). In case $F=\mathbb Z/p$ is a field of $p$ elements we 
use for linear complexity over $F$ the notation $\Psi_p$ rather than $\Psi_{\mathbb
Z/p}$. 

Linear complexity
is one of crusial for cryptography  properties: 
Pseudorandom generators
that produce sequences of low linear
complexity are not secure, since having relatively short  segment of output
sequence  
and solving a corresponding system of linear equations over $F$ a cryptoanalyst
could find  $a_0,a_1,\ldots,a_n$ and thus predict
with probability $1$ the rest of the members of the sequence. Of course, high
linear complexity per se does not guarantee security.  
\begin{thm}
\label{lincomp}
The linear complexity $\Psi_2(\mathcal S_j)$ of the $j$\textsuperscript{th} 
coordinate sequence $\mathcal S_j$ is exactly $2^j+1$.
\end{thm}
We need the following lemma: 
\begin{lem}
\label{le:lincomp}
Let $p$ be a prime, and let $\mathcal S$
be a purely periodic sequence
over $\mathbb Z/p$ of period
length exactly $p^{j+1}$. Then $\Psi _p(\mathcal S)>p^j$. 
\end{lem}
\begin{proof}[Proof of lemma \ref{le:lincomp}]
Since $p^{j+1}$ is the length of the period of the sequence $\mathcal S$, 
the polynomial $x^{p^{j+1}}-1$ over a field $\mathbb Z/p$ annihilates $\mathcal S$. 
Yet
$x^{p^{j+1}}-1=(x-1)^{p^{j+1}}$; thus, the minimal polynomial $m(x)$ of
$\mathcal S$ is of the form $(x-1)^r$, where $r\le p^{j+1}$. However, the
polynomial $x^{p^{j}}-1=(x-1)^{p^{j}}$ does not annihilate $\mathcal S$,
since otherwise the length of some period of $\mathcal S$ is a factor of $p^j$;
yet
$\mathcal S$ has no periods of length less than $p^{j+1}$ (see definition
\ref{def:strict}).
Hence, $\deg m(x)=r>p^j$, since
otherwise the polynomial $(x-1)^{p^{j}}$ annihilates $\mathcal S$. 
\end{proof}
\begin{proof}[Proof of the theorem \ref{lincomp}]
Since $s_{i+2^j}\equiv s_{i}+1\pmod 2$
for all $i=0,1,2,\ldots$ (see \ref{halfper}), the congruence 
$s_{i+1+2^{j}}+s_{i+2^j}+s_{i+1}+s_{i}\equiv
0\pmod 2$ holds for all  $i=0,1,2,\ldots$.  Hence, the polynomial
$x^{2^{j}+1}+x^{2^{j}}+x+1=(x+1)^{2^{j}+1}$ annihilates the $j$\textsuperscript{th} coordinate
sequence 
$\mathcal S_j=\{s_0,s_1,\dots\}$. Now
the assertion of \ref{lincomp} follows from \ref{le:lincomp}. 
\end{proof}
Theorem \ref{lincomp} could be generalized to the case of output sequences
of wreath products of automata. Namely, the following proposition holds.
\begin{prop}
\label{prop:WP:lincomp}
Let $\mathcal S=\{s_i\:i=0,1,2,\ldots\}$ be any of the sequences $\mathcal U_n$, 
$\mathcal X_n$,
$\mathcal W_n$, $\mathcal Y_n$, and $\mathcal Z$ defined, respectively,
in \ref{WP-even}, \ref{WP-even-trunc}, \ref{WP-odd}, \ref{le:WP-odd},
and \ref{thm:WP}. Then the linear complexity of the $(n-1)$\textsuperscript{th}
coordinate sequence 
$\delta_{n-1}(\mathcal S)=\{\delta_{n-1}(s_i)\:i=0,1,2,\ldots\}$ exceeds
$2^{n-1}$. 
\end{prop}
\begin{proof} Since the period length of the sequence $\delta_{n-1}(\mathcal S)$
is $2^n\ell$, where $\ell=2^m$ for $\mathcal S\in\{\mathcal U_n,
\mathcal X_n\}$, or $\ell=m$ otherwise
 (see corresponding statements), the polynomial $u(x)=x^{2^n\ell}-1=
(x^\ell-1)^{2^n}$ annihilates $\delta_{n-1}(\mathcal S)$. Thus, the minimal polynomial
$m(x)$ of $\delta_{n-1}(\mathcal S)$ is a factor of $u(x)$. On the other
hand $m(x)$ is not a factor of $w(x)=(x^\ell-1)^{2^{n-1}}$ since otherwise
the sequence $\delta_{n-1}(\mathcal S)$ has period of length $2^{n-1}\ell$;
however, this is impossible since the second half of the period of length
$2^n\ell$ of this sequence is a bitwise negation of the first half, see
\ref{note:halfper-odd}.
Since both polynomials $u(x)$, $w(x)$ have the same set of
roots in their splitting field, at least one of these roots is a root of
$m(x)$ with multiplicity exceeding $2^{n-1}$. Thus, $\deg m(x)>2^{n-1}$.
\end{proof}
Speaking formally, proposition \ref{prop:WP:lincomp} holds
for $\ell=1$ either, turning into theorem \ref{halfper} in this case. Thus,
we may say that the estimate of $\Psi_2(\delta_{n-1}(\mathcal S))$ given
by proposition \ref{prop:WP:lincomp} is sharp. However, it could be improved
for particular classes of $\ell$. For instance, if $\ell=2^m$, i.e.,
if $\mathcal S=\mathcal X_n$, then $\Psi_2(\delta_{n-1}(\mathcal S))=
2^{n-1}\ell+1$ in view of note \ref{WP-even-trunc} and theorem \ref{lincomp}.
Also, if $\ell=2^km_1$, where $m_1$ is odd, then the proof of proposition 
\ref{prop:WP:lincomp}
shows that $\Psi_2(\delta_{n-1}(\mathcal S))>2^{n-1+k}$ in this case. 

So it seems possible to improve significantly 
the estimate of linear complexity that gives proposition
\ref{prop:WP:lincomp} for various classes of wreath products 
described by \ref{WP-even}, \ref{WP-even-trunc}, \ref{WP-odd}, \ref{le:WP-odd},
and \ref{thm:WP}, i.e., for arbitrary
$\ell>1$. To do this now we have to run a bit ahead and
to use theorem \ref{thm:WP:AnyHalfPer}, which is proved further. With the
use of this theorem,
the general case could be reduced to the case $\ell>1$ odd. Namely, in view of 
theorem
\ref{thm:WP:AnyHalfPer}, every purely periodic binary sequence of period
length $2^n\ell$, $n>1$, such that  the second  half of the period of this
sequence is a bitwise negation of the first part of the period, could be considered
as $(n-1)$\textsuperscript{th} coordinate sequence of a certain wreath product
of automata that is described by theorem \ref{thm:WP}. Thus, if $\ell=2^km_1$,
where $m_1$ odd, this sequence in view of theorem \ref{thm:WP:AnyHalfPer}
could be considered as $(n-1+k)$\textsuperscript{th} coordinate sequence
of a suitable wreath product of automata mentioned in theorem \ref{thm:WP}
for $m=m_1$ odd. So we can assume that $\ell$ is odd.

Proceeding with this note and using  the congruence 
$\delta_{n-1}(s_{i+2^{n-1}\ell})\equiv
\delta_{n-1}(s_{i})+1\pmod 2$ (see \ref{note:halfper-odd}) we obtain that
the minimal polynomial $m_{n-1}(x)$ of the sequence $\delta_{n-1}(\mathcal S)$ is a factor of the polynomial 
\begin{multline*}
x^{2^{n-1}\ell+1}+x^{2^{n-1}\ell}
+x+1=\\ (x^\ell+1)^{2^{n-1}}(x+1)=(x^{\ell-1}+\cdots+x+1)^{2^{n-1}}
(x+1)^{2^{n-1}+1}.
\end{multline*} 
Thus, the root of multiplicity $>2^{n-1}$ of the proof of  
\ref{note:halfper-odd} is $1$ (since the polynomial $x^{\ell-1}+\cdots+x+1$
is a factor of $x^\ell-1$; yet $x^\ell-1$
has no roots of multiplicity $>1$ in its splitting field, as $\ell$ is
odd). Hence,
\begin{equation}
\label{eq:minpol} 
m_{n-1}(x)=v(x)(x+1)^{2^{n-1}+1},
\end{equation}
where $v(x)$ is a factor of $(x^{\ell-1}+\cdots+x+1)^{2^{n-1}}$. Thus,  
\begin{equation}
\label{eq:lincomp:sharp}
2^{n-1}\ell+1\ge\deg m_{n-1}(x)=\Psi_2(\delta_{n-1}(\mathcal S))\ge 2^{n-1}+1.
\end{equation} 
We shall show now that for $n>1$
the both bounds are sharp. 

Consider a finite sequence $\mathcal T$ of length $2^{n-1}\ell$ consisting of gaps and runs
(alternating blocks of $0$'s and $1$'s)
of length $2^{n-1}$ each. Take this sequence as the first half of a period
of a sequence $\mathcal S^{\prime}$, and take a bitwise negation $\hat{\mathcal T}$ 
of $\mathcal T$ as a second half of a period
of $\mathcal S^\prime$ (of course $\hat{\mathcal T}=(\mathcal
T)\XOR(2^{2^{n-1}\ell}-1)$, where we consider $\mathcal T$ as a base-$2$
expansion of a suitable rational integer $\gamma_{n-1}> 0$). 
Obviously, $\mathcal S^\prime$ is a purely periodic sequence
of period length $2^n\ell$, and the second half of its period is a bitwise
negation of the first half. Thus, as it is shown  
by theorem \ref{thm:WP:AnyHalfPer}, the sequence $\mathcal S^\prime$ could be 
outputted as $(n-1)$\textsuperscript{th}
coordinate sequence of a suitable wreath product of automata, which is
described by theorem \ref{thm:WP}. 
Yet obviously $\mathcal S^\prime$ is a sequence of gaps and runs of length $2^{n-1}$
each; thus, the exact period length of the sequence $\mathcal S^\prime$ is 
$2^n$. So linear complexity of $\mathcal S^\prime$ is $2^{n-1}+1$ (see the proof
of theorem \ref{lincomp}). 

Now we prove that the upper bound in \eqref{eq:lincomp:sharp} is also sharp.
Consider a sequence $\mathcal U$ of gaps and runs of length $2^{n-1}$ each,
and a purely periodic sequence $\mathcal V$ with period of length $2^{n-1}\ell$;
let this period consists of a run of length $2^{n-1}(\ell-1)$ followed
by a gap of length $2^{n-1}$. Let $m_{\mathcal U}(x), m_{\mathcal V}(x)$ be minimal polynomials
of corresponding sequences. 

Since $\mathcal U$ is a purely periodic sequence
with period length exactly $2^n$, and a second half of its period is a
bitwise negation of the first half, a polynomial $m_1(x)=x^{2^{n-1}+1}+x^{2^{n-1}}+x+1=
(x+1)^{2^{n-1}+1}$ annihilates $\mathcal U$ (see the argument above); so
$m_{\mathcal U}(x)$ is a factor of $m_1(x)$. However, the first $2^{n-1}$
overlapping $(2^{n-1})$-tuples considered as vectors of dimension $2^{n-1}$
over a field $\Z/2$ are obviously linearly independent. Thus, $\deg m_{\mathcal U}(x)>2^{n-1}$
(see \cite[Theorem 8.51]{LidNied}). Finally we conclude that $m_{\mathcal U}(x)=m_1(x)$.
A similar argument proves that $m_{\mathcal V}(x)=x^{2^{n-1}(\ell-1)}+x^{2^{n-1}(\ell-2)}+
\dots+x^{2^{n-1}}+1$. 

Now consider a sum $\mathcal R$ of these two sequences. i.e., $\mathcal
R=\mathcal U\XOR\mathcal V$. Obviously, $m_{\mathcal U}(x)$ and $m_{\mathcal V}(x)$ 
are coprime, since $1$ is the only root of $m_{\mathcal U}(x)$, yet $1$
is not a root of $m_{\mathcal V}(x)$ (recall $\ell$ odd). Thus, 
$m_{\mathcal U}(x)\cdot m_{\mathcal V}(x)$ is the minimal polynomial of
$\mathcal R$ (see \cite[Theorem 8.57]{LidNied}). Hence $\Psi_2(\mathcal
R)=2^{n-1}\ell+1$.

Since $\ell$ is odd, $\mathcal R$ is obviously a purely periodic sequence
of period length exactly $2^n\ell$, and the second half of the period is
a bitwise negation of its first half. Consequently, $\mathcal R$ is the
$(n-1)$\textsuperscript{th} coordinate sequence of a suitable wreath product of automata, which is
described by theorem \ref{thm:WP} (see \ref{thm:WP:AnyHalfPer}).

As a bonus we have that the exact period length $P$ of the  $(n-1)$\textsuperscript{th}
coordinate sequence $\delta_{n-1}(\mathcal S)$ for odd $\ell$ 
is a multiple of $2^{n}$: Since $x^P+1$
annihilates $\delta_{n-1}(\mathcal S)$, $m_{n-1}(x)$ is a factor of $x^P+1$. 
Yet $x^P+1=(x^s+1)^{2^t}=(x+1)^{2^t}(x^{s-1}+\dots+1)^{2^t}$, where $P=2^ts$,
$s$ odd, and $1$ is not a root of $x^{s-1}+\dots+1$ since $s$ is odd. 
Thus, necessarily $2^t\ge 2^{n-1}+1$ in view of \eqref{eq:minpol}. Hence,
$t\ge n$. So we conclude that $P=2^ns$; yet $P\le 2^n\ell$ since the output
sequence $\mathcal Z\bmod 2^n$ is purely periodic of period length exactly
$2^n\ell$ (see \ref{thm:WP}). Thus, $P=2^ns$, where $1\le s\le\ell$. As
demonstrate examples of sequences $\mathcal S^\prime$ and $\mathcal R$,
both extreme cases $s= 1$ and $s=\ell$ are possible.

We summarize the above considerations in the following 
\begin{thm}
\label{thm:lincomp:sharp} Let $\mathcal Z_j$, $j>0$, be the $j$\textsuperscript{th}
coordinate sequence of a wreath product of automata {\rm(described by any of
\ref{WP-even}, \ref{WP-even-trunc}, \ref{WP-odd}, \ref{le:WP-odd},
and \ref{thm:WP}: thus $\mathcal Z_j$ is a purely periodic binary sequence
of period length $2^{j+1}\ell$, where $\ell=2^m$ for wreath products described
by \ref{WP-even} or \ref{WP-even-trunc}, and $\ell=m$ otherwise\rm)}. 
Represent $\ell=2^kr$, where $r$
is odd.
Then the exact period length of $\mathcal Z_j$ is $2^{k+j+1}s$ for 
some $s\in\{1,2,\dots,r\}$, and both extreme cases $s=1$ and $s=r$ 
occur: for every sequence $s_1,s_2,\ldots$ over a set $\{1,r\}$
there exists a wreath product of automata such that the  period length 
of the  $j$\textsuperscript{th} coordinate sequence of its output is exactly
$2^{k+j+1}s_j$, $(j=1,2,\ldots)$.

Moreover, a linear complexity $\Psi_2(\mathcal Z_j)$ of the sequence $\mathcal
Z_j$ satisfies the following inequality:
$$2^{k+j}+1\le \Psi_2(\mathcal Z_j)\le 2^{k+j}r+1.$$
Both these bounds are sharp: 
For every sequence $t_1,t_2,\ldots$ over a set $\{1,r\}$ 
there exists a wreath product of automata such that the linear complexity
of the $j$\textsuperscript{th} coordinate sequence of its output is exactly
$ 2^{k+j}t_j+1$, $(j=1,2,\ldots)$.
\end{thm}
\begin{proof} Nearly everything is already done by the preceeding arguments.
We only note that in view of mentioned theorem \ref{thm:WP:AnyHalfPer},
we can choose coordinate sequences independently
one of another. That is, for each sequence of purely periodic binary sequences 
$\mathcal
Z_1, \mathcal Z_2, \dots$, such that period length of the  $j$\textsuperscript{th}
sequence $\mathcal Z_j$ $(j=1,2,\ldots)$ is $2^{j+1}\ell$, and the second
part of this period is a bitwise negation of the first part, there exist
a wreath product of automata, that satisfies \ref{thm:WP}, and such that
the $j$\textsuperscript{th} coordinate sequence of its output is exactly
$\mathcal Z_j$ for all $j=1,2,\ldots$.
\end{proof}


With the use of theorem \ref{halfper} it is possible to estimate two other
measures of complexity of the coordinate sequence, which were introduced
in \cite{Kl-Gor}: namely, 
{\it $2$-adic complexity} and {\it $2$-adic span}. 
Whereas linear complexity (also known
as {\it linear span}) is the number of cells in a linear
feedback shift register outputting a given sequence $\mathcal S$ over
$\mathbb Z/2$, the $2$-adic span is the number of cells in both memory and register
of a feedback with carry shift register (FCSR) that outputs $\mathcal S$,
and
the
$2$-adic complexity estimates the number of cells in the register of this
FCSR. To be more exact, the $2$-adic complexity $\Phi_2(\mathcal S)$ of the (eventually) periodic
sequence $\mathcal S=\{s_0,s_1,s_2,\ldots\}$ over $\mathbb Z/2$ is $\log_2(\Phi(u,v))$, 
where $\Phi(u,v)=\max\{|u|,|v|\}$ and $\frac{u}{v}\in\mathbb Q$ 
is the irreducible fraction
such that its $2$-adic expansion agrees with $\mathcal S$, that is, 
$\frac{u}{v}=s_0+s_12+s_22^2+\dots\in\mathbb Z_2$. The number of cells in the register
of FCSR producing $\mathcal S$ is then $\lceil\log_2(\Phi(u,v))\rceil$, 
the least rational integer not smaller than $\log_2(\Phi(u,v))$.
Thus, we only need to estimate $\Phi_2(\mathcal S)$.
\begin{thm}
\label{2-comp}
Let $\mathcal S_j=\{s_0,s_1,s_2,\dots\}$ be the $j$\textsuperscript{th} coordinate 
sequence. 
its $2$-adic complexity $\Phi_2(\mathcal S_j)$ is $\log_2\Big(\frac{2^{2^j}+1}{\gcd(2^{2^j}+1,\gamma+1)}\Big)$,
where $\gamma=s_0+s_12+s_22^2+\dots+s_{2^{j}-1}2^{2^{j}-1}$. 
\end{thm}
\begin{note*}
We note that $\gamma$ is a non-negative
rational integer, $0\le \gamma\le 2^{2^{j}}-1$; also we note that for each $\gamma$
of this range there exists an ergodic mapping such that the first half
of the period of the $j$\textsuperscript{th} 
coordinate sequence of the corresponding output is a base-$2$ expansion
of $\gamma$ (see \ref{AnyHalfPer}). Thus,
to find all possible values
of 2-adic complexity
of the $j$\textsuperscript{th} coordinate sequence one has to decompose
the $j$\textsuperscript{th} Fermat number $2^{2^j}+1$. It is known that
$j$\textsuperscript{th} Fermat number is prime for $0\le j\le 4$ and that
it
is composite for $5\le j\le 23$. For each Fermat number outside 
this range it is not known whether
it is prime or composite.
The complete decomposition of $j$\textsuperscript{th} Fermat number is not known
for $j>11$. Assuming for some $j\ge 2$ the $j$\textsuperscript{th} Fermat number 
is composite, 
all its factors are of the form $t2^{j+2}+1$, see e.g. \cite{Brent} for
further references. So, {\it the following bounds for $2$-adic
complexity $\Phi_2(\mathcal S_j)$ of the $j$\textsuperscript{th} coordinate sequence
$\mathcal S_j$ hold:
$$ j+3\le\lceil\Phi_2(\mathcal S_j)\rceil\le 2^j+1,$$
yet to prove whether the lower bound is sharp for a certain $j>11$, or whether  
$\lceil\Phi_2(\mathcal S_j)\rceil$ could be actually less
than $2^j+1$ for $j>23$ is as difficult as to decompose the $j$\textsuperscript{th} Fermat number
or, respectively, to determine whether the $j$\textsuperscript{th}
Fermat number 
is prime or composite.}   
\end{note*}
\begin{proof}[Proof of theorem \ref{2-comp}]
We only have to express $s_0+s_12+s_22^2+\dots$ as an irreducible fraction. Denote
$\gamma=s_0+s_12+s_22^2+\dots+s_{2^{j}-1}2^{2^{j}-1}$. Then
using 
the second identity of \eqref{eq:id} we in view of \ref{halfper} obtain that
$s_0+s_12+s_22^2+\dots+s_{2^{j+1}-1}2^{2^{j+1}-1}=\gamma+2^{2^j}(2^{2^j}-\gamma-1)=
\gamma^\prime$
and hence $s_0+s_12+s_22^2+\dots=
\gamma^\prime+\gamma^\prime 2^{2^{j+1}}+\gamma^\prime 2^{2\cdot 2^{j+1}}+
\gamma^\prime 2^{3\cdot
2^{j+1}}+\dots=\frac{\gamma+1}{2^{2^j}+1}-1$. 
This completes the proof in view of the definition of $2$-adic
complexity of a sequence. 
\end{proof}
\begin{note}
\label{note:WP:2comp}
Similar estimates of $\Phi_2(\delta_{n-1}(\mathcal S))$ could be obtained for the sequence
$\mathcal S\in\{
\mathcal W_n, \mathcal Y_n, \mathcal Z\}$ 
of  \ref{WP-odd}, \ref{le:WP-odd},
and \ref{thm:WP}, respectively (for $\mathcal S\in\{\mathcal U_n, \mathcal X_n\}$
of \ref{WP-even} and \ref{WP-even-trunc} this estimate is 
already given by \ref{2-comp} in view of \ref{WP-even-trunc}). In view
of \ref{note:halfper-odd} the argument of the proof of \ref{2-comp}
gives that the representation of the binary sequence $\delta_{n-1}(\mathcal S)$
as a $2$-adic integer is $\frac{\gamma+1}{2^{2^{n-1}m}+1}-1$, so we have
only to study a fraction 
$\frac{\gamma+1}{2^{2^{n-1}m}+1}$, 
where
$\gamma=s_0+s_12+s_22^2+\dots+s_{2^{n-1}m-1}2^{2^{n-1}m-1}$, and $m$
is of statements of \ref{WP-odd}, \ref{le:WP-odd},
and \ref{thm:WP}. 
Representing $m=2^km_1$ with $m_1>1$ odd, we can factorize
$2^{2^{n-1}m}+1=(2^{2^{n-1+k}}+1)(2^{2^{n-1+k}(m_1-1)}-2^{2^{n-1+k}(m_1-2)}
+\cdots-2^{2^{n-1+k}}+1)$, but the problem does not become much easier because
of the first multiplier. We omit further details.
\end{note}

Both theorems \ref{lincomp} and \ref{2-comp} show that all three measures
of complexity of a sequence (linear and $2$-adic spans and $2$-adic complexity)
are not too sensitive. For instance, assuming $f(x)=x+1$ to be a state transition
function and $0$ to be an initial state of the automaton $\mathfrak A^\prime$, 
we see that
values of both linear and $2$-adic complexity of  the 
$j$\textsuperscript{th} coordinate sequence
$\mathcal S_j$ of this automaton 
depend on $j$ exponentially:
$\Psi_2(\mathcal S_j)=\Phi_2(\mathcal S_j)=2^{j}+1$. However,
in this case $\mathcal S_j$ is merely a sequence of alternating blocks of $0$'s
and $1$'s of length $2^j$ each.

Looking through the proofs of the corresponding theorems it is easy to
observe that such
big figures for linear and $2$-adic complexity in the above example 
are due to a very
simple law the $j$\textsuperscript{th} coordinate sequence obeys: 
The second
half of the period is the bitwise negation of the first half (see \ref{halfper},
\ref{note:halfper-odd}). This means
that, intuitively,
the $j$\textsuperscript{th} coordinate sequence is as
complex as the first half of its period. Thus we have to understand what
sequences of length $2^j$ could be outputted as the first half of the
period of the $j$\textsuperscript{th} coordinate sequence, that is, what
values takes the rational integer $\gamma$ of \ref{2-comp}.

In other words, let $\gamma_j(f,z)\in\N_0$ be such a number that its base-$2$ expansion 
agrees with the first half
of the period of the $j$\textsuperscript{th} coordinate sequence produced
by the
automaton $\mathfrak A^\prime_j$, 
i.e., let
$$\gamma_j(f,z)=\delta_j(f^{(0)}(z))+2\delta_j(f^{(1)}(z))+
4\delta_j(f^{(2)}(z))+\dots+2^{2^j-1}\delta_j(f^{(2^j-1)}(z)).$$
Obviously, $0\le\gamma_j(f,z)\le 2^{2^j}-1$. A natural question arises:

{\slshape Given a compatible and ergodic mapping
$f\colon\mathbb Z_2\rightarrow\mathbb Z_2$ and a $2$-adic integer $z\in\mathbb
Z_2$, what infinite string $\gamma_0=\gamma_0(f,z),\gamma_1=\gamma_1(f,z),
\gamma_2=\gamma_2(f,z),\dots$ (where $\gamma_j\in\{0,1,\dots,2^{2^j}-1\}$
for
$j=0,1,2,\dots$) could be obtained?} 

The answer is:  {\slshape any one.} 

Namely, the
following theorem holds. 

\begin{thm}
\label{AnyHalfPer}
Let $\Gamma=\{\gamma_j\in\mathbb N_0\colon  j=0,1,2,\ldots\}$
be an arbitrary sequence of non-negative rational integers that satisfy
$0\le\gamma_j\le 2^{2^j}-1$ for $j=0,1,2,\ldots$ , then
there exist a compatible and ergodic mapping 
$f\colon\mathbb Z_2\rightarrow\mathbb Z_2$ and a $2$-adic integer 
$z\in\mathbb Z_2$ such that $\delta_j(z)=\delta_0(\gamma_j)$, 
$\delta_0(f^{(i)}(z))\equiv \gamma_0+i\pmod 2$, 
and
$$\delta_j(f^{(i)}(z))\equiv \delta_{i\bmod{2^j}}(\gamma_j)+
\biggl\lfloor\frac{\lfloor\log_2i\rfloor}{j}\biggr\rfloor\pmod 2$$ for all
$i,j\in\mathbb N$. 
\end{thm}
\begin{note*} The sequence 
$\Bigl\{\Bigl\lfloor\frac{\lfloor\log_2i\rfloor}{j}\Bigr\rfloor\bmod 2\: i=1,2,
\ldots\Bigr\}$ is merely a binary sequence of alternating gaps and runs (i.e., blocks
of consequtive $0$'s or $1$'s, respectively) of length
$2^j$ each.
\end{note*}
\begin{proof}[Proof of theorem \ref{AnyHalfPer}]
Put $z=z_0=\sum_{j=0}^{\infty}\delta_0(\gamma_j)2^j$ and 
$$z_i= (\gamma_0+i)\bmod 2+\sum_{j=1}^{\infty}\biggl(\biggl(
\delta_{i\bmod{2^j}}(\gamma_j)+
\biggl\lfloor\frac{\lfloor\log_2i\rfloor}{j}\biggr\rfloor\biggr)\bmod 2\biggr)\cdot
2^j$$
for $i=1,2,3,\ldots$ . Consider a sequence $Z=\{z_i\colon i=0,1,2,\ldots\}$.
Speaking informally, we are filling a table with countable infinite number of rows
and columns in such a way that the first $2^j$ entries of the $j$\textsuperscript{th}
column represent $\gamma_j$ in its base-2 expansion, and the other entries
of this column are obtained from these by applying recursive relation of theorem \ref{halfper}.
Then each $i$\textsuperscript{th} row of the table is a 2-adic canonical
representation of $z_i\in Z$.

We shall prove that $Z$ is a dense subset in $\mathbb Z_2$, and then
define $f$ on $Z$ in such a way that $f$ is compatible and ergodic on $Z$.
This will imply the assertion of the theorem. 

Proceeding along this way we claim that $Z\bmod 2^k = \mathbb Z/2^k$ for all $k=1,2,3,\ldots$,
i.e., a natural ring homomorphism $\bmod\, 2^k\colon z\mapsto z\bmod 2^k$ maps
$Z$ onto the residue ring $\mathbb Z/2^k$. Indeed, this trivially holds
for $k=1$. Assuming our claim holds for $k< m$ we prove it for $k=m$.
Given arbitrary $t\in\{0,1,\ldots,2^{m}-1\}$ there exists $z_i\in Z$ such
that $z_i\equiv t\pmod{2^{m-1}}$. If $z_i\not\equiv t\pmod{2^{m}}$ then
$\delta_{m-1}(z_i)\equiv\delta_{m-1}(t)+1\pmod 2$ and thus
$\delta_{m-1}(z_{i+2^{m-1}})\equiv\delta_{m-1}(t)\pmod 2$. However, 
$z_{i+2^{m-1}}\equiv z_i\pmod {2^{m-1}}$. Hence
$z_{i+2^{m-1}}\equiv t\pmod {2^m}$. 

A similar argument shows that for each $k\in\mathbb N$ 
the sequence $\{z_i\bmod 2^k\colon i=0,1,2,\ldots\}$
is purely periodic with period length $2^k$, and each $t\in\{0,1,\ldots,2^{k}-1\}$
occurs at the period exactly once (in particular, all members of $Z$ are
pairwise distinct 2-adic integers). Moreover, $i\equiv i^{\prime}\pmod{2^k}$
iff $z_{i}\equiv z_{i^{\prime}}\pmod{2^k}$. Consequently, $Z$ is dence
in $\mathbb Z_2$ since for each $t\in\mathbb Z_2$ and each $k\in\mathbb
N$ there exists $z_i\in Z$ such that $\|z_i-t\|_2\le 2^{-k}$. Moreover, if
we define $f(z_i)=z_{i+1}$ for all $i=0,1,2,\ldots$ then 
$\|f(z_i)-f(z_{i^{\prime}})\|_2=\|z_{i+1}-z_{i^{\prime}+1}\|_2=
\|(i+1)-(i^{\prime}+1)\|_2=\|i-i^{\prime}\|_2=\|z_i-z_{i^{\prime}}\|_2$.
Hence, $f$ is well defined and compatible on $Z$; it follows that the continuation
of $f$ to the whole space $\mathbb Z_2$ is compatible. Yet $f$ is transitive
modulo $2^k$ for each $k\in\mathbb N$, so its continuation is ergodic.
\end{proof}
Theorem \ref{AnyHalfPer} could be extended to coordinate
sequences of wreath products of automata (see Section \ref{sec:Constr}), i.e., to 
the sequences 
$\delta_j(\mathcal Z)=\{\delta_j(x_i)\:i=0,1,2,\ldots\}$, where 
$\mathcal Z=\{x_i\:i=0,1,2,\ldots\}$ is a recurrence sequence
over $\Z_2$ defined in \ref{thm:WP}. Speaking loosely, {\slshape each first half
of a period of each $i$\textsuperscript{th} $(i\ge 1)$ coordinate sequence 
of wreath products of automata
could be arbitrary and independent of others}. Now we give a formal statement
and a proof of it.  

Recall that $\delta_j(\mathcal Z)$
is a purely periodic binary sequence of period length $2^{j+1}m$, and the
second half of the period is a bitwise negation of its first half, see
\ref{note:halfper-odd}. Thus, the sequence $\delta_j(\mathcal Z)$ could
be identified with a rational number (which will be denoted by the same
symbol $\delta_j(\mathcal Z)$) such that its canonical $2$-adic representation
is $\delta_j(x_0)+\delta_j(x_1)2+\delta_j(x_2)2^2+\dots$. Hence in view
of note \ref{note:WP:2comp},
\begin{equation}
\label{eq:num:coord} 
\frac{2^{2^{j}m}-\gamma_j}{2^{2^{j}m}+1}=\delta_j(\mathcal Z),
\end{equation}
where
$\gamma_j=\delta_j(x_0)+\delta_j(x_1)2+\delta_j(x_2)2^2+\dots+\delta_j(x_{2^{j}m-1})2^{2^{j}m-1}$, and $m$ and
$x_i$ are
of the statement of \ref{thm:WP}. In other
words,
$\gamma_j\in\N_0$ is such a number that its base-$2$ expansion agrees with
the first $2^jm$ terms of the sequence $\{\delta_j(x_i)\:i=0,1,2,\ldots\}$,
where $x_{i+1}=g_{i\bmod m}(x_i)$, and $\mathcal G=\{g_0,\ldots,g_{m-1}\}$
is a finite sequence of compatible measure preserving mappings of $\Z_2$ onto itself,
see \ref{thm:WP}. Thus, $\gamma_j\in\{0,1,\ldots,2^{2^jm}-1\}$, and $\gamma_j$
depends on $x_0$ and on $\mathcal G$. Yet an arbitrary purely periodic
sequence of period length $2^{j+1}m$ such that the second half of its period
is a bitwise negation of the first half (the latter could be considered
as a base-$2$ expansion of rational integer $\gamma_j$), being treated
as a $2$-adic reresentation of a rational number could be represented as 
\eqref{eq:num:coord} (see the proof of \ref{note:WP:2comp}). So we wonder
what sequences of such kind could be represented by coordinate sequences
of wreath products of automata described by theorem \ref{thm:WP}. 

In other
words, to each sequence $\mathcal Z$ described by theorem \ref{thm:WP}
we associate a sequence $\Gamma(\mathcal Z)=\{\gamma_0,\gamma_1,\ldots\}$
of non-negative raional integers $\gamma_j$ such that $0\le\gamma_j\le
2^{2^{j}m}-1$ iff \eqref{eq:num:coord} holds for all $j=0,1,2,\ldots$. Now
we take an arbitrary sequence $\Gamma$ of this type and wonder
whether this sequence could be associated to some sequence $\mathcal Z$
described by theorem \ref{thm:WP}. Generally speaking, the answer is {\slshape no}, since 
according to \ref{thm:WP} the
sequence $\delta_0(\mathcal F)$ is purely periodic with period length {\slshape
exactly} $2m$. However, a purely periodic sequence $\mathcal S$ of period length 
$2^nm$ such that the second half of its period is a bitwise negation of
the first half, i.e., such that $\mathcal S$ could be represented in a
form \eqref{eq:num:coord} as  $\mathcal S=\frac{2^{2m}-\gamma_0}{2^{2m}+1}$
for suitable $0\le\gamma_0\le 2^{2m}-1$, {\slshape not necessrily has
exact period length} $2^nm$ (see note \ref{note:halfper-odd}). However, according
to \ref{note:halfper-odd}, senior coordinate sequences $\delta_j(\mathcal
Z)$ $(j\ge 1)$ could have exact periods smaller than $2^{j+1}m$. So it
is reasonable to ask whether an arbitrary sequence $\Gamma =\{\gamma_1,\gamma_2,\ldots\}$
of non-negative rational integers $\gamma_j$ such that $0\le\gamma_j\le
2^{2^{j}m}-1$ corresponds in the above meaning to a certain sequence
$\mathcal Z$ described by theorem \ref{thm:WP}. In this case the answer
is {\slshape yes}. Namely, the following theorem holds.  
\begin{thm}
\label{thm:WP:AnyHalfPer}
Let $m> 1$ be a rational integer, and let $\Gamma=\{\gamma_0,\gamma_1,\dots\}$ 
be an arbitrary sequence over $\N_0$
such that $\gamma_j\in\{0,1,\ldots,2^{2^jm}-1\}$ for
all $j=0,1,2,\dots$. Then there exist  a finite sequence 
$\mathcal G=\{g_0,\ldots,g_{m-1}\}$
of compatible  measure preserving mappings of $\Z_2$ onto itself and a
$2$-adic integer $x_0\in\Z_2$ such that $\mathcal G$ satisfies conditions
of theorem \ref{thm:WP}, and $\delta_j(\mathcal Z)$ satisfies \eqref{eq:num:coord}
for all $j=1,2,\dots$,
where the recurrence sequence $\mathcal Z=\{x_0,x_1,\ldots\in\Z_2\}$ is
defined by the recurrence relation $x_{i+1}=g_{i\bmod m}(x_i)$, $(i=0,1,2,\dots)$.
\end{thm}
\begin{proof} According to \ref{ergBool}, 
a mapping $g_i\:\Z_2\>\Z_2$ is compatble and measure preserving iff each
$\delta_j(g_i(x))$ is a Boolean polynomial in Boolean veriables $\chi_0=\delta_0(x),
\chi_1=\delta_1(x),\dots$ that is linear with respect to $\chi_j$, i.e.,
$\delta_j(g_i(x))$
could be represented as
$$\delta_j(g_i(x))=\chi_j+\varphi_j^i(\chi_0,\dots,\chi_{j-1}),$$
where $\varphi_j^i=\varphi_j^i(\chi_0,\dots,\chi_{j-1})$ is an arbitrary Boolean polynomial
in Boolean variables $\chi_0,\dots,\chi_{j-1}$. Thus, a compatible
and measure preserving mapping
$g_i$ is completely determined by a sequence $\varphi_0^i,\varphi_1^i,\dots$ of
corresponding Boolean polynomials. So,  given a sequence $\Gamma$ we have
to determine $x_0\in\N_0$ and a family $\{\varphi_j^i\: i=0,1,\ldots,m-1; j=0,1,2,\ldots\}$
of Boolean functions such that the respective measure preserving mappings
$g_k$ $(k=0,1,\ldots,m-1)$
satisfy theorem \ref{thm:WP} and that $\delta_j(\mathcal Z)$ satisfies \eqref{eq:num:coord}
for all $j=1,2,\dots$,
where the recurrence sequence $\mathcal Z=\{x_0,x_1,\ldots\in\Z_2\}$ is
defined by the recurrence relation $x_{i+1}=g_{i\bmod m}(x_i)$, $(i=0,1,2,\dots)$.

To start with, we set $x_0=\delta_0(\gamma_0)+\delta_0(\gamma_1)\cdot 2+
\delta_0(\gamma_2)\cdot 2^2+\dots\in\Z_2$. Further we describe an inductive
procedure to determine $\varphi_j^i$ successively for j=0,1,2,\ldots. 

For $j=0$ we fix arbitrary 
$g_0(0)=\varphi_0^0,\dots,g_{m-1}(0)=\varphi_0^{m-1}\in\{0,1\}$ that satisfy
conditions (1) and (2) of theorem \ref{thm:WP}. Note that thus we have
determined all the mappings $g_i$ $(i=0,1,\dots,m-1)$ modulo 2.
Note also
that a recurrence sequence 
$\mathcal X_0=\{\xi_0^0,\xi_0^1,\dots\}$ defined by relations 
$\xi_0^0=x_0\bmod 2$,  $\xi_{k+1}^0= g_{k\bmod m}(\xi_k^0)\bmod 2$ is
a purely periodic sequence over $\Z/2=\{0,1\}$ with period length exactly
$2m$, that each element of $\Z/2$ occurs at the period exactly $m$ times, and
that $\xi_{k+m}^0\equiv\xi_{k}^0+1\pmod2$
(see the very beginning of the proof of \ref{le:WP-odd}).

Suppose that we have already determined Boolean polynomials $\varphi_j^i$
for $j=0,1,\dots,n-1$, $i=0,1,\dots,m-1$ in such a way that all the members
of a recurrence sequence $\mathcal X_{n-1}=\{\xi_0^{n-1},\xi_1^{n-1},\dots\}$
defined by relations $\xi_{0}^{n-1}=x_0\bmod 2^n$, 
$\xi_{k+1}^{n-1}= g_{k\bmod m}(\xi_k^{n-1})\bmod 2^n$, satisfy a congruence
$\delta_j(\xi_{k+2^{n-1}m}^{n-1})\equiv\delta_j(\xi_{k}^{n-1})+1\pmod{2}$
for all 
$j=0,1,\dots,n-1$ and $k=0,1,2,\ldots$. Note that then 
easy induction on $j$ (which actually is already done during the proof
of claim (3) of lemma \ref{le:WP-odd}) shows that for any $k$
\begin{equation}
\label{eq:WP:AnyHalfPer0} 
|\{\xi_{k+sm}^{n-1}\:s=0,1,\dots, 2^{n}-1\}|=2^n.
\end{equation} 
Hence, $\mathcal X_{n-1}$ is a purely periodic
sequence over $\Z/2^n$  
of period length exactly $2^nm$, with each element of $\Z/2^n$ occuring
at the period exactly $m$ times. Now we define Boolean polynomials $\varphi_n^i$
for $i=0,1,\dots,m-1$.

For a Boolean polynomial $\varphi$ in Boolean
variables $\chi_0,\dots,\chi_s$ and for $z\in\Z_2$ denote $\varphi(z)=
\varphi(\delta_0(z),\dots,\delta_s(z))$. Proceeding with this notation,
set 
\begin{equation}
\label{eq:WP:AnyHalfPer1}
\varphi_n^{k\bmod m}(\xi^{n-1}_{k})\equiv\delta_k(\gamma_n)+\delta_{k+1}(\gamma_n)
\pmod 2,
\end{equation}
for $k=0,2,\dots,2^nm-2$. Set also
\begin{equation}
\label{eq:WP:AnyHalfPer2}
\varphi_n^{m-1}(\xi^{n-1}_{2^nm-1})\equiv\delta_{2^nm-1}(\gamma_n)+\delta_0(\gamma_n)+1\pmod2.
\end{equation}
Note that in view of  \eqref{eq:WP:AnyHalfPer1} and \eqref{eq:WP:AnyHalfPer0}
the Boolean functions $\varphi_n^{i}$ of $n$ variables (and
whence, corresponding Boolean polynomials) for $i=0,1,\dots,m-2$ are well
defined; Also, the Boolean polynomial $\varphi_n^{m-1}$ is well defined
in view of \eqref{eq:WP:AnyHalfPer2},
\eqref{eq:WP:AnyHalfPer1}, and \eqref{eq:WP:AnyHalfPer0}.

Consider now a recurrence sequence $\mathcal E_n=\{\varepsilon_k\:k=0,1,2,\dots\}$ 
over $\Z/2$ defined by relations $\varepsilon_0=\delta_0(\gamma_n)$, 
$\varepsilon_{k+1}=\varepsilon_k+\varphi_n^{k\bmod m}(\xi_k^{n-1})\pmod
2$. In view of \eqref{eq:WP:AnyHalfPer1} one has $\varepsilon_k=\delta_k(\gamma_n)$
for $k=0,2,\dots,2^nm-1$, and $\varepsilon_{2^nm}\equiv\delta_0(\gamma_n)+1\pmod
2$ in view of \eqref{eq:WP:AnyHalfPer2}. Yet 
$\mathcal X_{n-1}$ is a purely periodic
sequence over $\Z/2^n$  
of period length exactly $2^nm$; proceeding with this we obtain succesively
in view of \eqref{eq:WP:AnyHalfPer2} and
\eqref{eq:WP:AnyHalfPer1}:
\begin{align*}
&\varepsilon_{2^nm}\equiv\delta_0(\gamma_n)+1\pmod 2,&\ldots&,
&{}&\varepsilon_{2^{n}m+(2^{n}m-1)}\equiv\delta_{2^nm-1}(\gamma_n)+1\pmod 2,\\
&\varepsilon_{2\cdot 2^{n}m}\equiv\delta_{0}(\gamma_n)\pmod 2,&\ldots&,
&{}&\varepsilon_{2\cdot 2^{n}m+(2^{n}m-1)}\equiv\delta_{2^nm-1}(\gamma_n)\pmod 2,\\
&\varepsilon_{3\cdot 2^{n}m}\equiv\delta_{0}(\gamma_n)+1\pmod 2,&\ldots&
\end{align*}
Note that in view of the definition of $\varepsilon_k$ one has
$$\varepsilon_{2^nm}=\delta_0(\gamma_n)+\sum_{k=0}^{2^nm-1}\varphi_n^{k\bmod
m}(\xi_k^{n-1}).$$
But the sum in the right hand side must be $1$ modulo $2$ since 
$\varepsilon_{2^nm}\equiv\delta_0(\gamma_n)+1\pmod 2$, as it was proved
above. So, in view of \eqref{eq:WP:AnyHalfPer0} one has
$$\sum_{k=0}^{2^nm-1}\varphi_n^{k\bmod m}(\xi_k^{n-1})\equiv
\sum_{i=0}^{m-1}\sum_{\xi\in\Z/2^n}\varphi_n^{i}(\xi)\equiv 1\pmod 2. $$
With the note that $\sum_{\xi\in\Z/2^n}\varphi_n^{i}(\xi)$ is just a weight
of a Boolean polynomial $\varphi_n^{i}$, we conclude that an odd number of Boolean
polymomials of $\varphi_n^{0},\ldots,\varphi_n^{m-1}$ must be of odd weight
(cf. conditions of lemma \ref{le:WP-odd}).

Now setting $\xi^n_k=\xi^{n-1}_k+2^n\cdot\varepsilon_k$ 
for $k=0,1,2,\dots$ we obtain a sequense 
$\mathcal X_{n}=\{\xi_0^{n},\xi_1^{n},\dots\}$ over $\Z/2^{n+1}$ 
such that members of $\mathcal X_{n}$ satisfy the following relations
\begin{align*} 
&\qquad\xi_{0}^{n}=x_0\bmod 2^{n+1},\\ 
&\qquad\xi_{k+1}^{n}= g_{k\bmod m}(\xi_k^{n})\bmod 2^{n+1},\\
&\qquad\delta_j(\xi_{k+2^{n}m}^{n})\equiv\delta_j(\xi_{k}^{n})+1\pmod{2}
\end{align*}
for all 
$j=0,1,\dots,n$ and $k=0,1,2,\ldots$. Moreover, $\mathcal X_{n}$ is a purely
periodic sequence with period length $2^{n+1}m$ (in view of the third of preceeding
congruences, since the sequence $\mathcal X_{n-1}$ is purely periodic with period length
exactly $2^nm$ by the above assumption), and each element of $\Z/2^{n+1}$
occurs at the period exactly $2^{n+1}m$ times. Finally, 
$\delta_n(\mathcal X_{n})=\{\varepsilon_0,\varepsilon_1,\ldots\}=\frac{2^{2^{n}m}-\gamma_n}{2^{2^{n}m}+1}$.

With the use of this inductive procedure we construct for $n=1,2,\ldots$
well defined 
mappings $g_i$ modulo $2^{n+1}$ $(i=0,1,\ldots,m-1)$ 
that are compatible and bijective modulo $2^{n+1}$; moreover, a corresponding
recurrence sequence $\mathcal X_n$ defined by relation 
$x_{i+1}=g_{i\bmod m}(x_i)\bmod
2^{n+1}$ satisfy \eqref{eq:num:coord} for $j=1,\ldots,n$. The mappings
$g_i$ satisfy condition (3) of \ref{thm:WP} for $k=1,2,\ldots,n+1$ since,
as it was noted above, 
the odd number of Boolean
polymomials of $\varphi_k^{0},\ldots,\varphi_k^{m-1}$ are of odd weight
for all $k=1,2,\ldots,n$. From the definition of $g_i$ modulo 2 it follows
that  these mappings $g_i$ satisfy conditions (1) and
(2) of \ref{thm:WP}.
This completes
the proof in view of the notices that were made at the very beginning of
it.
\end{proof}
\subsection*{Distribution of $k$-tuples} In this subsection we study a
distribution of overlapping binary $k$-tuples in output sequences of automata
introduced above. As it was shown, an output sequence of any of these automata with
output alphabet $\{0,1,2,\ldots, 2^n-1\}=\Z/2^n$ is strictly uniformly
distributed as a sequence over $\Z/2^n$. That is, it is purely periodic,
and each element of $\Z/2^n$
occurs at the period the same number of times. However, one could consider
the same sequence as a binary sequence, and ask what is a distribution
of $n$-tuples in such a sequence. {\slshape Strict uniform distribution of an
arbitrary sequence $\mathcal T$
as a sequence over $\Z/2^n$ does not necessarily imply uniform distribution
of overlapping $n$-tuples, if this sequence is considered as a binary sequence!} 

For instance, let $\mathcal T$ be the following strictly uniformly
distributed sequence over $\Z/4$ with perid length exactly $4$:
$\mathcal T=023102310231\ldots$. Then its representation as a binary sequence
is $\mathcal T=000111100001111000011110\ldots$ (recall that according to
our conventions in Section \ref{Prelm} we write senior bits right, and not left;
i.e., $2=01$, $1=10$, etc.) Obviously, when we consider $\mathcal T$ as
a sequence over $\Z/4$, then each number of $\{0,1,2,3\}$ occurs in the
sequence with the same frequency $\frac{1}{4}$. Yet if we consider $\mathcal
T$ as a binary sequence, then $00$ (as well as $11$) occurs in this sequence with
frequency $\frac{3}{8}$, whereas $01$ (and $10$) occurs with frequency $\frac{1}{8}$.
Thus, the sequence $\mathcal T$ is uniformly distributed over $\Z/4$, and
it is not uniformly distributed over $\Z/2$.

In this subsection we show that such an effect does not take
place for output sequences of automata described in \ref{WP-even}, \ref{WP-even-trunc}, \ref{WP-odd}, \ref{le:WP-odd},
and \ref{thm:WP}: {\slshape Considering any of these sequences as
a binary sequence, a distribution
of $k$-tuples is uniform, for all $k\le n$}. Now we state this property more formally.

Consider a (binary) {\it $n$-cycle} 
$C=(\varepsilon_0\varepsilon_1\dots \varepsilon_{n-1})$; 
that is, an oriented
graph with vertexes $\{a_0,a_1,\ldots, a_{n-1}\}$ and edges 
$$\{(a_0,a_1),(a_1,a_2),\ldots, (a_{n-2},a_{n-1}),(a_{n-1},a_0)\},$$ 
where
each vertex $a_j$ is labelled with $\varepsilon_j\in\{0,1\}$, $j=0,1,\dots,n-1$.
(Note that then $(\varepsilon_0\varepsilon_1\dots \varepsilon_{n-1})=
(\varepsilon_{n-1}\varepsilon_0\dots \varepsilon_{n-2})=\ldots$, etc.).

Clearly, each purely periodic sequence $\mathcal S$ over $\Z/2$ with period 
$\alpha_0\ldots\alpha_{n-1}$
of length $n$
could be related to a binary $n$-cycle $C(\mathcal S)=(\alpha_0\ldots\alpha_{n-1})$.
Conversly, to each binary $n$-cycle $(\alpha_0\ldots\alpha_{n-1})$ we could
relate $n$ purely periodic binary sequences of period length $n$: They
are $n$ shifted versions of the sequence
$$\alpha_0\ldots\alpha_{n-1}\alpha_0\ldots\alpha_{n-1}\ldots,$$
that is
\begin{align*}
&\alpha_1\ldots\alpha_{n-1}\alpha_0\alpha_1\ldots\alpha_{n-1}\alpha_0\ldots,\\
&\alpha_2\ldots\alpha_{n-1}\alpha_0\alpha_1\alpha_2\ldots\alpha_{n-1}\alpha_0\alpha_1\ldots,\\
&\ldots\qquad\ldots\qquad\ldots\\
&\alpha_{n-1}\alpha_0\alpha_1\alpha_2\ldots\alpha_{n-2}\alpha_{n-1}\alpha_0\alpha_1\alpha_2\ldots\alpha_{n-2}\ldots
\end{align*}

Further, {\it a $k$-chain in a binary $n$-cycle}  
$C$ is a
binary string $\beta_0\dots\beta_{k-1}$, $k<n$, that satisfies the following
condition: There exists $j\in\{0,1,\ldots,n-1\}$ such that $\beta_i=\varepsilon_{(i+j)\bmod
n}$ for $i=0,1,\ldots, k-1$. Thus, a $k$-chain
is just a string of length
$k$ of labels that corresponds to a chain of length $k$ in a graph $C$.

We call a binary $n$-cycle $C$ {\it $k$-full}, if each $k$-chain
occurs in the graph $C$ the same number $r>0$ of times.

Clearly, if $C$ is $k$-full, then $n=2^kr$. For instance, a well-known
De Bruijn sequence is an $n$-full $2^n$-cycle, 
see e.g. \cite{MrH}
for further references. Clearly enough that a $k$-full $n$-cycle is $(k-1)$-full:
Each $(k-1)$-chain occurs in $C$ exactly $2r$ times, etc. Thus, if an $n$-cycle
$C(\mathcal S)$ is $k$-full, then each $m$-tuple (where $1\le m\le k$) occurs in
the sequence $\mathcal S$ with the same probability (limit frequency) $\frac{1}{2^m}$.
That is, the sequence $\mathcal S$ is {\it $k$-distributed}, see
\cite[Section 3.5, Definition D]{Knuth}.
\begin{defn} A purely periodic binary sequence $\mathcal S$ with period length
exactly $N$ is said to be {\it
strictly $k$-distributed} iff a corresponding $N$-cycle $C(\mathcal S)$
is $k$-full.
\end{defn}

Thus, if a sequence $\mathcal S$ is strictly $k$-distributed, then it is
strictly $s$-distributed, for all positive $s\le k$.

 A $k$-distribution is a good ``indicator
of randomness" of an infinite sequence: The larger $k$, the better the
sequence, i.e., ``more random". The best case is when a sequence is $k$-distibuted
for all $k=1,2,\ldots$. Such sequences are called $\infty$-distributed.
Obviuosly, a periodic sequence can not be $\infty$-distributed.

On the other hand, a periodic sequence is just an infinite repetition of a finite
sequence, the period. A common requirement in applications is that the
period length must be large, and the whole period is never used in practice. For instance,
in cryptography normally a relatively small part of a period is used. 
So we
are interested of ``how random" is a finite sequence, namely, the period.
Of course, it seems very reasonable to consider a period of length $n$ as an $n$-cycle 
and to study a distribution
of $k$-tuples in $n$-cycle; for instance,
if this $n$-cycle is $k$-full, the distribution of $k$-tuples is strictly
uniform. However, other approaches also exist.

In \cite[Section 3.5, Definition Q1]{Knuth} there is considered the following 
``indicator of randomness"
of a finite sequence over a finite alphabet $A$ (we formulate the corresponding
definition for $A=\{0,1\}$): A finite binary sequence 
$\varepsilon_0\varepsilon_1\dots \varepsilon_{N-1}$
of length $N$ is said to be random, iff
\begin{equation}
\label{eq:Q1}
\bigg|\frac{\nu(\beta_0\ldots\beta_{k-1})}{N}-\frac{1}{2^k}\bigg|\le\frac{1}{\sqrt
N}
\end{equation}
for all $0<k\le\log_2N$, where $\nu(\beta_0\ldots\beta_{k-1})$ is the number
of occurences of a binary word $\beta_0\ldots\beta_{k-1}$ in a binary word
$\varepsilon_0\varepsilon_1\dots \varepsilon_{N-1}$. If a finite sequence
is random in a sence of this Definition Q1 of \cite{Knuth}, we shall say
that it has {\it a property} Q1, or {\it satisfies} Q1. We shall also
say that an {\it infinite periodic sequence satisfy} Q1 iff its exact
period satisfies Q1.
Note that, constrasting to the case of strict $k$-distribution, which implies
strict $(k-1)$-distribution, 
it is not enough to demonstrate only
that \eqref{eq:Q1}
holds for $k=\lfloor\log_2N\rfloor$ to prove a finite sequence of length $N$ 
satisfies Q1:
For instance, a sequence $1111111100000111$ satisfies \eqref{eq:Q1} for
$k=\lfloor\log_2n\rfloor=4$, and does not satisfy \eqref{eq:Q1} for $k=3$.
Note that an analogon of property Q1 for odd prime $p$ could be stated in an obvious
way. 

Now we are able to state the following
\begin{thm}
\label{thm:distr}
Let a sequence $\mathcal Z$ over $\Z/2^n$ be any of output sequences of
wreath products of automata {\rm(described in \ref{WP-even}, \ref{WP-even-trunc}, \ref{WP-odd}, \ref{le:WP-odd},
and \ref{thm:WP}; hence $\mathcal Z$ is a purely periodic sequence
of period length $2^n\ell$, where $\ell=2^m$ for wreath products described
by \ref{WP-even} or \ref{WP-even-trunc}, and $\ell=m$ otherwise)} or, in
particular, of a congruential generator of a maximum period length {\rm(this 
corresponds to the case $\ell=m=1$)}. Let $\mathcal Z^\prime$ be a
binary representation of  $\mathcal Z$ {\rm (hence $\mathcal Z^\prime$ is
a purely periodic binary sequence of period length exactly $2^n\ell n$)}.
Then 
the sequence $\mathcal Z^\prime$ is strictly $n$-distributed.

Moreover, if $\mathcal Z^\prime$ is a binary output sequence of a 
congruential generator of a maximum period length, then this sequence satisfies
{\rm Q1}.
\end{thm}
\begin{proof} The sequence $\mathcal Z=z_0z_1\ldots$ 
is a recurrence sequence over $\{0,1,\ldots,n-1\}$ that satisfy the following
recurrence relation:
$$z_{i+1}=f_i(z_i)\bmod 2^n \qquad (i=0,1,2,\ldots),$$
where $f_i$ is compatible and measure preserving mapping of $\Z_2$ onto itself. Here
and further in the proof we assume that subscript $i$ of $f$ is always
reduced
modulo $\ell$ for $\ell>1$ and is empty symbol for $\ell=1$ (the latter
case corresponds to congruential generator of a maximum period length with
state transition function $f\bmod 2^n$, where $f$ is ergodic). Let
$\mathcal Z^\prime=\zeta_0\zeta_1\ldots$ be a binary representation of
the sequence $\mathcal Z$. Take
an arbitrary binary word $\mathbf b=\beta_0\beta_1\ldots\beta_{n-1}$, $\beta_j\in\{0,1\}$,
and for $k\in\{0,1,\ldots, n-1\}$ denote 
$$\nu_k(\mathbf b)=|\{r\: 0\le r<2^n\ell n;
\ r\equiv k\pmod n;\
\zeta_r\zeta_{r+1}\ldots\zeta_{r+n-1}=\beta_0\beta_1\ldots\beta_{n-1}\}|$$
Obviously, $\nu_0(\mathbf b)$ is the number of occurences of a rational
integer $z$ with base-$2$ expansion $\beta_0\beta_1\ldots\beta_{n-1}$ at
the exact period of the sequence $\mathcal Z$. Hence, $\nu_0(\mathbf b)=\ell$
since the sequence $\mathcal Z$ is strictly uniformly distributed modulo
$2^n$. Now consider $\nu_k(\mathbf b)$ for $0<k<n$.

Fix $k\in\{1,2\ldots,n-1\}$ and let $r=k+tn$. As all $f_i$ are compatible,  then 
$\zeta_r\zeta_{r+1}\ldots\zeta_{r+n-1}=\beta_0\beta_1\ldots\beta_{n-1}$
holds if and only if the following two relations hold simultaneously:
\begin{align}
\label{eq:distr1}
&\zeta_{tn+k}\zeta_{tn+k+1}\ldots\zeta_{tn+n-1}=\beta_0\beta_1\ldots\beta_{n-k-1}
\\ \label{eq:distr2}
&f_{t}(\overline{\zeta_{tn}\zeta_{tn+1}\ldots\zeta_{tn+k-1}})\equiv
\overline{\beta_{n-k}\beta_{n-k+1}\ldots\beta_{n-1}}\pmod{2^k}.
\end{align}
Here $\overline{\gamma_0\gamma_1\ldots\gamma_s}=\gamma_0+\gamma_1\cdot
2+\dots+\gamma_s\cdot 2^s$ for $\gamma_0,\gamma_1,\ldots,\gamma_s\in\{0,1\}$
is a rational integer with base-$2$ expansion $\gamma_0\gamma_1\ldots\gamma_s$.

We consider  a case $\ell=1$ first; so $f_{t}=f$. Then for a given 
$\mathbf b=\beta_0\beta_1\ldots\beta_{n-1}$
congruence \eqref{eq:distr2} has exactly one solution 
$\overline{\alpha_0\alpha_1\dots\alpha_{k-1}}$ modulo $2^k$, since
$f$
is ergodic, whence, bijective modulo $2^k$.
Thus,
in view of \eqref{eq:distr1} and \eqref{eq:distr2} we conclude that 
$\zeta_r\zeta_{r+1}\ldots\zeta_{r+n-1}=\beta_0\beta_1\ldots\beta_{n-1}$
holds if and only if 
\begin{equation}
\label{eq:distr3}
\zeta_{s}\zeta_{s+1}\ldots\zeta_{s+n-1}=
\alpha_0\alpha_1\dots\alpha_{k-1}\beta_0\beta_1\ldots\beta_{n-k-1},
\end{equation}
where $s=tn$. Yet there 
exists exactly one 
$s\equiv 0\pmod n$, $0\le s< 2^nn$ such that  \eqref{eq:distr3} holds, 
since every element of
$\Z/2^n$ occurs at the period of $\mathcal Z$ exactly once. We conclude
now that
if $\ell=1$ then
$\nu_k(\mathbf b)=1$ for all $k\in\{0,1,\ldots, n-1\}$; thus, $\nu(\mathbf b)=
\sum_{j=0}^{n-1}\nu_j(\mathbf b)=n$ for all $\mathbf b$. This means that
$(2^nn)$-cycle 
$C(\mathcal Z^{\prime})$ is $n$-full, whence, the sequence $\mathcal Z^{\prime}$
is strictly $n$-distributed.

A similar argument is applied to the case $\ell>1$. Namely, 
for a given $j\in\{0,1,\ldots,\ell-1\}$ consider those $r=k+tn<2^n\ell
n$ where $t\equiv j\pmod \ell$ and denote 
$$\nu_k^j(\mathbf b)=|\{r\: 0\le r<2^n\ell n;
\ r=k+tn;\ t\equiv j\pmod\ell;\
\zeta_r\zeta_{r+1}\ldots\zeta_{r+n-1}=\mathbf b\}|.$$
Now $\zeta_r\zeta_{r+1}\ldots\zeta_{r+n-1}=\beta_0\beta_1\ldots\beta_{n-1}$
holds if and only if \eqref{eq:distr3} holds, where 
$\overline{\alpha_0\alpha_1\dots\alpha_{k-1}}$ is a unique solution of
congruence \eqref{eq:distr2} modulo $2^k$. This solution exists since all
$f_j$ are measure preserving, see theorem \ref{thm:WP}. Yet \eqref{eq:distr3}
is equivalent to the condition
$$z_t=\overline{\alpha_0\alpha_1\dots\alpha_{k-1}\beta_0\beta_1\ldots\beta_{n-k-1}},$$
where $t\in\{j,j+\ell,\ldots,j+(2^n-1)\ell\}$. But in view of claim (3) of
lemma \ref{le:WP-odd} for a given 
$\overline{\alpha_0\alpha_1\dots\alpha_{k-1}\beta_0\beta_1\ldots\beta_{n-k-1}}$
there exist exactly one $t\in\{j,j+\ell,\ldots,j+(2^n-1)\ell\}$ such that the
latter equality holds. So we conclude that $\nu_k^j(\mathbf b)=1$, hence
$\nu_k(\mathbf b)=\sum_{j=0}^{\ell-1}\nu_k^j(\mathbf b)=\ell$, and finally
$\nu(\mathbf b)=
\sum_{k=0}^{n-1}\nu_k(\mathbf b)=n\ell$ for all $\mathbf b$.
This completes the proof of the first assertion of the theorem.

To prove the second assertion note that we return to the case $\ell=1$;
hence, in view of the first assertion every $m$-tuple 
for $1\le m\le n$ occurs at the $2^nn$-cycle $C(\mathcal Z^\prime)$ exactly
$2^{n-m}n$ times. Thus, every such $m$-tuple occurs $2^{n-m}n-c$ times
at the finite binary sequence 
$\hat{\mathcal Z}=\hat z_0\hat z_1\ldots\hat z_{2^n-1}$, where
$\hat z$ for $z\in\{0,1,\ldots,2^n-1\}$ is an $n$-bit sequence that agrees
with base-$2$ expansion of $z$. Note that $c$ depends on the $m$-tuple, yet
$0\le c\le m-1$ for every $m$-tuple. Easy algebra shows that \eqref{eq:Q1}
holds for these $m$-tuples. 

Now to prove that $\mathcal Z^\prime$ satisfies Q1 
we have only to demonstrate that \eqref{eq:Q1} holds for $m$-tuples with
$m=n+d$, where $0<d\le\log_2n$. We claim that such an $m$-tuple occurs at
the sequence $\hat{\mathcal Z}$ not more than $n$ times.

Indeed, in this case 
$\zeta_r\zeta_{r+1}\ldots\zeta_{r+n+d-1}=\beta_0\beta_1\ldots\beta_{n+d-1}$
holds iff besides the two relations \eqref{eq:distr1} and \eqref{eq:distr2} the following
extra congruence holds:
$$f(\overline{\zeta_{tn}\zeta_{tn+1}\ldots\zeta_{tn+k-1}\beta_0\beta_1\ldots\beta_{d-1}})
\equiv
\overline{\beta_{n-k}\beta_{n-k+1}\ldots\beta_{n+d-1}}\pmod{2^{k+d}},$$
where $k=r\bmod n$. Yet this extra congruence may or may not have a solution
in unknowns $\zeta_{tn},\zeta_{tn+1},\ldots,\zeta_{tn+k-1}$; this depends on $\beta_0\beta_1\ldots\beta_{n+d-1}$.
But if such a solution exists, it is unique for a given $k\in\{0,1,\ldots,n-1\}$, since $f$
is ergodic, whence, bijective modulo $2^s$ for all $s=1,2,\ldots$.
This proves our claim. Now easy exercise in inequalities shows that \eqref{eq:Q1}
holds in this case, thus completing the proof of the theorem.
\end{proof}
\begin{note}
\label{note:distr} 
The first asssertion of theorem \ref{thm:distr} remains true for {\it wreath
products of truncated automata}, i.e. for the sequence $\mathcal F$ of corollary
\ref{cor:WP}, where $F_j(x)=\big\lfloor\frac{x}{2^{n-k}}\big\rfloor\bmod 2^k$, 
$j=0,1,\ldots,\ell-1$, a truncation of $n-k$ low order bits. Namely, {\it
a binary representation $\mathcal F^\prime$ of the sequence $\mathcal F$
is a purely periodic strictly $k$-distributed binary sequence of period
length exactly $2^n\ell k$.}

The second assertion of theorem \ref{thm:distr} holds for arbitrary prime
$p$. Namely, {\it a base-$p$ representation of an output sequence 
of a congruential generator over $\Z/p^n$
of a maximum period length is strictly $n$-distributed sequence over $\Z/p$
of period length exactly $p^nn$, which satisfies Q1}.

Moreover, the first assertion of \ref{thm:distr} holds 
for truncated congruential generators with output
function $F(x)=\big\lfloor\frac{x}{p^{n-k}}\big\rfloor\bmod p^k$. Namely, {\it
a base-$p$
representation of  an output sequence of a truncated congruential generator
over $\Z/p^n$ of a maximum period length is a purely periodic strictly $k$-distributed
sequence over $\Z/p$ of period length exactly $p^nk$}.

The second assertion for this generator holds whenever $2+p^k>kp^{n-k}$;
thus, {\it  one could truncate $\le\big(\frac{n}{2}-\log_p\frac{n}{2}\big)$ lower order digits 
without affecting property Q1}.

All these statements could be proved by slight modifications of the
proof of theorem \ref{thm:distr}. We omit details.
\end{note}


\section
{Some cryptanalysis}
\label{sec:Predict} 
A main goal of this section is to demonstrate
that with the use of constructions described in Section \ref{sec:Constr}
it is possible to design stream ciphers such that the problem of their
key recovery is intractable up to some plausible conjectures. 

%

Consider a ``known plaintext" attack. That is, a cryptanalyst obtains
a plaintext and a corresponding encrypted text and tries to recover a key. 
Since the encryption with
stream cipher is just bitwise XORing of a plaintext with a binary
output sequence of a generator, a cryptanalyst obtains an output sequence
and try to recover a key. Note that the constructions we considered above
enables one to make both the initial state, state transition function and
output function to be key-dependent, so in general a cryptanalst has to
recover a key from a known recurrence sequence $\{y_s,y_{s+1},\ldots\}$
that corresponds to the recurrence law $x_{i+1}=f_i(x_i)\bmod 2^n$, 
$y_{i+1}=g_i(x_i)$. Thus, in general a cryptoanalyst has to recover an
initial state $x_0$, a family of state transition functions $\{f_j\}$,
a family of output functions $\{g_j\}$, and the order these state transition
and output functions
are used while producing the output sequence.

Of course, an analysis in such a general form is senseless. On the one
hand it is obvious that
nothing can be recovered in case $f_i$ and $g_i$ are arbitrary mappings
that satisfy conditions of \ref{cor:WP}, and no extra information is known
to a cryptoanalyst.
On the other hand, it is obvious that there exist degenerate cases
that everything can be easily recovered even without extra information
available. 

For instance, let $m=4k-1$;   put $f_i(x)=x+1$ if $i\in\{0,1,\ldots,m-1\}$ is odd, and
put $f_i(x)=1\oplus(x+1)$ for even $i\in\{0,1,\ldots,m-1\}$. Let
all $g_i=\lfloor\frac{x}{2}\rfloor\bmod 2^n$ be truncations of the least significant
bit. Note that this case satisfies conditions of \ref{cor:WP}; thus, the
corresponding output sequence modulo $2^n$ is purely periodic of period
length $2^nm$, and each element of $\Z/2^n$ occurs at the period exactly
twice. Yet the structure of the output sequence is so specific (exact description
of it could easily be obtained by a reader) that it is absolutely no problem
to break such a scheme.

Thus, one can say nothing definite on  how strong
are generators considerd in the paper against even a single attack without
considering a concrete scheme. We are not going to study concrete schemes in
this paper, yet we demonstrate by a corresponding example that among the
generators we study there exist ones that are provably strong
against certain attacks, say, against a known plaintext attack.

To describe such an example we have to make some preliminary assumptions. 
Choose
(randomly and independently) $k$ Boolean polynomials 
$$\psi_i(\chi_0,\ldots,\chi_{n-1}) \qquad (i=0,1,\ldots,k-1)$$
in $n$ Boolean variables $\chi_0,\ldots,\chi_{n-1}$ each, such that the
number of non-zero monomials in each $\psi_i$ is a polynomial in $n$ ($k$
could
be fixed, or could be a polynomial in $n$ either). Consider
a mapping $F\:\Z/2^n\>\Z/2^k$ defined by
$$F(\chi_0,\ldots,\chi_{n-1})=\psi_0(\chi_0,\ldots,\chi_{n-1})+\dots
+\psi_{k-1}(\chi_0,\ldots,\chi_{n-1})2^{k-1},$$
where $\chi_j=\delta_j(x)$ for $x\in\Z/2^n$. We conjecture that {\slshape 
this function $F$ could be considered as one-way}, that is, 
one could invert it (i.e., find an $F$-preimage in case it exists)
only with negligible in $n$
probability. Note that to
find any $F$-preimage, i.e. to solve an equation $F(x)=y$ in unknown $x$
one has to solve a system of $k$ Boolean equations in $n$ variables.
However, {\slshape to determine whether a given system of $k$ Boolean polynomials
in $n$ variables have a common zero is an $NP$-complete problem}, 
see e.g. \cite[Appendix
A, Section A7.2, Problem ANT-9]{GJ}. So, at our view, the conjecture  that 
the function $F$ is one-way is as plausible as the one concerning any other
``candidate to one-wayness" (for the short list of the latter see e.g.
\cite{Gold}): Nobody today can solve a system of Boolean equations even
if it is known that a solution exists (unless the system is of some special
form).

Proceeding with this plausible conjecture, 
to each Boolean polynomial $\psi_i$, $i=0,1,2,\ldots,k-1$ 
we relate a mapping $\Psi_i\:\Z_2\>\Z_2$
in the following way: $\Psi_i(x)=\psi_i(\delta_0(x),\ldots,\delta_{n-1}(x))\in\{0,1\}
\subset\Z_2$. 
Now to each above mapping $F$ we relate a mapping 
$$f_F(x)=(1+x)\oplus(2^{n+1}\Psi_0(x)+2^{n+2}\Psi_1(x)+\dots+ 2^{n+k}\Psi_{k-1}(x))$$
of $\Z_2$ onto itself. 

By the way, despite it is not very important, note that this mapping is a
composition of bitwise logical and arithmetic operations: To a monomial
$\chi_{r_1}\cdots\chi_{r_s}$, where $r_1,\ldots,r_s\in\{0,1,\ldots,n-1\}$,
$r_1<\ldots<r_s$ we relate a binomial coefficient $\binom{x}{2^{r_1}+\cdots+ 2^{r_s}}$,
then to a Boolean polynomial we relate a sum of corresponding binomial
coefficients. For instance, to the Boolean polynomial $\psi=1+\chi_0+\chi_0\chi_1+
\chi_1\chi_3$ 
we relate an integer valued polynomial $1+x+\binom{x}{3}+\binom{x}{10}$.
Since 
$$\binom{x}{2^{r_1}+\cdots +2^{r_s}}\equiv \delta_{r_1}(x)\cdots\delta_{r_s}(x)\pmod
2$$
in view of Lucas' congruence\footnote{$\binom{n}{m}\equiv\binom{n_0}{m_0}\cdots
\binom{n_s}{m_s}\pmod p$, where $n=n_0+\cdots+n_sp^s$, $m=m_0+\cdots+m_sp^s$
are base-$p$ expansions of, respectively, $n$ and $m$; $p$ prime.},  $\Psi_j(x)\equiv P_j(x)\pmod 2$, where $P_j(x)$
is a polynomial over a field of rational integers $\Q$ 
that corresponds to the Boolean polynomial $\psi_j$
in the above sence. Thus, $\Psi_j(x)=P_j(x)\AND 1$, and the result follows.

Clearly,
$$
\delta_j(f_F(x))=\begin{cases}
1\oplus\delta_0(x),\qquad\text{if $j=0$;}\\
\delta_j(x)\oplus\delta_0(x)\cdots\delta_{j-1}(x),\qquad\text{if $0<j\le n$;}\\
\delta_j(x)\oplus\delta_0(x)\cdots\delta_{j-1}(x)\oplus
\psi_{j-n-1}(\delta_0(x),\dots,\delta_{n-1}(x)),\text{otherwise.}
\end{cases}$$
In view of \ref{ergBool} the mapping $f_F\:\Z_2\>\Z_2$ is compatible and
ergodic for any choice of Boolean polynomials $\psi_0,\ldots,\psi_{k-1}$.

Consider a truncated congruential generator 
$$\mathfrak F=\langle\Z/2^{n+k+1},\Z/2^k,f_F\bmod 2^{n+k+1},g, x_0\rangle,$$ 
where 
$g(x)=\lfloor\frac{x}{2^{n+1}}\rfloor\bmod2^k$, a truncation of $n+1$ low
order bits of $x$. Since the state transition function is transitive and
the output function is equiprobable, the output sequence of this generator
is purely periodic with period length exactly $2^{n+k+1}$, and each element
of $\Z/2^k$ occurs at the period exactly $2^{n+1}$ times.

Let $x_0\in\{0,1,\ldots,2^n-1\}$ be a key; in other words, the key length
of a stream cipher is $n$, and we always take a key $z\in\{0,1,\ldots,2^n-1\}$
as an initial state (a seed). Thus, senior $k+1$ bits of an initial state are
always zero. The key $z$ is the only information that is not known to a
cryptanalyst. Everything else, i.e., $n$, $k$, $f_F$, and $g$ are known,
as well as  the first $m$ members of the output sequence $\{y_i\}$ of the automaton.


Since $\delta_0(x)\cdots\delta_{j-1}(x)=1$ iff $x\equiv -1\pmod
{2^j}$,  the first $m$ members of the output sequence  with probability  $1-\epsilon$
(where  $\epsilon$ is negligible if $m$ is a polynomial in $n$) are:
\begin{align*}
&y_{0}=\Psi_0(z)+2\Psi_1(z)+\dots+ 2^{k-1}\Psi_{k-1}(z)=F(z);\\
&\ldots\ \ldots\ \ldots\ \ldots\ \ldots\ \ldots\ \ldots\ \ldots\ \ldots\\
&y_{m-1}=\Psi_0(z+m-1)+\dots+ 2^{k-1}\Psi_{k-1}(z+m-1)=F(z+m-1).
\end{align*}
To find $z$ a cryptanalist may solve any of the above equations; he could
do it with negligible probability of success, since $F$ is one-way. On
the other hand, an assumption that a cryptanalist could find $z$ with non-negligible probability
means that he could invert $F$ with non-negligible probability
(see the first of the above equations). This contradicts our conjecture
that $F$ is one-way. Thus, the problem of key recovery of this scheme is
intractable up to the conjecture that $F$ is one-way.
\begin{note*} This construction could be extended to counter-dependent generators
in an obvious way. We also note that the restriction the state transition
function of the above generator is $1+x$ modulo $2^{n+1}$ 
is imposed 
only to make the idea of the construction more transparent: 
It is possible to construct a corresponding
stream cipher, which is provably secure against a known plaintext attack,
without this assumption.
\end{note*}

\end{document}